\documentclass[12pt,draftclsnofoot,onecolumn]{IEEEtran}

\usepackage{cite}

\ifCLASSINFOpdf
\usepackage[pdftex]{graphicx}
\usepackage{epstopdf}
\else
\usepackage[dvips]{graphicx}
\fi

\usepackage{etoolbox}
\usepackage{float}
\usepackage{array}
\usepackage{amsmath}

\usepackage[subrefformat=parens]{subfig}

\usepackage{bm}
\usepackage{amssymb}
\usepackage{multirow}
\usepackage{multicol}
\usepackage{xtab}
\usepackage{tabularx}
\usepackage{color}
\usepackage{dsfont}
\usepackage{optidef}
\usepackage{textgreek}

\usepackage{dsfont}
\usepackage{textcomp}

\usepackage{amsthm}

\usepackage[nolist]{acronym}
\makeatletter
\newif\if@in@acrolist
\AtBeginEnvironment{acronym}{\@in@acrolisttrue}
\newrobustcmd{\LU}[2]{\if@in@acrolist#1\else#2\fi}
\newcommand{\ACF}[1]{{\@in@acrolisttrue\acf{#1}}}
\makeatother

\usepackage[allbordercolors={0 1 0},bookmarksopen=true,bookmarksnumbered=true]{hyperref}

\begin{document}
	
\title{A Tb/s Indoor MIMO Optical Wireless Backhaul System Using VCSEL Arrays}
	
\author{Hossein~Kazemi,~\IEEEmembership{Member,~IEEE,}
	Elham~Sarbazi,~\IEEEmembership{Member,~IEEE,}
	Mohammad~Dehghani~Soltani,
	Taisir~E.~H.~El-Gorashi,
	Jaafar~M.~H.~Elmirghani,~\IEEEmembership{Fellow,~IEEE,}
	Richard~V.~Penty,~\IEEEmembership{Senior Member,~IEEE,}
	Ian~H.~White,
	Majid~Safari,~\IEEEmembership{Senior Member,~IEEE,}
	and~Harald~Haas,~\IEEEmembership{Fellow,~IEEE}
	\thanks{This work was financially supported by the Engineering and Physical Research Council (EPSRC) under grant EP/S016570/1 `Terabit Bidirectional Multi-User Optical Wireless System (TOWS) for 6G LiFi'. This work was partially presented at the IEEE 31st Annual International Symposium on Personal, Indoor and Mobile Radio Communications (PIMRC), Aug.--Sep. 2020 \cite{Kazemi7}.\\
	\indent H. Kazemi (corresponding author), E. Sarbazi and H. Haas are with the LiFi Research and Development Center (LRDC), Department of Electronic and Electrical Engineering, University of Strathclyde, Glasgow G1 1RD, UK (email: \{h.kazemi; e.sarbazi; harald.haas\}@strath.ac.uk). M. Dehghani Soltani and M. Safari are with the Institute for Digital Communications, University of Edinburgh, Edinburgh EH9 3FD, UK (email: \{m.dehghani; majid.safari\}@ed.ac.uk). T. E. H. El-Gorashi and J. M. H. Elmirghani are with the School of Electronic and Electrical Engineering, University of Leeds, Leeds LS2 9JT, UK (email: \{j.m.h.elmirghani; t.e.h.elgorashi\}@leeds.ac.uk). R. V. Penty and I. H. White are with the Department of Engineering, University of Cambridge, Cambridge CB2 1PZ, UK (email: \{rvp11; ihw3\}@cam.ac.uk).}
}

\maketitle

\vspace{-1.55cm}

\begin{abstract}
In this paper, the design of a multiple-input multiple-output (MIMO) optical wireless communication (OWC) link based on vertical cavity surface emitting laser (VCSEL) arrays is systematically carried out with the aim to support data rates in excess of $1$ Tb/s for the backhaul of sixth generation (6G) indoor wireless networks. The proposed design combines direct current optical orthogonal frequency division multiplexing (DCO-OFDM) and a spatial multiplexing MIMO architecture. For such an ultra-high-speed line-of-sight (LOS) OWC link with low divergence laser beams, maintaining alignment is of high importance. In this paper, two types of misalignment error between the transmitter and receiver are distinguished, namely, radial displacement error and orientation angle error, and they are thoroughly modeled in a unified analytical framework assuming Gaussian laser beams, resulting in a generalized misalignment model (GMM). The derived GMM is then extended to MIMO arrays and the performance of the MIMO-OFDM OWC system is analyzed in terms of the aggregate data rate. Novel insights are provided into the system performance based on computer simulations by studying various influential factors such as beam waist, array configuration and different misalignment errors, which can be used as guidelines for designing short range Tb/s MIMO OWC systems.
\end{abstract}


\begin{acronym}[]
	\acro{2D}{\LU{T}{t}wo \LU{D}{d}imensional}
	\acro{3D}{\LU{T}{t}hree \LU{D}{d}imensional}
	\acro{5G}{\LU{F}{f}ifth \LU{G}{g}eneration}
	\acro{6G}{\LU{S}{s}ixth \LU{G}{g}eneration}
	\acro{AP}{\LU{A}{a}ccess \LU{P}{p}oint}
	\acro{AEL}{\LU{A}{a}llowable \LU{E}{e}xposure \LU{L}{l}imit}
	\acro{AF}{\LU{A}{a}mplify-and-\LU{F}{f}orward}
	\acro{ACO{-}OFDM}{\LU{A}{a}symmetrically \LU{C}{c}lipped \LU{O}{o}ptical OFDM}
	\acro{AOA}{\LU{A}{a}ngle-\LU{O}{o}f-\LU{A}{a}rrival}
	\acro{APC}{\LU{A}{a}daptive \LU{P}{p}ower \LU{C}{c}ontrol}
	\acro{AR}{\LU{A}{a}ugmented \LU{R}{r}eality}
	\acro{ARPC}{\LU{A}{a}verage \LU{R}{r}ate \LU{P}{p}ower \LU{C}{c}ontrol}
	\acro{ASE}{\LU{A}{a}rea \LU{S}{s}pectral \LU{E}{e}fficiency}
	\acro{ASPC}{\LU{A}{a}verage SINR \LU{P}{p}ower \LU{C}{c}ontrol}
	\acro{AWGN}{\LU{A}{a}dditive \LU{W}{w}hite Gaussian \LU{N}{n}oise}
	\acro{AWGR}{\LU{A}{a}rrayed \LU{W}{w}aveguide \LU{G}{g}rating \LU{R}{r}outer}
	\acro{BBO}{\LU{B}{b}ackhaul \LU{B}{b}ottleneck \LU{O}{o}ccurrence}
	\acro{BER}{\LU{B}{b}it \LU{E}{e}rror \LU{R}{r}atio}
	\acro{BPSK}{\LU{B}{b}inary \LU{P}{p}hase \LU{S}{s}hift \LU{K}{k}eying}
	\acro{BS}{\LU{B}{b}ase \LU{S}{s}tation}
	\acro{CBS}{\LU{C}{c}ell-based \LU{B}{b}andwidth \LU{S}{s}cheduling}
	\acro{CCDF}{\LU{C}{c}omplementary \LU{C}{c}umulative \LU{D}{d}istribution \LU{F}{f}unction}
	\acro{CCI}{\LU{C}{c}o-\LU{C}{c}hannel \LU{I}{i}nterference}
	\acro{CDF}{\LU{C}{c}umulative \LU{D}{d}istribution \LU{F}{f}unction}
	\acro{CFFR}{\LU{C}{c}ooperative FFR}
	\acro{CLT}{\LU{C}{c}entral \LU{L}{l}imit \LU{T}{t}heorem}
	\acro{CoMP}{\LU{C}{c}oordinated \LU{M}{m}ulti-\LU{P}{p}oint}
	\acro{CP}{\LU{C}{c}yclic \LU{P}{p}refix}
	\acro{CSI}{\LU{C}{c}hannel \LU{S}{s}tate \LU{I}{i}nformation}
	\acro{CPC}{\LU{C}{c}ompound \LU{P}{p}arabolic \LU{C}{c}oncentrator}
	\acro{DAS}{\LU{D}{d}istributed \LU{A}{a}ntenna \LU{S}{s}ystem}
	\acro{DSL}{\LU{D}{d}igital \LU{S}{s}ubscriber \LU{L}{l}ine}
	\acro{DC}{\LU{D}{d}irect \LU{C}{c}urrent}
	\acro{DCO{-}OFDM}{\LU{D}{d}irect \LU{C}{c}urrent-biased \LU{O}{o}ptical \LU{O}{o}rthogonal \LU{F}{f}requency \LU{D}{d}ivision \LU{M}{m}ultiplexing}
	\acro{DDO{-}OFDM}{\LU{D}{d}irect \LU{D}{d}etection \LU{O}{o}ptical OFDM}
	\acro{DF}{\LU{D}{d}ecode-and-\LU{F}{f}orward}
	\acro{DMT}{\LU{D}{d}iscrete \LU{M}{m}ultitone}
	\acro{DSP}{\LU{D}{d}igital \LU{S}{s}ignal \LU{P}{p}rocessing}
	\acro{DWDM}{\LU{D}{d}ense \LU{W}{w}avelength \LU{D}{d}ivision \LU{M}{m}ultiplexing}
	\acro{EMI}{\LU{E}{e}lectromagnetic \LU{I}{i}nterference}
	\acro{eU{-}OFDM}{\LU{E}{e}nhanced \LU{U}{u}nipolar OFDM}
	\acro{FDE}{\LU{F}{f}requency \LU{D}{d}omain \LU{E}{e}qualization}
	\acro{FEC}{\LU{F}{f}orward \LU{E}{e}rror \LU{C}{c}orrection}
	\acro{FF}{\LU{F}{f}ill \LU{F}{f}actor}
	\acro{FFR}{\LU{F}{f}ractional \LU{F}{f}requency \LU{R}{r}euse}
	\acro{FFT}{\LU{F}{f}ast Fourier \LU{T}{t}ransform}
	\acro{FOV}{\LU{F}{f}ield \LU{O}{o}f \LU{V}{v}iew}
	\acro{FPC}{\LU{F}{f}ixed \LU{P}{p}ower \LU{C}{c}ontrol}
	\acro{FR}{\LU{F}{f}ull \LU{R}{r}euse}
	\acro{FRF}{\LU{F}{f}requency \LU{R}{r}euse \LU{F}{f}actor}
	\acro{FR{-}VL}{\LU{F}{f}ull \LU{R}{r}euse \LU{V}{v}isible \LU{L}{l}ight}
	\acro{FSO}{\LU{F}{f}ree \LU{S}{s}pace \LU{O}{o}ptical}
	\acro{FTTB}{\LU{F}{f}iber-\LU{T}{t}o-\LU{T}{t}he-\LU{B}{b}uilding}
	\acro{FTTH}{\LU{F}{f}iber-\LU{T}{t}o-\LU{T}{t}he-\LU{H}{h}ome}
	\acro{FTTP}{\LU{F}{f}iber-\LU{T}{t}o-\LU{T}{t}he-\LU{P}{p}remises}
	\acro{GMM}{\LU{G}{g}eneralized \LU{M}{m}isalignment \LU{M}{m}odel}
	\acro{IB{-}VL}{\LU{I}{i}n-\LU{B}{b}and \LU{V}{v}isible \LU{L}{l}ight}
	\acro{ICI}{\LU{I}{i}nter-\LU{C}{c}ell \LU{I}{i}nterference}
	\acro{IM{-}DD}{\LU{I}{i}ntensity \LU{M}{m}odulation and \LU{D}{d}irect \LU{D}{d}etection}
	\acro{i.i.d.}{\LU{I}{i}ndependent and \LU{I}{i}dentically \LU{D}{d}istributed}
	\acro{IFFT}{\LU{I}{i}nverse \LU{F}{f}ast Fourier \LU{T}{t}ransform}
	\acro{IoT}{Internet of Things}
	\acro{IR}{\LU{I}{i}nfrared}
	\acro{ISI}{\LU{I}{i}nter-\LU{S}{s}ymbol \LU{I}{i}nterference}
	\acro{JTDF}{\LU{J}{j}oint \LU{T}{t}ransmission with \LU{D}{d}ecode-and-\LU{F}{f}orward}
	\acro{LAN}{\LU{L}{l}ocal \LU{A}{a}rea \LU{N}{n}etwork}
	\acro{LED}{\LU{L}{l}ight \LU{E}{e}mitting \LU{D}{d}iode}
	\acro{LiFi}{\LU{L}{l}ight \LU{F}{f}idelity}
	\acro{LOS}{\LU{L}{l}ine-\LU{O}{o}f-\LU{S}{s}ight}
	\acro{LPF}{\LU{L}{l}ow \LU{P}{p}ass \LU{F}{f}ilter}
	\acro{LSSB}{\LU{L}{l}ower \LU{S}{s}ingle-\LU{S}{s}ide \LU{B}{b}and}
	\acro{LTE}{\LU{L}{l}ong-\LU{T}{t}erm \LU{E}{e}volution}
	\acro{MAC}{\LU{M}{m}edium \LU{A}{a}ccess \LU{C}{c}ontrol}
	\acro{MC}{\LU{M}{u}ulti-\LU{C}{c}arrier}
	\acro{MIMO}{\LU{M}{m}ultiple-\LU{I}{i}nput \LU{M}{m}ultiple-\LU{O}{o}utput}
	\acro{MHP}{\LU{M}{m}ost \LU{H}{h}azardous \LU{P}{p}osition}
	\acro{MPE}{\LU{M}{m}aximum \LU{P}{p}ermissible \LU{E}{e}xposure}
	\acro{MSE}{\LU{M}{m}ean \LU{S}{s}quare \LU{E}{e}rror}
	\acro{MSPC}{\LU{M}{m}aximum SINR \LU{P}{p}ower \LU{C}{c}ontrol}
	\acro{MMSE}{\LU{M}{m}inimum \LU{M}{m}ean \LU{S}{s}quare \LU{E}{e}rror}
	\acro{mmWave}{\LU{M}{m}illimeter \LU{W}{w}ave}
	\acro{NPC}{\LU{N}{n}o \LU{P}{p}ower \LU{C}{c}ontrol}
	\acro{LAN}{\LU{L}{l}ocal \LU{A}{a}rea \LU{N}{n}etwork}
	\acro{NLOS}{\LU{N}{n}on-\LU{L}{l}ine-\LU{O}{o}f-\LU{S}{s}ight}
	\acro{NMSE}{\LU{N}{n}ormalized \LU{M}{m}ean \LU{S}{s}quare \LU{E}{e}rror}
	\acro{NODF}{\LU{N}{n}on-\LU{O}{o}rthogonal \LU{D}{d}ecode-and-\LU{F}{f}orward}
	\acro{OFDM}{\LU{O}{o}rthogonal \LU{F}{f}requency \LU{D}{d}ivision \LU{M}{m}ultiplexing}
	\acro{OFDMA}{\LU{O}{o}rthogonal \LU{F}{f}requency \LU{D}{d}ivision \LU{M}{m}ultiple \LU{A}{a}ccess}
	\acro{OOK}{\LU{O}{o}n-\LU{O}{o}ff \LU{K}{k}eying}
	\acro{OWC}{\LU{O}{o}ptical \LU{W}{w}ireless \LU{C}{c}ommunication}
	\acro{PAM}{\LU{P}{p}ulse \LU{A}{a}mplitude \LU{M}{m}odulation}
	\acro{PAPR}{\LU{P}{p}eak-to-\LU{A}{a}verage \LU{P}{p}ower \LU{R}{r}atio}
	\acro{PD}{\LU{P}{p}hotodetector}
	\acro{PDF}{\LU{P}{p}robability \LU{D}{d}ensity \LU{F}{f}unction}
	\acro{PE}{\LU{P}{p}ower \LU{E}{e}fficiency}
	\acro{PHY}{\LU{P}{p}hysical \LU{L}{l}ayer}
	\acro{PLC}{\LU{P}{p}ower \LU{L}{l}ine \LU{C}{c}ommunication}
	\acro{PMF}{\LU{P}{p}robability \LU{M}{m}ass \LU{F}{f}unction}
	\acro{PoE}{\LU{P}{p}ower-over-Ethernet}
	\acro{P{-}OFDM}{\LU{P}{p}olar OFDM}
	\acro{PON}{\LU{P}{p}assive \LU{O}{o}ptical \LU{N}{n}etwork}
	\acro{PPP}{\LU{P}{p}oisson \LU{P}{p}oint \LU{P}{p}rocess}
	\acro{PSD}{\LU{P}{p}ower \LU{S}{s}pectral \LU{D}{d}ensity}
	\acro{PTP}{\LU{P}{p}oint-\LU{T}{t}o-\LU{P}{p}oint}
	\acro{PTMP}{\LU{P}{p}oint-\LU{T}{t}o-\LU{M}{m}ulti-\LU{P}{p}oint}
	\acro{QAM}{\LU{Q}{q}uadrature \LU{A}{a}mplitude \LU{M}{m}odulation}
	\acro{QoS}{\LU{Q}{q}uality of \LU{S}{s}ervice}
	\acro{QPSK}{\LU{Q}{q}uadrature \LU{P}{p}hase \LU{S}{s}hift \LU{K}{k}eying}
	\acro{RF}{\LU{R}{r}adio \LU{F}{f}requency}
	\acro{RGB}{\LU{R}{r}ed-\LU{G}{g}reen-\LU{B}{b}lue}
	\acro{RHS}{\LU{R}{r}ight \LU{H}{h}and \LU{S}{s}ide}
	\acro{RIN}{\LU{R}{r}elative \LU{I}{i}ntensity \LU{N}{n}oise}
	\acro{RMS}{\LU{R}{r}oot \LU{M}{m}ean \LU{S}{s}quare}
	\acro{RoF}{\LU{R}{r}adio-over-\LU{F}{f}iber}
	\acro{RTP}{\LU{R}{r}elative \LU{T}{t}otal \LU{P}{p}ower}
	\acro{SC}{\LU{S}{s}ingle \LU{C}{c}arrier}
	\acro{SE}{\LU{S}{s}pectral \LU{E}{e}fficiency}
	\acro{SEE{-}OFDM}{\LU{S}{s}pectral and \LU{E}{e}nergy \LU{E}{e}fficient OFDM}
	\acro{SINR}{\LU{S}{s}ignal-to-\LU{I}{i}nterference-plus-\LU{N}{n}oise \LU{R}{r}atio}
	\acro{SISO}{\LU{S}{s}ingle \LU{I}{i}nput \LU{S}{s}ingle \LU{O}{o}utput}
	\acro{SM}{\LU{S}{s}patial \LU{M}{m}ultiplexing}
	\acro{SMF}{\LU{S}{s}ingle \LU{M}{m}ode \LU{F}{f}iber}
	\acro{SNR}{\LU{S}{s}ignal-to-\LU{N}{n}oise \LU{R}{r}atio}
	\acro{SSB{-}OFDM}{\LU{S}{s}ingle-\LU{S}{s}ide \LU{B}{b}and OFDM}
	\acro{SVD}{\LU{S}{s}ingular \LU{V}{v}alue \LU{D}{d}ecomposition}
	\acro{TEM}{\LU{T}{t}ransverse \LU{E}{e}lectromagnetic \LU{M}{m}ode}
	\acro{TIA}{\LU{T}{t}ransimpedance \LU{A}{a}mplifier}
	\acro{UB}{\LU{U}{u}nlimited \LU{B}{b}ackhaul}
	\acro{UBS}{\LU{U}{u}ser-based \LU{B}{b}andwidth \LU{S}{s}cheduling}
	\acro{UE}{\LU{U}{u}ser \LU{E}{e}quipment}
	\acro{UHD}{\LU{U}{u}ltra-\LU{H}{h}igh-\LU{D}{d}efinition}
	\acro{USSB}{\LU{U}{u}pper \LU{S}{s}ingle-\LU{S}{s}ide \LU{B}{b}and}
	\acro{VCSEL}{\LU{V}{v}ertical \LU{C}{c}avity \LU{S}{s}urface \LU{E}{e}mitting \LU{L}{l}aser}
	\acro{VPPM}{\LU{V}{v}ariable \LU{P}{p}ulse \LU{P}{p}osition \LU{M}{m}odulation}
	\acro{VL}{\LU{V}{v}isible \LU{L}{l}ight}
	\acro{VLC}{\LU{V}{v}isible \LU{L}{l}ight \LU{C}{c}ommunication}
	\acro{VR}{\LU{V}{v}irtual \LU{R}{r}eality}
	\acro{WDM}{\LU{W}{w}avelength \LU{D}{d}ivision \LU{M}{m}ultiplexing}
	\acro{WiFi}{\LU{W}{w}ireless \LU{F}{f}idelity}
	\acro{WLAN}{\LU{W}{w}ireless \LU{L}{l}ocal \LU{A}{a}rea \LU{N}{n}etwork}
	\acro{WOC}{\LU{W}{w}ireless \LU{O}{o}ptical \LU{C}{c}ommunication}
\end{acronym}

\section{Introduction}\label{sec1}
The proliferation of Internet-enabled premium services such as 4K and 8K \ac{UHD} video streaming, immersion into \ac{VR} or \ac{AR} with \ac{3D} stereoscopic vision, holographic telepresence and multi-access edge computing will extremely push wireless connectivity limits in years to come \cite{Giordani}. These technologies will require an unprecedented system capacity above $1$ Tb/s for real-time operation, which is one of the key performance indicators of the future \ac{6G} wireless systems \cite{Calvanese}.

The achievability of ultra-high transmission rates of Tb/s has been addressed in the literature for both wired and wireless systems \cite{Shrestha,Idler,Petrov}. Targeting \ac{DSL} applications, in \cite{Shrestha}, Shrestha \textit{et al.} have used a two-wire copper cable as a multi-mode waveguide for \ac{MIMO} transmission and experimentally measured the received power for signals with $200$ GHz bandwidth. They have predicted that aggregate data rates of several Tb/s over a twisted wire pair are feasible at short distances of $\leq10$~m by using \ac{DMT} modulation and vector coding. In \cite{Idler}, the authors have reported successful implementation of a $1$ Tb/s super channel over a $400$ km optical \ac{SMF} link based on \ac{QAM} with probabilistic constellation shaping. In \cite{Petrov}, Petrov \textit{et al.} have elaborated on a roadmap to actualize last meter indoor broadband wireless access in the terahertz band, i.e. $0.1$--$10$ THz, in order to enable Tb/s connectivity between the wired backbone infrastructure and personal wireless devices.

The feasibility of Tb/s data rates has been actively studied for outdoor point-to-point \ac{FSO} communications \cite{Ciaramella,Parca,Poliak1}. In \cite{Ciaramella}, Ciaramella \textit{et al.} have presented experimental results for a terrestrial \ac{FSO} link achieving a net data rate of $1.28$ Tb/s over a distance of $212$ m using $32$ \ac{WDM} channels centered at $1550$ nm and direct detection. In \cite{Parca}, Parca \textit{et al.} have demonstrated a transmission rate of $1.6$ Tb/s over a hybrid fiber and \ac{FSO} system with a total distance of $4080$ m based on polarization multiplexing, $16$ \ac{WDM} channels and coherent detection. In \cite{Poliak1}, Poliak \textit{et al.} have set up a field experiment emulating the uplink \ac{FSO} transmission in ground-to-geostationary satellite communications under adverse atmospheric turbulence conditions, corroborating a throughput of $1.72$ Tb/s over $10.45$ km with the aid of $40$ \ac{WDM} \ac{OOK} channels and active beam tracking at the receiver.

Beam-steered \ac{IR} light communication is an \ac{OWC} technology primarily tailored to provide multi-Gb/s data rates per user by directing narrow \ac{IR} laser beams at mobile devices. In \cite{Koonen3}, Koonen \textit{et al.} have proposed a wavelength-controlled beam steering technique by using a \ac{2D} high port-count \ac{AWGR} array and a lens. The authors have demonstrated $112$ Gb/s $4$-\ac{PAM} transmission with $2.5$ m reach based on an $80$-ports \ac{AWGR}, anticipating a total throughput of about $8.9$ Tb/s when all the beams are in use. In \cite{Sun}, Sun \textit{et al.} have proposed an alternative optical design employing a fiber port array and a transceiver lens to provide full beam coverage within the desired area without beam steering. The authors have shown that this system supports $10$ Gb/s single-user transmission rates and over $1$ Tb/s multi-user sum rates.

Indoor laser-based optical wireless access networks can generate aggregate data rates beyond $1$ Tb/s \cite{Sarbazi1}. Such ultra-high-speed indoor access networks impose a substantial overhead on the backhaul capacity, and a cost-effective backhaul solution is a major challenge. In this paper, a high-capacity wireless backhaul system is designed based on laser-based \ac{OWC} to support aggregate data rates of at least $1$ Tb/s for backhaul connectivity in next generation Tb/s indoor networks. While \ac{FSO} systems suffer from outdoor channel impairments such as weather-dependent absorption loss and atmospheric turbulence, short range laser-based \ac{OWC} under stable and acclimatized conditions of indoor environments potentially enhances the signal quality. This way, the need for bulky \ac{FSO} transceivers equipped with expensive subsystems to counteract outdoor effects is eliminated. Moreover, the aforementioned \ac{FSO} systems use \ac{DWDM} to deliver Tb/s data rates, which significantly increases the cost and complexity of the front-end system.

Different from \ac{WDM} \ac{FSO} systems, in this paper, a single wavelength is used to achieve a data rate of $\geq1$ Tb/s by means of \acp{VCSEL}. The choice of \acp{VCSEL} for the optical wireless system design is motivated by the fact that, among various types of laser diodes, \acp{VCSEL} are one of the strongest contenders to fulfil this role due to several important features of them, including \cite{KIga,Larsson}: 1) a high modulation bandwidth of $\geq10$ GHz; 2) a high power conversion efficiency of $>50\%$; 3) cost-efficient fabrication by virtue of their compatibility with large scale integration processes; 4) possibility for multiple devices to be densely packed and precisely arranged as \ac{2D} arrays. These attributes make \acp{VCSEL} appealing to many applications such as optical networks, highly parallel optical interconnects and laser printers, to name a few \cite{Liu}. Single mode \acp{VCSEL}, which are the focus of this paper, generate an output optical field in the fundamental \ac{TEM} (i.e. $\text{TEM}_{00}$ mode), resulting in a Gaussian profile on the transverse plane, in that the optical power is maximum at the center of the beam spot and it decays exponentially with the squared radial distance from the center \cite{Saleh}.

For \ac{LOS} \ac{OWC} links, accurate alignment between the transmitter and receiver is a determining factor of the system performance and reliability. In principle, two types of misalignment may occur in the link: radial displacement between the transmitter and receiver, and orientation angle error at the transmitter or receiver side. Modeling of the Gaussian beam misalignment has been addressed in the context of terrestrial \ac{FSO} systems such as the works of Farid and Hranilovic, for \ac{SISO} \cite{Farid1} and \ac{MIMO} \cite{Farid2} links. The \ac{FSO} transceiver equipment is commonly installed on the rooftops of high-rise buildings and hence random building sways due to wind loads and thermal expansions cause a pointing error in the transmitter orientation angle with independent and identical random components in elevation and horizontal directions \cite{Arnon}. The works in \cite{Farid1,Farid2,Arnon} implicitly base their modeling methodology on the assumption of treating the effect of this angle deviation at the transmitter (with a typical value of $1$ mrad) as a radial displacement of the beam spot position at the receiver (typically located at $1$ km distance from the transmitter). By contrast, in \cite{Huang}, Huang and Safari, through applying a small angle approximation, have modeled the receiver-induced \ac{AOA} misalignment again as a radial displacement of the optical field pattern on the \ac{PD} plane. In \cite{Poliak2,Poliak3}, Poliak \textit{et al.} have presented a link budget model for \ac{FSO} systems in an effort to incorporate misalignment losses for Gaussian beams individually, including lateral displacement, tilt of the transmitter and tilt of the receiver. Nonetheless, the effect of these tilt angles has been simplified by a lateral displacement. In \cite{Azzolin}, Azzolin \textit{et al.} have proposed an exact analytical solution in terms of the Marcum Q-function for the geometric and misalignment loss in \ac{SISO} \ac{FSO} links in the presence of only radial displacement of the Gaussian beam spot at the circular aperture of the receiver.

Although the problem of misalignment constitutes an important practical challenge for designing Tb/s indoor optical wireless links as a key enabler for ultra-high-speed networks of the future, it has not been widely studied yet. For short range indoor \ac{OWC} systems with compact \acp{PD}, to minimize the geometric loss, the beam spot size is required to be relatively small, comparable to the size of a \ac{PD}, in which case angular misalignment can significantly influence the link performance, independent of the radial displacement error. In a previous work \cite{Kazemi7}, the authors have presented preliminary results to study the effect of only displacement error on the performance of indoor Tb/s \ac{MIMO} \ac{OWC}. However, such high performance indoor \ac{OWC} links are essentially prone to any type of misalignment errors. To the best of the authors' knowledge, there is a lack of a comprehensive and analytically tractable model of the link misalignment for laser-based \ac{OWC} systems inclusive of orientation angle errors at the transmitter and receiver sides as well as the radial displacement error. This paper puts forward the modeling and design of a spatial multiplexing \ac{MIMO} \ac{OWC} system based on \ac{DCO{-}OFDM} and \ac{VCSEL} arrays to unlock Tb/s data rates with single mode \acp{VCSEL}. The contributions of this paper are concisely given as follows:

\begin{itemize}
	\item An in-depth analytical modeling of the misalignment for \ac{SISO} optical wireless channels with Gaussian beams is presented. Thereupon, a \ac{GMM} is derived where radial displacement and orientation angle errors at both the transmitter and receiver sides are all taken into consideration in a unified manner. This model is verified by computer simulations using a commercial optical design software, Zemax OpticStudio.
	\item The \ac{GMM} derivation is extended to \ac{MIMO} \ac{OWC} systems with arbitrary configurations for transmitter and receiver arrays. The geometric modeling of the \ac{VCSEL} and \ac{PD} arrays is explicated by highlighting critical design parameters such as array size, spacing between array elements and the size of \acp{PD}.
	\item The spatial multiplexing \ac{MIMO}-OFDM transceiver under consideration is elucidated and the received \ac{SINR} and aggregate data rate are analyzed. It is shown that treating an angular pointing error of the transmitter as a radial displacement is a special case of the \ac{GMM} and a tight analytical approximation of the \ac{MIMO} channel \ac{DC} gains is derived for this case.
\end{itemize}

The remainder of the paper is organized as follows. In Section~\ref{sec2}, the \ac{SISO} channel model for a perfectly aligned \ac{VCSEL}-based \ac{OWC} system is described. In Section~\ref{sec3}, the detailed analytical modeling framework for the generalized misalignment of the \ac{SISO} channel is established. In Section~\ref{sec4}, the design and analysis of the \ac{MIMO}-OFDM \ac{OWC} system using \ac{VCSEL} and \ac{PD} arrays is presented, including the incorporation of the \ac{GMM} in the \ac{MIMO} channel model. In Section~\ref{sec5}, numerical results are provided. In Section~\ref{sec6}, concluding remarks are drawn and a number of possible directions are suggested for the future research.

\begin{figure}[!t]
	\centering
	\includegraphics[width=0.55\linewidth]{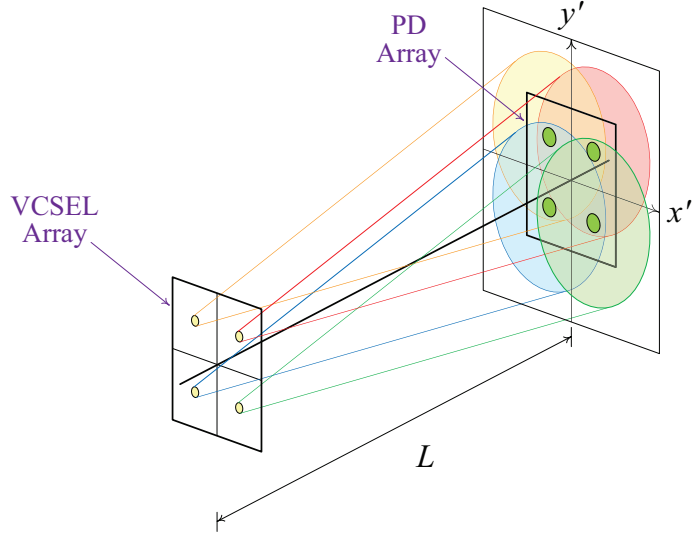}
	\caption{A $4\times4$ OWC system using a $2\times2$ VCSEL array and a $2\times2$ PD array.}
	\label{fig_3_1}
	\vspace{-20pt}
\end{figure}

\section{VCSEL-based Optical Wireless Channel}\label{sec2}
A \ac{MIMO} \ac{OWC} system can be realized by means of an array of \acp{VCSEL}. Fig.~\ref{fig_3_1} illustrates a $4\times4$ \ac{MIMO} system configuration, comprising a $2\times2$ \ac{VCSEL} array and a $2\times2$ \ac{PD} array, which are perfectly aligned to each other. There is a directed \ac{LOS} link from every \ac{VCSEL} to its corresponding \ac{PD} as represented by an exclusive color. This section deals with the channel model for a \ac{VCSEL}-based \ac{SISO} optical wireless link as a building block for the \ac{MIMO} system.

\subsection{Gaussian Beam Propagation}
The wavefront of the Gaussian beam is initially planar at the beam waist and then expanding in the direction of propagation. The wavefront radius of curvature at distance $z$ from the transmitter is characterized by \cite{Saleh}:
\begin{equation}\label{eq_2_1}
R(z) = z\left[1+\left(\frac{\pi w_0^2}{\lambda z}\right)^{\!2}\right],
\end{equation}
where $w_0$ is the waist radius; and $\lambda$ is the laser wavelength. Also, the radius of the beam spot, which is measured at the $\frac{1}{e^2}$ normalized intensity contour on the transverse plane, takes the following form \cite{Saleh}:
\begin{equation}\label{eq_2_2}
w(z) = w_0\sqrt{1+\left(\frac{z}{z_\mathrm{R}}\right)^{\!2}},
\end{equation}
where $z_\mathrm{R}$ is the Rayleigh range. It is defined as the distance at which the beam radius is extended by a factor of $\sqrt{2}$, i.e. $w(z_\mathrm{R})=\sqrt{2}w_0$. In this case, the beam spot has an area twice that of the beam waist. The Rayleigh range is related to $w_0$ and $\lambda$ via \cite{Saleh}:
\begin{equation}\label{eq_2_3}
z_\mathrm{R} = \frac{\pi w_0^2}{\lambda}.
\end{equation}
From \eqref{eq_2_2}, $w(z)$ for $z\gg z_\mathrm{R}$ approaches the asymptotic value:
\begin{equation}\label{eq_2_4}
w(z) \approx \frac{w_0z}{z_\mathrm{R}} = \frac{\lambda z}{\pi w_0},
\end{equation}
thus varying linearly with $z$. Therefore, the circular beam spot in far field is the base of a cone whose vertex lies at the center of the beam waist with a divergence half-angle:
\begin{equation}\label{eq_2_5}
\theta = \tan^{-1}\left(\frac{w(z)}{z}\right) \approx \frac{w(z)}{z} \approx \frac{\lambda}{\pi w_0}.
\end{equation}

The spatial distribution of a Gaussian beam along its propagation axis is described by the intensity profile on the transverse plane. By using Cartesian coordinates, the intensity distribution at distance $z$ from the transmitter at the point $(x,y)$ is given by \cite{Saleh}:
\begin{equation}\label{eq_2_6}
I(x,y,z) = \frac{2P_\mathrm{t}}{\pi w^2(z)}\exp\left(-\frac{2\rho^2(x,y)}{w^2(z)}\right).
\end{equation}
where $P_\mathrm{t}$ is the transmitted optical power; and $\rho(x,y)$ is the Euclidean distance of the point $(x,y)$ from the center of the beam spot.

\subsection{Channel DC Gain}
As observed from \eqref{eq_2_6}, the fundamental property of the Gaussian beam propagation is that the intensity distribution is given on the transverse plane as a circularly symmetric function centered about the beam axis. For a general link configuration, the transmitter and receiver are oriented toward arbitrary directions in the \ac{3D} space with $\mathbf{n}_\mathrm{t}$ and $\mathbf{n}_\mathrm{r}$ denoting their normal vectors, respectively. In this case, the received optical power $P_\mathrm{r}$ is obtained by integrating \eqref{eq_2_6} over the effective area of the \ac{PD} \cite{Kahn1}. To this end, the \ac{PD} area is projected onto the transverse plane by using the cosine of the normal vector of the \ac{PD} plane with respect to the beam propagation axis, which is equal to $\mathbf{n}_\mathrm{t}\cdot\mathbf{n}_\mathrm{r}$. The \ac{DC} gain of an \ac{IM{-}DD} channel is defined as the ratio of the average optical power of the received signal to that of the transmitted signal. Therefore, the \ac{DC} gain of the channel is calculated as follows:
\begin{equation}\label{eq_2_7}
H_0 = \frac{P_\mathrm{r}}{P_\mathrm{t}} = \iint_{(x,y)\in\mathcal{A}}\frac{2}{\pi w^2(z)}\exp\left(-\frac{2\rho^2(x,y)}{w^2(z)}\right)\mathbf{n}_\mathrm{t}\cdot\mathbf{n}_\mathrm{r}dxdy,
\end{equation}
where $\mathcal{A}$ denotes the \ac{PD} area; and the factor $\mathbf{n}_\mathrm{t}\cdot\mathbf{n}_\mathrm{r}$ accounts for Lambert's cosine law. Throughout the paper, a circular \ac{PD} of radius $r_\mathrm{PD}$ is assumed for which $\mathcal{A}=\{(x,y)\in\mathbb{R}^2|~x^2+y^2\leq r_\mathrm{PD}^2\}$.

\begin{figure}[!t]
	\centering
	\includegraphics[width=0.49\linewidth]{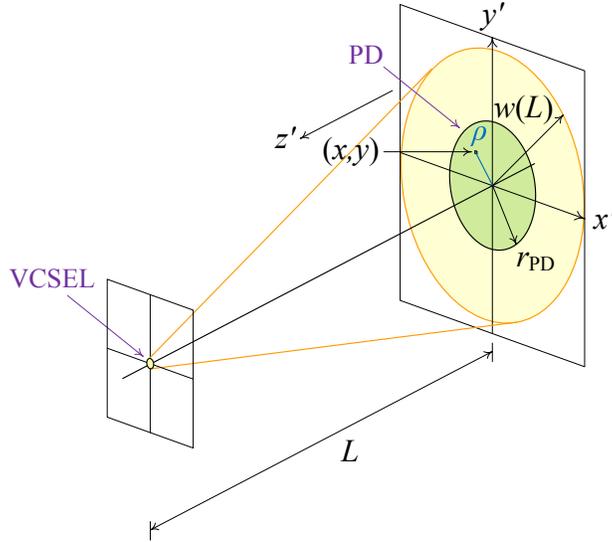}
	\captionsetup{width=0.9\linewidth}
	\caption{Link configuration of the SISO OWC system with perfect alignment.}
	\label{fig1}
	\vspace{-20pt}
\end{figure}

Fig.~\ref{fig1} illustrates a \ac{SISO} \ac{OWC} system in a directed \ac{LOS} configuration with perfect alignment. In this case, the beam waist plane is parallel to the \ac{PD} plane so that $\mathbf{n}_\mathrm{t}\cdot\mathbf{n}_\mathrm{r}=1$ and the center of the beam spot is exactly located at the center of the \ac{PD}. Hence, $\rho(x,y)$ in \eqref{eq_2_6} is equal to $r=\sqrt{x^2+y^2}$ on the \ac{PD} plane. From \eqref{eq_2_7}, for a link distance of $z=L$, the \ac{DC} gain of the channel becomes:
\begin{equation}\label{eq_2_8}
H_0 = \int_{0}^{2\pi}\int_{0}^{r_\mathrm{PD}}\frac{2}{\pi w^2(L)}\exp\left(-\frac{2r^2}{w^2(L)}\right)rdrd\theta = 1-\exp\left(-\frac{2r_\mathrm{PD}^2}{w^2(L)}\right),
\end{equation}
where $w^2(L)=w_0^2\left(1+\frac{L^2}{z_\mathrm{R}^2}\right)$.

\section{Generalized Misalignment Modeling}\label{sec3}
This section establishes a mathematical framework for the analytical modeling of misalignment errors for the \ac{SISO} optical wireless channel discussed in Section~\ref{sec2}. In the following, two cases of displacement error and orientation angle error are presented.

\begin{figure}[!t]
	\centering
	\includegraphics[width=0.49\linewidth]{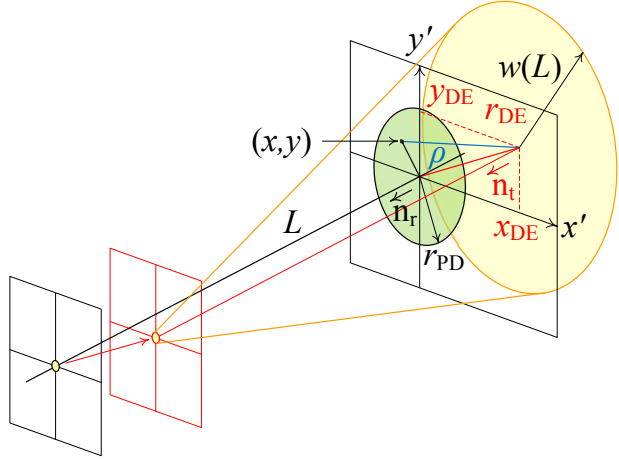}
	\captionsetup{width=0.9\linewidth}
	\caption{Misalignment geometry for the displacement error $r_\mathrm{DE}$ between the transmitter and receiver with components $x_\mathrm{DE}$ and $y_\mathrm{DE}$ along the $x'$ and $y'$ axes.}
	\label{fig5}
	\vspace{-20pt}
\end{figure}

\subsection{Displacement Error}
A displacement error between the transmitter and receiver causes the center of the beam spot at the PD plane to deviate radially, relative to the center of the PD, which is equivalent to the radial displacement in \cite{Farid1}. In this case, $\mathbf{n}_\mathrm{t}\cdot\mathbf{n}_\mathrm{r}=1$. The magnitude of the error is represented by $r_\mathrm{DE}=\sqrt{x_\mathrm{DE}^2+y_\mathrm{DE}^2}$, where $x_\mathrm{DE}$ and $y_\mathrm{DE}$ correspond to the error components along the $x'$ and $y'$ axes, as shown in Fig.~\ref{fig5}. It can be observed that the intensity value depends on the axial distance $z$ between the beam waist and the \ac{PD} plane where $z=L$, and the Euclidean distance $\rho$ from the center of the beam spot to the coordinates $(x,y)$. It follows that:
\begin{equation}\label{DE_eq1}
\rho^2(x,y) = (x-x_\mathrm{DE})^2+(y-y_\mathrm{DE})^2.
\end{equation}
Substituting \eqref{DE_eq1} in \eqref{eq_2_7}, the \ac{DC} gain of the channel turns into:
\begin{equation}\label{DE_eq2}
H_0(x_\mathrm{DE},y_\mathrm{DE}) = \int_{-r_\mathrm{PD}}^{r_\mathrm{PD}}\int_{-\sqrt{r_\mathrm{PD}^2-y^2}}^{\sqrt{r_\mathrm{PD}^2-y^2}}\frac{2}{\pi w^2(L)}\exp\left(-\frac{2\left[(x-x_\mathrm{DE})^2+(y-y_\mathrm{DE})^2\right]}{w^2(L)}\right)dxdy.
\end{equation}

\begin{figure}[!t]
	\centering
	\subfloat[\label{fig3a} Azimuth and elevation angle components $\phi_\mathrm{a}$ and $\phi_\mathrm{e}$ of the orientation angle error $\phi$ at the transmitter side.]{\includegraphics[width=0.49\linewidth]{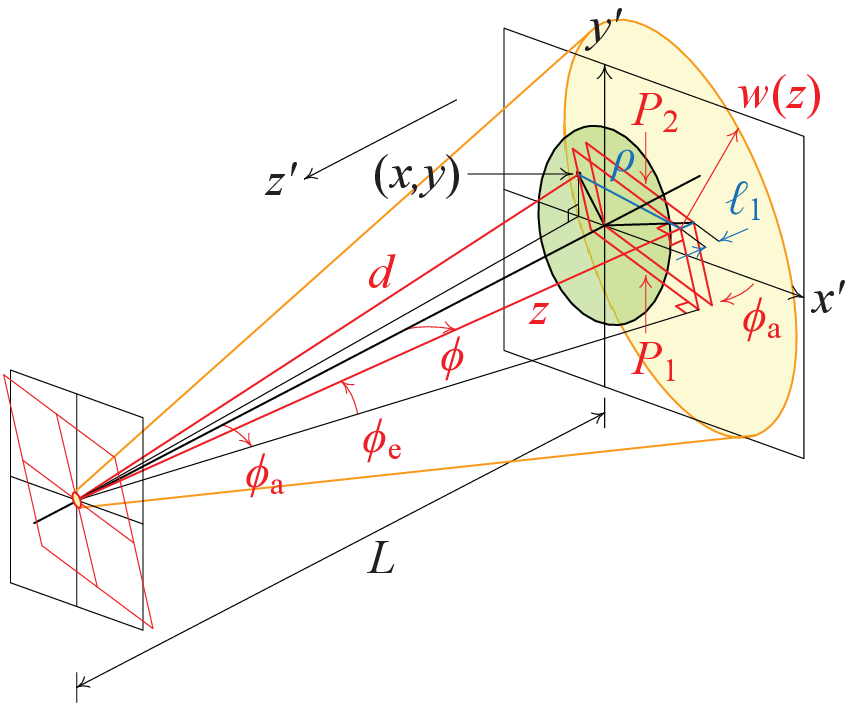}}
	\hspace{5pt}
	\subfloat[\label{fig3b} Azimuth and elevation angle components $\psi_\mathrm{a}$ and $\psi_\mathrm{e}$ of the orientation angle error $\psi$ at the receiver side.]{\includegraphics[width=0.49\linewidth]{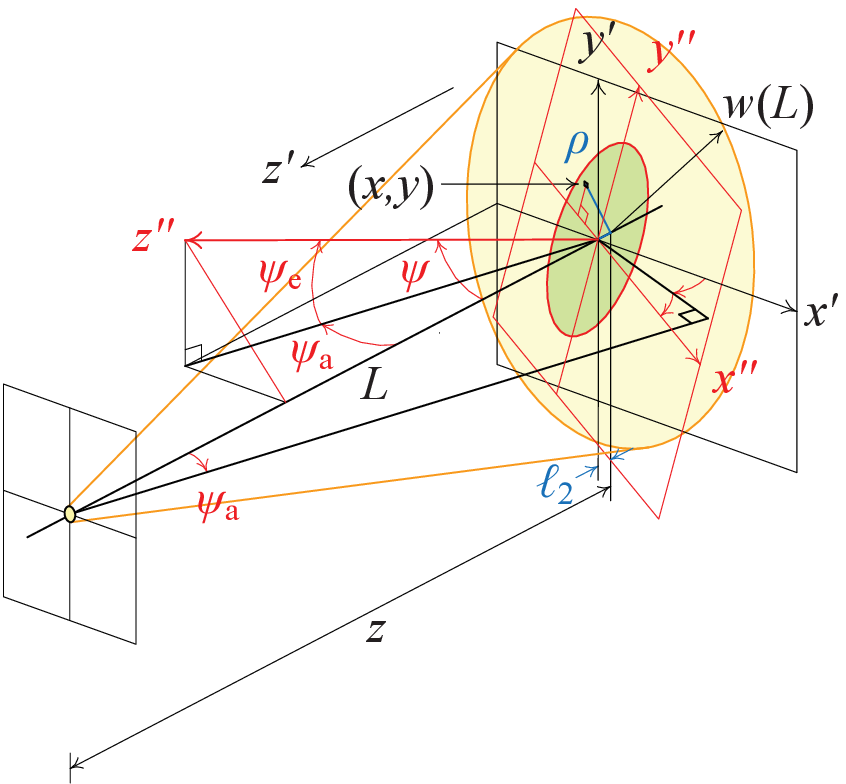}}
	\captionsetup{width=0.9\linewidth}
	\caption{Misalignment geometry for the orientation error at the transmitter and receiver.}
	\label{fig3}
	\vspace{-20pt}
\end{figure}

\subsection{Orientation Angle Error}
An orientation error occurs when the transmitter or receiver has a non-zero tilt angle with respect to the alignment axis. Orientation angles of the transmitter and receiver, denoted by $\phi$ and $\psi$, respectively, entail arbitrary and independent directions in the \ac{3D} space. Note that the transmitter and receiver orientation errors are jointly modeled, though they are separately depicted in Fig.~\ref{fig3} to avoid intricate geometry. Furthermore, the angles $\phi$ and $\psi$ are decomposed into azimuth and elevation components in the \ac{3D} space using the rotation convention shown in Fig.~\ref{fig3}.

The aim is to determine the intensity at a given point $(x,y)$ on the \ac{PD} surface based on \eqref{eq_2_6}. To elaborate, according to Fig.~\ref{fig3}, consider the family of concentric, closed disks perpendicular to the beam axis, with their centers lying on the beam axis. Among them, the one with a circumference intersecting the point $(x,y)$ on the \ac{PD} plane is the basis for analysis. This particular disk is referred to as \textit{principal disk} hereinafter, which is drawn as a yellow disk in Fig.~\ref{fig3}. The variables to be characterized are the axial distance $z$ between the beam waist and the center of the principal disk and the Euclidean distance $\rho$ of the point $(x,y)$ to the beam axis, i.e. the radius of the principal disk.

The \ac{PD}'s $x''y''z''$ coordinate system is rotated with respect to the reference coordinate system $x'y'z'$ as shown in Fig.~\subref*{fig3b}. Based on Euler angles with clockwise rotations, the $x'y'z'$ system is transformed into the $x''y''z''$ system by rotating first about the $y'$ axis through an angle $-\psi_\mathrm{a}$, then about the $x''$ axis through an angle $-\psi_\mathrm{e}$, using the following rotation matrices:
\begin{subequations}\label{OAER_eq8}
\begin{align}
\mathbf{R}_{y'}(\beta) &= 
\left[\begin{matrix}
\cos\beta & 0 & \sin\beta \\
0 & 1 & 0 \\ 
-\sin\beta & 0 & \cos\beta
\end{matrix}\right], \label{OAER_eq8a}\\
\mathbf{R}_{x''}(\alpha) &= 
\left[\begin{matrix}
1 & 0 & 0 \\
0 & \cos\alpha & -\sin\alpha \\
0 & \sin\alpha & \cos\alpha
\end{matrix}\right], \label{OAER_eq8b}
\end{align}
\end{subequations}
for $\beta=-\psi_\mathrm{a}$ and $\alpha=-\psi_\mathrm{e}$. The desired point $[x~y~0]^\top$ is given in the $x''y''z''$ system. The projected coordinates of this point in the $x'y'z'$ system is obtained as:
\begin{equation}\label{OAER_eq5}
\left[\begin{matrix}
u \\ v \\ w
\end{matrix}\right] =
\mathbf{R}_{y'}(-\psi_\mathrm{a})\mathbf{R}_{x''}(-\psi_\mathrm{e})\left[\begin{matrix}
x \\ y \\ 0
\end{matrix}\right] = \left[\begin{matrix}
x\cos(\psi_\mathrm{a})+y\sin(\psi_\mathrm{a})\sin(\psi_\mathrm{e}) \\ y\cos(\psi_\mathrm{e}) \\ x\sin(\psi_\mathrm{a})-y\cos(\psi_\mathrm{a})\sin(\psi_\mathrm{e})
\end{matrix}\right].
\end{equation}

The axial distance $z$ consists of two segments including the projection of $L$ onto the beam axis and the additive length $\ell$:
\begin{equation}\label{OAETR_eq2}
z = L\cos(\phi_\mathrm{e})\cos(\phi_\mathrm{a})+\ell.
\end{equation}
From Fig.~\subref*{fig3a}, there are two parallel planes indicated by $P_1$ and $P_2$ so that the additive length $\ell$ can be found as the distance between $P_1$ and $P_2$, which is subsequently derived in \eqref{OAETR_eq7}. These planes are perpendicular to the beam axis such that $P_1$ passes through the point  $[u~v~w]^\top$ in the $x'y'z'$ system and $P_2$ crosses the origin. The normal vector of both planes is:
\begin{equation}\label{OAET_eq5}
\mathbf{n}_\mathrm{t} = -\cos(\phi_\mathrm{e})\sin(\phi_\mathrm{a})\mathbf{n}_{x'}-\sin(\phi_\mathrm{e})\mathbf{n}_{y'}+\cos(\phi_\mathrm{e})\cos(\phi_\mathrm{a})\mathbf{n}_{z'},
\end{equation}
where $\mathbf{n}_{x'}$, $\mathbf{n}_{y'}$ and $\mathbf{n}_{z'}$ represent unit vectors for $x'$, $y'$ and $z'$ axes, respectively. Let $\mathbf{n}_\mathrm{t}=[a~b~c]^\top$ to simplify notation, where $\|\mathbf{n}_\mathrm{t}\|=\sqrt{a^2+b^2+c^2}=1$. It follows that:
\begin{subequations}\label{OAETR_eq6}
	\begin{align}
	&P_1:\quad a(x'-u)+b(y'-v)+c(z'-w)=0 \label{OAETR_eq6a}\\
	&P_2:\quad ax'+by'+cz'=0 \label{OAETR_eq6b}
	\end{align}
\end{subequations}
The additive length $\ell$ in \eqref{OAETR_eq2} is derived by finding the distance from the origin to $P_1$, resulting in:
\begin{equation}\label{OAETR_eq7}
\ell = -au-bv-cw.
\end{equation}
Combining \eqref{OAETR_eq7} with \eqref{OAER_eq5} and \eqref{OAET_eq5}, and using trigonometric identities, yields:
\begin{equation}\label{OAETR_eq1}
\ell = x\cos(\phi_\mathrm{e})\sin(\phi_\mathrm{a}-\psi_\mathrm{a})+y(\sin(\psi_\mathrm{e})\cos(\phi_\mathrm{e})\cos(\phi_\mathrm{a}-\psi_\mathrm{a})+\cos(\psi_\mathrm{e})\sin(\phi_\mathrm{e})).
\end{equation}

The squared radius of the principal disk illustrated in Fig.~\subref*{fig3a} is given by:
\begin{equation}\label{OAETR_eq4}
\rho^2 = d^2-z^2,
\end{equation}
where:
\begin{equation}\label{OAETR_eq3}
d^2 = (L-w)^2+u^2+v^2.
\end{equation}
Substituting $u$, $v$ and $w$ from \eqref{OAER_eq5} into \eqref{OAETR_eq3}, and simplifying, leads to:
\begin{equation}\label{OAETR_eq9}
d^2 = L^2+x^2+y^2+2L(-x\sin(\psi_\mathrm{a})+y\cos(\psi_\mathrm{a})\sin(\psi_\mathrm{e})).
\end{equation}
The last piece required to complete the analysis of the channel gain based on \eqref{eq_2_7} is the calculation of the inner product of $\mathbf{n}_\mathrm{t}$ and $\mathbf{n}_\mathrm{r}$. From Fig.~\subref*{fig3b}, the normal vector to the \ac{PD} surface is:
\begin{equation}\label{OAER_eq6}
\mathbf{n}_\mathrm{r} = -\cos(\psi_\mathrm{e})\sin(\psi_\mathrm{a})\mathbf{n}_{x'}+\sin(\psi_\mathrm{e})\mathbf{n}_{y'}+\cos(\psi_\mathrm{e})\cos(\psi_\mathrm{a})\mathbf{n}_{z'}.
\end{equation}
By using \eqref{OAET_eq5} and \eqref{OAER_eq6}, the cosine of the planar angle between the surface normal and the beam axis is obtained as follows:
\begin{equation}\label{OAETR_eq8}
\mathbf{n}_\mathrm{t}\cdot\mathbf{n}_\mathrm{r} = \cos(\phi_\mathrm{e})\cos(\psi_\mathrm{e})\cos(\phi_\mathrm{a}-\psi_\mathrm{a})-\sin(\phi_\mathrm{e})\sin(\psi_\mathrm{e}).
\end{equation}
By combining \eqref{eq_2_2}, \eqref{OAETR_eq2}, \eqref{OAETR_eq1}, \eqref{OAETR_eq4} and \eqref{OAETR_eq9}, the \ac{DC} gain of the channel, denoted by $H_0(\phi_\mathrm{a},\phi_\mathrm{e},\psi_\mathrm{a},\psi_\mathrm{e})$, can be evaluated based on \eqref{eq_2_7} and \eqref{OAETR_eq8} when using:
\begin{equation}\label{OAETR_eq5}
\begin{aligned}
w^2(z) = w_0^2\Big(1+z_\mathrm{R}^{-2}\big[&L\cos(\phi_\mathrm{e})\cos(\phi_\mathrm{a})+x\cos(\phi_\mathrm{e})\sin(\phi_\mathrm{a}-\psi_\mathrm{a})+\\
&y(\sin(\psi_\mathrm{e})\cos(\phi_\mathrm{e})\cos(\phi_\mathrm{a}-\psi_\mathrm{a})+\cos(\psi_\mathrm{e})\sin(\phi_\mathrm{e}))\big]^2\Big),
\end{aligned}
\end{equation}
\begin{equation}\label{OAETR_eq10}
\begin{aligned}
\rho^2(x,y) =\ &L^2+x^2+y^2+2L(-x\sin(\psi_\mathrm{a})+y\cos(\psi_\mathrm{a})\sin(\psi_\mathrm{e}))-\big[L\cos(\phi_\mathrm{e})\cos(\phi_\mathrm{a})+\\
&x\cos(\phi_\mathrm{e})\sin(\phi_\mathrm{a}-\psi_\mathrm{a})+y(\sin(\psi_\mathrm{e})\cos(\phi_\mathrm{e})\cos(\phi_\mathrm{a}-\psi_\mathrm{a})+\cos(\psi_\mathrm{e})\sin(\phi_\mathrm{e}))\big]^2.
\end{aligned}
\end{equation}

\subsection{Unified Misalignment Model}\label{sec3_3}
In order to unify displacement and orientation errors, after the transmitter is rotated, it is shifted to the point $[x_\mathrm{DE}~y_\mathrm{DE}~L]^\top$ in the $x'y'z'$ system. Referring to the parallel planes $P_1$ and $P_2$ in \eqref{OAETR_eq6}, $P_2$ now intersects the point $[x_\mathrm{DE}~y_\mathrm{DE}~0]^\top$. Therefore, $\ell=-au'-bv'-cw'$ from \eqref{OAETR_eq7}, such that $u'=u-x_\mathrm{DE}$, $v'=v-y_\mathrm{DE}$ and $w'=w$. Consequently, the squared radius of the principal disk is determined by using \eqref{OAETR_eq4} in conjunction with \eqref{OAETR_eq2} and $d^2=(L-w')^2+{u'}^2+{v'}^2$ from \eqref{OAETR_eq3}. Altogether, the generalized channel gain $H_0(x_\mathrm{DE},y_\mathrm{DE},\phi_\mathrm{a},\phi_\mathrm{e},\psi_\mathrm{a},\psi_\mathrm{e})$ is computed based on \eqref{eq_2_7} and \eqref{OAETR_eq8} through the use of:
\begin{equation}\label{OAETR_eq11}
\begin{aligned}
w^2(z) = w_0^2\Big(1+z_\mathrm{R}^{-2}\big[&L\cos(\phi_\mathrm{e})\cos(\phi_\mathrm{a})+x\cos(\phi_\mathrm{e})\sin(\phi_\mathrm{a}-\psi_\mathrm{a})+\\
&y(\sin(\psi_\mathrm{e})\cos(\phi_\mathrm{e})\cos(\phi_\mathrm{a}-\psi_\mathrm{a})+\cos(\psi_\mathrm{e})\sin(\phi_\mathrm{e}))-\\
&x_\mathrm{DE}\cos(\phi_\mathrm{e})\sin(\phi_\mathrm{a})-y_\mathrm{DE}\sin(\phi_\mathrm{e})\big]^2\Big),
\end{aligned}
\end{equation}
\begin{equation}\label{OAETR_eq12}
\begin{aligned}
\rho^2&(x,y) = (L-x\sin(\psi_\mathrm{a})+y\cos(\psi_\mathrm{a})\sin(\psi_\mathrm{e}))^2+(x\cos(\psi_\mathrm{a})+y\sin(\psi_\mathrm{a})\sin(\psi_\mathrm{e})-x_\mathrm{DE})^2+\\
&(y\cos(\psi_\mathrm{e})-y_\mathrm{DE})^2-\big[L\cos(\phi_\mathrm{e})\cos(\phi_\mathrm{a})+x\cos(\phi_\mathrm{e})\sin(\phi_\mathrm{a}-\psi_\mathrm{a})+\\
&y(\sin(\psi_\mathrm{e})\cos(\phi_\mathrm{e})\cos(\phi_\mathrm{a}-\psi_\mathrm{a})+\cos(\psi_\mathrm{e})\sin(\phi_\mathrm{e}))-x_\mathrm{DE}\cos(\phi_\mathrm{e})\sin(\phi_\mathrm{a})-y_\mathrm{DE}\sin(\phi_\mathrm{e})\big]^2.
\end{aligned}
\end{equation}

\begin{figure}[!t]
	\centering
	\subfloat[$K\times K$ VCSEL array]
	{\includegraphics[width=0.45\textwidth]{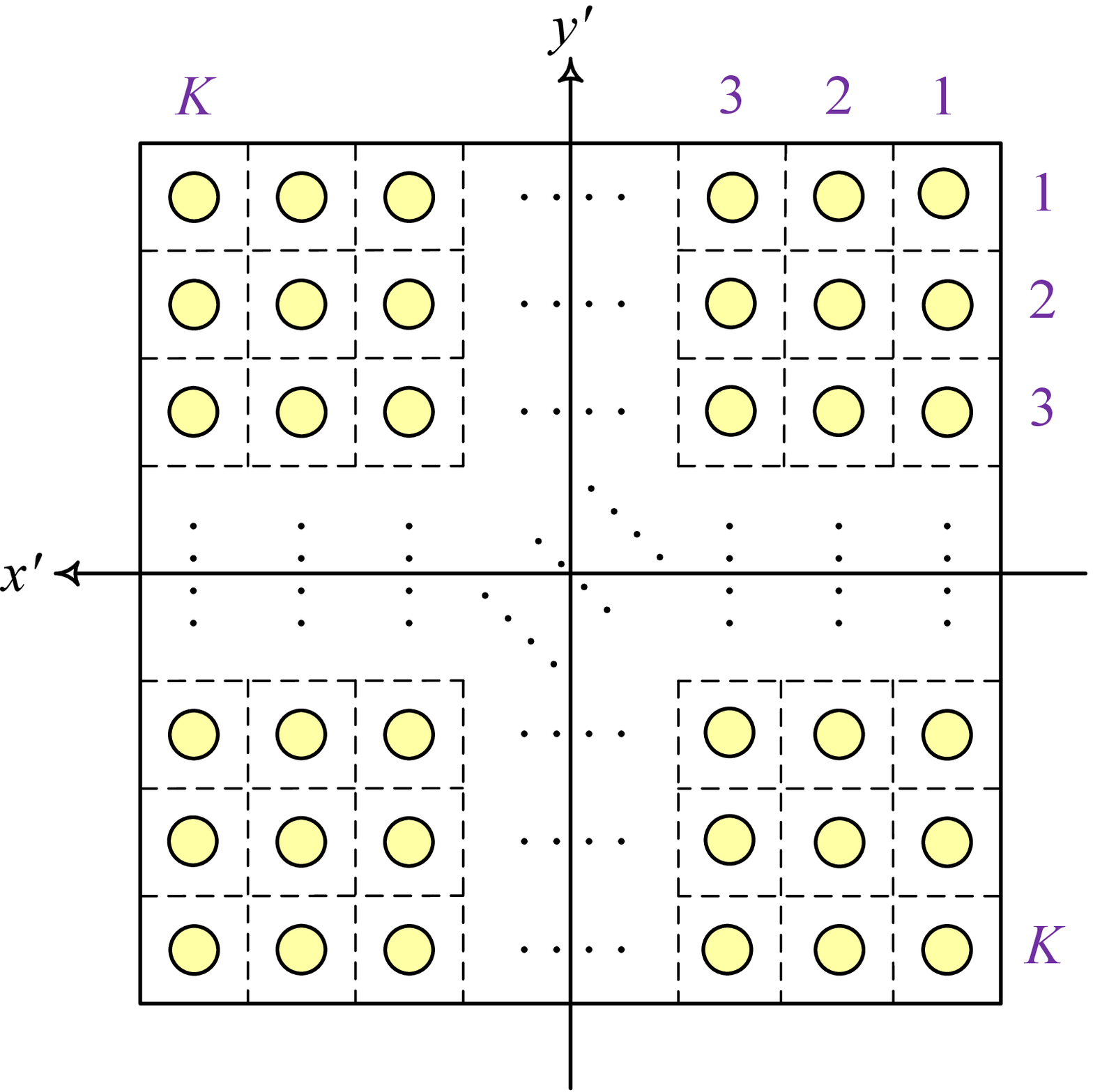}\label{fig_3_2a}}
	\hspace{5pt}
	\subfloat[$K\times K$ PD array]
	{\includegraphics[width=0.45\textwidth]{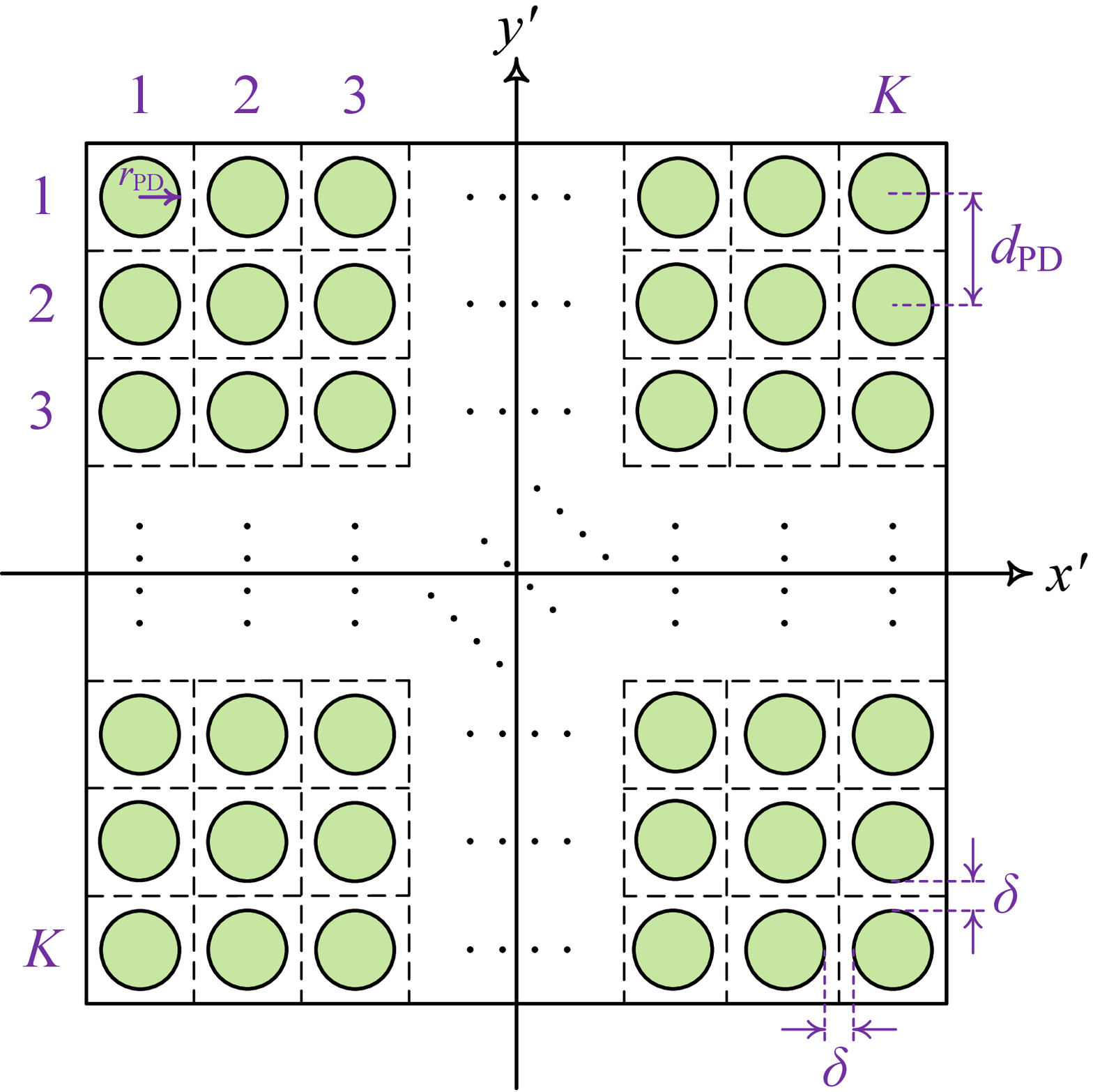}\label{fig_3_2b}}
	\caption{Structure of a $K\times K$ VCSEL array and a $K\times K$ PD array, forming an $N_\mathrm{t}\times N_\mathrm{r}$ MIMO OWC system where $N_\mathrm{t}=N_\mathrm{r}=K^2$.}
	\label{fig_3_2}
	\vspace{-20pt}
\end{figure}

\section{MIMO OWC System Using VCSEL Arrays}\label{sec4}

\subsection{Structure of Arrays and MIMO Channel}\label{sec4_1}
The array structure shown in Fig.~\ref{fig_3_1} is extended to a $K\times K$ square, forming an $N_\mathrm{t}\times N_\mathrm{r}$ \ac{MIMO} \ac{OWC} system where\footnote{The assumption of $N_\mathrm{t}=N_\mathrm{r}$ is only used for convenience of the presentation and it is not a necessary requirement. In fact, for the same array structure shown in Fig.~\ref{fig_3_2}, the receiver array can be designed such that $N_\mathrm{r}\geq N_\mathrm{t}$ as discussed in Section~\ref{sec5}.} $N_\mathrm{t}=N_\mathrm{r}=K^2$. Fig.~\ref{fig_3_2} depicts a $K\times K$ \ac{VCSEL} array and a $K\times K$ \ac{PD} array on the $x'y'$ plane. The gap between adjacent elements of the \ac{PD} array is controlled by $\delta>0$, which is referred to as \textit{inter-element spacing} hereinafter. For those \acp{PD} that are close to the edges of the array, there is a margin of $\frac{\delta}{2}$ with respect to the edges. The center-to-center distance for neighboring \acp{PD} along rows or columns of the array is:
\begin{equation}\label{eq_3_6}
d_\mathrm{PD}=2r_\mathrm{PD}+\delta.
\end{equation}
The side length for each array is $W=Kd_\mathrm{PD}$, leading to array dimensions of $W\times W$.

The \ac{MIMO} channel is identified by an $N_\mathrm{r}\times N_\mathrm{t}$ matrix of \ac{DC} gains for all transmission paths between the transmitter and receiver arrays:
\begin{equation}\label{eq_3_1}
\mathbf{H}_0 = 
\begin{bmatrix}
H_{11} & H_{12} & \cdots & H_{1N_\mathrm{t}} \\
H_{21} & H_{22} & \cdots & H_{2N_\mathrm{t}} \\
\vdots & \vdots & \ddots & \vdots \\
H_{N_\mathrm{r}1} & H_{N_\mathrm{r}2} & \cdots & H_{N_\mathrm{r}N_\mathrm{t}}
\end{bmatrix},
\end{equation}
where the entry $H_{ij}$ corresponds to the link from $\text{VCSEL}_j$ to $\text{PD}_i$. For the array structure shown in Fig.~\ref{fig_3_2}, the elements are labeled by using a \textit{single index} according to their row and column indices. This way, for an $N_\mathrm{r}\times N_\mathrm{t}=K^2\times K^2$ array, the \ac{VCSEL} (resp. \ac{PD}) situated at the $(m,n)$th entry of the matrix for $m,n\in\{1,2,\dots,K^2\}$ is denoted by $\text{VCSEL}_i$ (resp. $\text{PD}_i$) where $i=(m-1)K+n$. Let $[\check{x}_i~\check{y}_i~\check{z}_i]^\top$ and $[\hat{x}_i~\hat{y}_i~\hat{z}_i]^\top$ be the coordinates of the $i$th element of the \ac{VCSEL} and \ac{PD} arrays, respectively, in the $x'y'z'$ system, for $i\in\{1,2,\dots,K^2\}$. Under perfect alignment, $\check{x}_i=\hat{x}_i=x_i$, $\check{y}_i=\hat{y}_i=y_i$, $\check{z}_i=L$ and $\hat{z}_i=0$. Here, $(x_i,y_i)$ are \ac{2D} coordinates of the $i$th element on each array. From Fig.~\ref{fig_3_2}, it is straightforward to show that:
\begin{subequations}\label{eq_3_8}
	\begin{align}
	x_i &= \left(-\frac{K-1}{2}+n-1\right)d_\mathrm{PD}, \\
	y_i &= \left(\frac{K-1}{2}-m+1\right)d_\mathrm{PD},
	\end{align}
\end{subequations}
where $m=\lceil\frac{i}{K}\rceil$ and $n=i-\left(\lceil\frac{i}{K}\rceil-1\right)K$, with $\lceil q\rceil$ denoting the smallest integer that satisfies $\lceil q\rceil\geq q$. In this case, evaluating $H_{ij}$ based on \eqref{eq_2_7} leads to:
\begin{equation}\label{eq_3_3}
H_{ij} = \int_{-r_\mathrm{PD}}^{r_\mathrm{PD}}\int_{-\sqrt{r_\mathrm{PD}^2-y^2}}^{\sqrt{r_\mathrm{PD}^2-y^2}}\frac{2}{\pi w^2(L)}\exp\left(-2\frac{(x-x_i+x_j)^2+(y-y_i+y_j)^2}{w^2(L)}\right)dxdy.
\end{equation}

\begin{figure}[!t]
	\noindent
	\centering
	\includegraphics[width=0.85\linewidth]{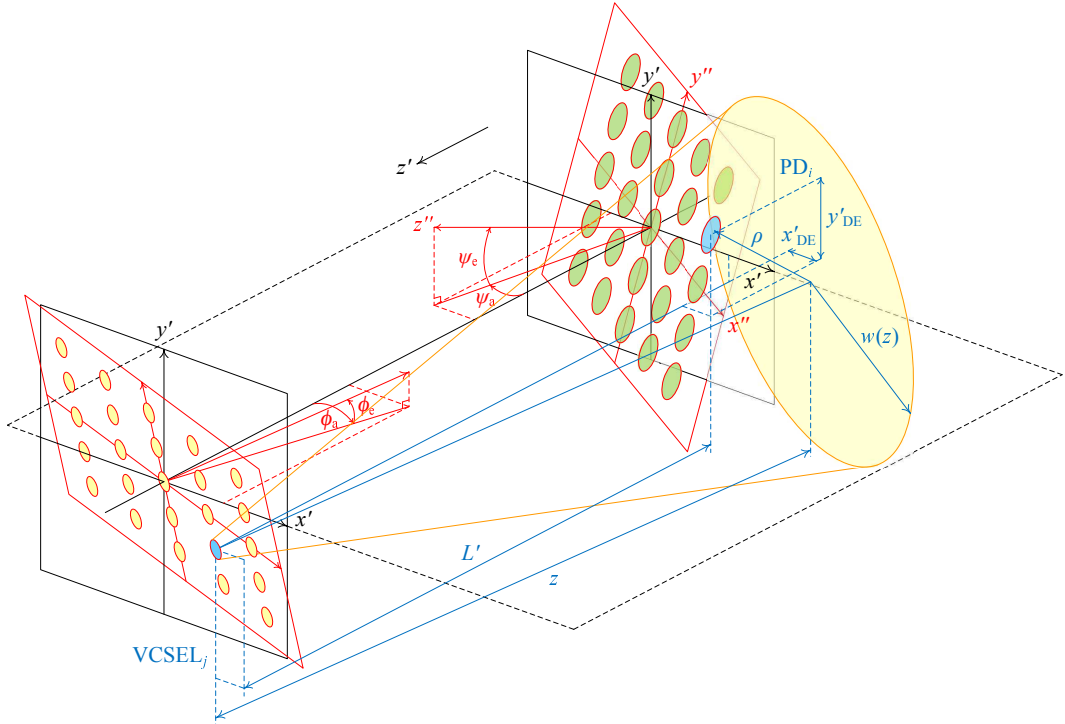}
	\caption{Orientation angle errors of the transmitter and receiver in a MIMO system configuration. The parameters of the LOS channel between $\text{VCSEL}_j$ and $\text{PD}_i$ are highlighted in blue.}
	\label{fig_3_4}
\end{figure}

\subsection{Generalized Misalignment of Arrays}\label{sec4_2}
Under the generalized misalignment, the whole transmitter and receiver arrays are affected by both displacement and orientation errors. To characterize the \ac{MIMO} channel with the generalized misalignment, the entries of the \ac{MIMO} channel matrix can be derived by applying the procedure described in Section~\ref{sec3_3} for the \ac{SISO} channel\footnote{In consideration of the receiver optics, a non-imaging optical design based on ideal \acp{CPC} can be used. Such a receiver structure would not alter the spatial distribution of the beam spots at the receiver array, since an ideal lossless \ac{CPC} transfers all the optical power collected at the entrance aperture to the exit aperture \cite{Kahn1}. Therefore, the analytical framework proposed for the generalized misalignment modeling in Section~\ref{sec3} would still be valid for computing the optical power incident on each receiver element due to the fact that all the integrations can be equally evaluated over the circular area of the entrance aperture of \acp{CPC}. The optimal design of non-imaging receivers is, however, beyond the scope of this paper and it is subject to a separate study.}. Fig.~\ref{fig_3_4} depicts the \ac{MIMO} link configuration where the transmitter and receiver arrays are rotated according to their respective orientation angles. Note that Fig.~\ref{fig_3_4} does not include radial displacement for simplicity. The corresponding parameters of the \ac{SISO} link between $\text{VCSEL}_j$ and $\text{PD}_i$ are highlighted in blue, as shown in Fig.~\ref{fig_3_4}. Considering the generalized misalignment, the \ac{VCSEL} array is first rotated by an angle $\phi$ and then its center is radially displaced relative to the center of the receiver array. The coordinates of $\text{VCSEL}_j$ in the reference coordinate system $x'y'z'$ are:
\begin{equation}\label{eq_3_9}
	\left[\begin{matrix}
		\check{x}_j \\ \check{y}_j \\ \check{z}_j
	\end{matrix}\right] =
	\mathbf{R}_{y'}(-\phi_\mathrm{a})\mathbf{R}_{x''}(\phi_\mathrm{e})\left[\begin{matrix}
		x_j \\ y_j \\ 0
	\end{matrix}\right]+
	\left[\begin{matrix}
		x_\mathrm{DE} \\ y_\mathrm{DE} \\ L
	\end{matrix}\right],
\end{equation}
where $\mathbf{R}_{y'}(-\phi_\mathrm{a})$ and $\mathbf{R}_{x''}(\phi_\mathrm{e})$ are given by \eqref{OAER_eq8} for $\beta=-\phi_\mathrm{a}$ and $\alpha=\phi_\mathrm{e}$. Also, after the receiver array undergoes a rotation by an angle $\psi$, the coordinates of $\text{PD}_i$ in the $x'y'z'$ system are:
\begin{equation}\label{eq_3_10}
	\left[\begin{matrix}
		\hat{x}_i \\ \hat{y}_i \\ \hat{z}_i
	\end{matrix}\right] =
	\mathbf{R}_{y'}(-\psi_\mathrm{a})\mathbf{R}_{x''}(-\psi_\mathrm{e})\left[\begin{matrix}
		x_i \\ y_i \\ 0
	\end{matrix}\right],
\end{equation}
where $\mathbf{R}_{y'}(-\psi_\mathrm{a})$ and $\mathbf{R}_{x''}(\psi_\mathrm{e})$ are given by \eqref{OAER_eq8} for $\beta=-\psi_\mathrm{a}$ and $\alpha=-\psi_\mathrm{e}$. To calculate the channel gain between $\text{VCSEL}_j$ and $\text{PD}_i$, denoted by $H_{ij}(x_\mathrm{DE},y_\mathrm{DE},\phi_\mathrm{a},\phi_\mathrm{e},\psi_\mathrm{a},\psi_\mathrm{e})$, using \eqref{eq_2_7}, the parameters $w(z)$ and $\rho(x,y)$ are evaluated based on \eqref{OAETR_eq11} and \eqref{OAETR_eq12} by substituting the link distance $L'=\check{z}_j-\hat{z}_i$ for $L$, and the displacement components $x'_\mathrm{DE}=\check{x}_j-\hat{x}_i$ and $y'_\mathrm{DE}=\check{y}_j-\hat{y}_i$ for $x_\mathrm{DE}$ and $y_\mathrm{DE}$, respectively. This exact procedure is referred to as the \ac{MIMO} \ac{GMM} for brevity.

\subsection{Approximation of the MIMO GMM}\label{sec4_3}
The computation of the \ac{MIMO} \ac{GMM} as described above entails numerical integrations. In the following, approximate analytical expressions of the \ac{MIMO} channel gain $H_{ij}$ are derived for two special cases of radial displacement and orientation error at the transmitter. Then, the relation between them for a small angle error is elaborated. The area of a circular \ac{PD} of radius $r_\mathrm{PD}$ is approximated by an equivalent square of side length $a_\mathrm{PD}=\sqrt{\pi}r_\mathrm{PD}$ with the same area.
\subsubsection{Radial Displacement}
In this case, $\check{x}_j-\hat{x}_i=x_j+x_\mathrm{DE}-x_i$, $\check{y}_j-\hat{y}_i=y_j+y_\mathrm{DE}-y_i$ and $\check{z}_j-\hat{z}_i=L$. Therefore, $z=L$, $\mathbf{n}_\mathrm{t}\cdot\mathbf{n}_\mathrm{r}=1$ and: 
\begin{equation}\label{eq_A_8}
	\rho^2(x,y)=(x+x_i-x_j-x_\mathrm{DE})^2+(y+y_i-y_j-y_\mathrm{DE})^2.
\end{equation}
From \eqref{eq_2_7}, $H_{ij}$ is then written as:
\begin{equation}\label{eq_A_1}
	H_{ij} \approx \int_{-\frac{a_\mathrm{PD}}{2}}^{\frac{a_\mathrm{PD}}{2}}\int_{-\frac{a_\mathrm{PD}}{2}}^{\frac{a_\mathrm{PD}}{2}}\frac{2}{\pi w^2(L)}\exp\left(-2\frac{(x+x_i-x_j-x_\mathrm{DE})^2+(y+y_i-y_j-y_\mathrm{DE})^2}{w^2(L)}\right)dxdy,
\end{equation}
which can be derived as follows:
\begin{equation}\label{eq_A_2}
	\begin{aligned}
		H_{ij} \approx \frac{1}{4}&\left[\mathrm{erf}\left(\frac{\sqrt{\pi}r_\mathrm{PD}+2(x_i-x_j-x_\mathrm{DE})}{\sqrt{2}w(L)}\right)+\mathrm{erf}\left(\frac{\sqrt{\pi}r_\mathrm{PD}-2(x_i-x_j-x_\mathrm{DE})}{\sqrt{2}w(L)}\right)\right]\times\\
		&\left[\mathrm{erf}\left(\frac{\sqrt{\pi}r_\mathrm{PD}+2(y_i-y_j-y_\mathrm{DE})}{\sqrt{2}w(L)}\right)+\mathrm{erf}\left(\frac{\sqrt{\pi}r_\mathrm{PD}-2(y_i-y_j-y_\mathrm{DE})}{\sqrt{2}w(L)}\right)\right],
	\end{aligned}
\end{equation}
where $\mathrm{erf}(t) = \frac{2}{\sqrt{\pi}}\int_{0}^{t}e^{-s^2}ds$ is the error function.

\subsubsection{Orientation Error of the Transmitter}
For the case of orientation angle error at the transmitter, the use of \eqref{eq_3_9} and \eqref{eq_3_10} leads to $\check{x}_j-\hat{x}_i=x_j\cos(\phi_\mathrm{a})-y_j\sin(\phi_\mathrm{e})\sin(\phi_\mathrm{a})-x_i$, $\check{y}_j-\hat{y}_i=y_j\cos(\phi_\mathrm{e})-y_i$ and $\check{z}_j-\hat{z}_i=L+x_j\sin(\phi_\mathrm{a})+y_j\sin(\phi_\mathrm{e})\cos(\phi_\mathrm{a})$. After simplifying, the parameters $w^2(z)$ and $\rho^2(x,y)$ are obtained as:
\begin{equation}\label{eq_A_3}
	w^2(z) = w_0^2\left(1+\frac{\left[L\cos(\phi_\mathrm{e})\cos(\phi_\mathrm{a})+(x+x_i)\cos(\phi_\mathrm{e})\sin(\phi_\mathrm{a})+(y+y_i)\sin(\phi_\mathrm{e})\right]^2}{z_\mathrm{R}^2}\right),
\end{equation}
\begin{equation}\label{eq_A_4}
	\begin{aligned}
		&\rho^2(x,y) = \left[(x+x_i)\cos(\phi_\mathrm{a})-x_j-L\sin(\phi_\mathrm{a})\right]^2+\left[(y+y_i)\cos(\phi_\mathrm{e})-y_j-L\sin(\phi_\mathrm{e})\cos(\phi_\mathrm{a})\right]^2+ \\
		&(x+x_i)\sin(\phi_\mathrm{e})\sin(\phi_\mathrm{a})\left[2L\sin(\phi_\mathrm{e})\cos(\phi_\mathrm{a})+(x+x_i)\sin(\phi_\mathrm{e})\sin(\phi_\mathrm{a})+2y_j-2(y+y_j)\cos(\phi_\mathrm{e})\right].
	\end{aligned}
\end{equation}
Considering that $x+x_i\ll L$ and $y+y_i\ll L$ hold, for sufficiently small values of $\phi_\mathrm{a}$ and $\phi_\mathrm{e}$, $(x+x_i)\cos(\phi_\mathrm{e})\sin(\phi_\mathrm{a})+(y+y_i)\sin(\phi_\mathrm{e})\ll L\cos(\phi_\mathrm{e})\cos(\phi_\mathrm{a})$, which gives rise to:
\begin{equation}
	L\cos(\phi_\mathrm{e})\cos(\phi_\mathrm{a})+(x+x_i)\cos(\phi_\mathrm{e})\sin(\phi_\mathrm{a})+(y+y_i)\sin(\phi_\mathrm{e})\approx L\cos(\phi_\mathrm{e})\cos(\phi_\mathrm{a}).
\end{equation}
This approximation means the axial distance variation of the slightly tilted beam spot over the \ac{PD} surface is ignored due to its small size. Hence, \eqref{eq_A_3} is simplified to $w^2(z) \approx w^2(L\cos(\phi_\mathrm{e})\cos(\phi_\mathrm{a}))$. Besides, when $\phi_\mathrm{a}$ and $\phi_\mathrm{e}$ are small enough, in the right hand side of \eqref{eq_A_4}, the last term is deemed negligible compared to the first two terms from the factor $\sin(\phi_\mathrm{e})\sin(\phi_\mathrm{a})\ll1$. Consequently, using $\mathbf{n}_\mathrm{t}\cdot\mathbf{n}_\mathrm{r}=\cos(\phi_\mathrm{e})\cos(\phi_\mathrm{a})$, the integration in \eqref{eq_2_7} is approximated by:
\begin{equation}\label{eq_A_5}
	\begin{aligned}
		&H_{ij} \approx \int_{-\frac{a_\mathrm{PD}}{2}}^{\frac{a_\mathrm{PD}}{2}}\int_{-\frac{a_\mathrm{PD}}{2}}^{\frac{a_\mathrm{PD}}{2}}\frac{2\cos(\phi_\mathrm{e})\cos(\phi_\mathrm{a})dxdy}{\pi w^2(L\cos(\phi_\mathrm{e})\cos(\phi_\mathrm{a}))}\times \\ &\exp\left(-2\frac{\left[(x+x_i)\cos(\phi_\mathrm{a})-x_j-L\sin(\phi_\mathrm{a})\right]^2+\left[(y+y_i)\cos(\phi_\mathrm{e})-y_j-L\sin(\phi_\mathrm{e})\cos(\phi_\mathrm{a})\right]^2}{w^2(L\cos(\phi_\mathrm{e})\cos(\phi_\mathrm{a}))}\right).
	\end{aligned}
\end{equation}
A closed form solution of \eqref{eq_A_5} is readily derived as follows:
\begin{equation}\label{eq_A_7}
	\begin{aligned}
		H_{ij} \approx \frac{1}{4}&\bigg[\mathrm{erf}\left(\frac{\sqrt{\pi}r_\mathrm{PD}\cos(\phi_\mathrm{a})+2\left[x_i\cos(\phi_\mathrm{a})-x_j -L\sin(\phi_\mathrm{a})\right]}{\sqrt{2}w(L\cos(\phi_\mathrm{e})\cos(\phi_\mathrm{a}))}\right)+ \\
		&\mathrm{erf}\left(\frac{\sqrt{\pi}r_\mathrm{PD}\cos(\phi_\mathrm{a})-2\left[x_i\cos(\phi_\mathrm{a})-x_j-L\sin(\phi_\mathrm{a})\right]}{\sqrt{2}w(L\cos(\phi_\mathrm{e})\cos(\phi_\mathrm{a}))}\right)\bigg]\times\\
		&\bigg[\mathrm{erf}\left(\frac{\sqrt{\pi}r_\mathrm{PD}\cos(\phi_\mathrm{e})+2\left[y_i\cos(\phi_\mathrm{e})-y_j-L\sin(\phi_\mathrm{e})\cos(\phi_\mathrm{a})\right]}{\sqrt{2}w(L\cos(\phi_\mathrm{e})\cos(\phi_\mathrm{a}))}\right)+ \\
		&\mathrm{erf}\left(\frac{\sqrt{\pi}r_\mathrm{PD}\cos(\phi_\mathrm{e})-2\left[y_i\cos(\phi_\mathrm{e})-y_j-L\sin(\phi_\mathrm{e})\cos(\phi_\mathrm{a})\right]}{\sqrt{2}w(L\cos(\phi_\mathrm{e})\cos(\phi_\mathrm{a}))}\right)\bigg].
	\end{aligned}
\end{equation}
Note that \eqref{eq_A_7} essentially represents \eqref{eq_A_2} for $x_\mathrm{DE}=L\sin(\phi_\mathrm{a})$ and $y_\mathrm{DE}=L\sin(\phi_\mathrm{e})\cos(\phi_\mathrm{a})$. This means an orientation angle error of the transmitter array with azimuth and elevation components of $\phi_\mathrm{a}$ and $\phi_\mathrm{e}$ produces an effect equivalent to horizontal and vertical displacements of $L\sin(\phi_\mathrm{a})$ and $L\sin(\phi_\mathrm{e})\cos(\phi_\mathrm{a})$, respectively. The approximations in \eqref{eq_A_2} and \eqref{eq_A_7} are verified in Appendix and it is shown that they are highly accurate as long as the beam spot size at the receiver is in the order of or larger than the \ac{PD} size.

\begin{figure}[!t]
	\centering
	\includegraphics[width=0.85\linewidth]{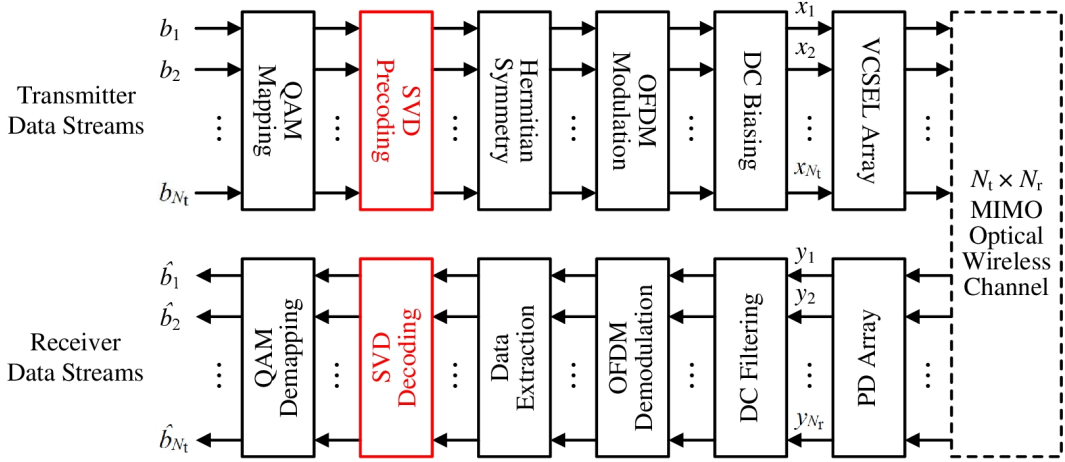}
	\caption{MIMO-OFDM transceiver architecture.}
	\label{fig_3_3}
	\vspace{-20pt}
\end{figure}

\subsection{Spatial Multiplexing MIMO-OFDM}\label{sec4_4}
\subsubsection{Transceiver Architecture}
In this paper, a spatial multiplexing \ac{MIMO}-OFDM system is used to maximize the transmission rate by sending independent data streams over the \ac{MIMO} optical wireless channel. To ensure a high spectral efficiency for transmitting each data stream, the standard \acs{DCO{-}OFDM} technique along with adaptive \ac{QAM} is used \cite{Gonzalez1}. In this case, the spatial overlapping between the beam spots of the \acp{VCSEL} at the \ac{PD} array causes crosstalk in the \ac{MIMO} channel. Fig.~\ref{fig_3_3} depicts a simplified block diagram of the \ac{MIMO}-OFDM transceiver architecture. The \ac{SVD} precoding and decoding stages shown in Fig.~\ref{fig_3_3} can be applied provided the \ac{MIMO} \ac{CSI} is known at both the transmitter and receiver. First, it is assumed that this is not the case to avoid the overhead associated with the \ac{CSI} estimation and feedback, and the transceiver system is described without the use of \ac{SVD}.

At the transmitter, the $N_\mathrm{t}$ input binary data streams are individually mapped to a sequence of complex \ac{QAM} symbols. With a digital realization of the OFDM modulation and demodulation by way of $N_\mathrm{FFT}$-point \ac{IFFT} and \ac{FFT}, respectively, the resulting sequences are buffered into blocks of size $N_\mathrm{t}\times N_\mathrm{QAM}$. They are loaded onto the $N_\mathrm{QAM}$ data-carrying subcarriers of the $N_\mathrm{t}$ OFDM frames in positive frequencies, where $N_\mathrm{QAM}=\frac{N_\mathrm{FFT}}{2}-1$. For baseband OFDM transmission in \ac{IM{-}DD} systems, it is necessary for the time domain signal to be real-valued. To this end, for each OFDM frame, the number of symbols is extended to $N_\mathrm{FFT}$ according to a Hermitian symmetry and the \ac{DC} and Nyquist frequency subcarriers are zero-padded before the \ac{IFFT} operation. Also, in order to comply with the non-negativity constraint of \ac{IM{-}DD} channels, a proper \ac{DC} level is added in the time domain to obtain a positive signal \cite{Gonzalez1}. Let $\mathbf{x}(t) = [x_1(t)~x_2(t)~\cdots~x_{N_\mathrm{t}}(t)]^\top$ be the vector of instantaneous optical powers emitted by the \acp{VCSEL} at time sample $t$ for $t=0,1,\dots,N_\mathrm{FFT}-1$. It is given by:
\begin{equation}\label{eq_3_7}
\mathbf{x}(t) = \sqrt{P_\mathrm{elec}}\mathbf{s}(t)+x_\mathrm{DC}\mathbf{1},
\end{equation}
where $P_\mathrm{elec}$ is the average electrical power of each OFDM symbol; $\mathbf{s}(t)=\left[s_1(t)~s_2(t)~\cdots~s_{N_\mathrm{t}}(t)\right]^\top$ is the vector of the normalized discrete time OFDM samples; $x_\mathrm{DC}=P_\mathrm{t}$ is the \ac{DC} bias with $P_\mathrm{t}$ representing the average optical power per \ac{VCSEL}; and $\mathbf{1}\in\mathbb{R}^{N_\mathrm{t}\times1}$ is an all-ones vector. The finite dynamic range of the \acp{VCSEL} determines the available peak-to-peak swing for their modulating OFDM signal. The envelope of the unbiased OFDM signal follows a zero mean real Gaussian distribution for $N_\mathrm{FFT}\geq64$ \cite{Dardari}. The choice of $P_\mathrm{t}=3\sqrt{P_\mathrm{elec}}$ guarantees that $99.7\%$ of the signal variations remains undistorted, thereby effectively discarding the clipping noise \cite{Dimitrov2}. Thus, the average power of the OFDM signal assigned to each \ac{VCSEL} is $P_\mathrm{elec}=\frac{1}{9}P_\mathrm{t}^2$.

At the receiver array, after filtering out the \ac{DC} component and perfect sampling, the vector of received photocurrents is obtained. Let $\tilde{\mathbf{X}}_{k}\in\mathbb{R}^{N_\mathrm{t}\times1}$ be the vector of symbols modulated on the $k$th subcarrier in the frequency domain. After the \ac{FFT} operation, the received symbols are extracted from the data-carrying subcarriers and then they are demodulated using maximum likelihood detection. The vector of received signals on the $k$th subcarrier for $k=0,1,\dots,N_\mathrm{FFT}-1$ is written in the form:
\begin{equation}\label{eq_3_16}
\mathbf{Y}_k = R_\mathrm{PD}\sqrt{P_\mathrm{elec}}\mathbf{H}_k\tilde{\mathbf{X}}_k+\mathbf{Z}_k,
\end{equation}
where $R_\mathrm{PD}$ is the PD responsivity; $\mathbf{H}_k$ is the frequency response of the \ac{MIMO} channel; and $\mathbf{Z}_k$ is the \ac{AWGN} vector. Note that without \ac{SVD} processing, $N_\mathrm{r}=N_\mathrm{t}$ holds, in which case $\mathbf{Y}_k\in\mathbb{R}^{N_\mathrm{t}\times1}$ and $\mathbf{H}_k\in\mathbb{R}^{N_\mathrm{t}\times N_\mathrm{t}}$. Considering strong \ac{LOS} components when using laser beams with low divergence, the channel is nearly flat for which $\mathbf{H}_k=\mathbf{H}_0$ $\forall k$, where $\mathbf{H}_0$ refers to \eqref{eq_3_1}. The $i$th element of the noise vector comprises thermal noise and shot noise of the $i$th branch of the receiver and the \ac{RIN} caused by all the \acp{VCSEL} which depends on the average received optical power \cite{KChang}. The total noise variance is given by:
\begin{equation}\label{eq_3_5}
\sigma_i^2 = \frac{4\kappa T}{R_\mathrm{L}}BF_\mathrm{n}+2q\left(\sum\nolimits_{j=1}^{N_\mathrm{t}}R_\mathrm{PD}H_{ij}P_\mathrm{t}\right)B+\mathrm{RIN}\left(\sum\nolimits_{j=1}^{N_\mathrm{t}}\left(R_\mathrm{PD}H_{ij}P_\mathrm{t}\right)^2\right)B,
\end{equation}
where $\kappa$ is the Boltzmann constant; $T$ is temperature in Kelvin; $R_\mathrm{L}$ is the load resistance; $B$ is the single-sided bandwidth of the system; $F_\mathrm{n}$ is the noise figure of the \ac{TIA}; $q$ is the elementary charge; and $\mathrm{RIN}$ is defined as the mean square of instantaneous power fluctuations divided by the squared average power of the laser source \cite{KChang}. Based on \eqref{eq_3_16}, the received \ac{SINR} per subcarrier for the $i$th link is derived as follows:
\begin{equation}\label{eq_4_1}
\gamma_i = \frac{R_\mathrm{PD}^2H_{ii}^2P_\mathrm{elec}}{\sum_{j\neq i}R_\mathrm{PD}^2H_{ij}^2P_\mathrm{elec}+\sigma_i^2}.
\end{equation}

\subsubsection{SVD Processing}
When the \ac{CSI} is available at the transmitter and receiver, the \ac{MIMO} channel can be transformed into a set of parallel independent subchannels by means of \ac{SVD} of the channel matrix in the frequency domain. The use of \ac{SVD} leads to the capacity achieving architecture for spatial multiplexing \ac{MIMO} systems \cite{Tse}. The \ac{SVD} of $\mathbf{H}_k\in\mathbb{R}^{N_\mathrm{r}\times N_\mathrm{t}}$, with $N_\mathrm{r}\geq N_\mathrm{t}$, is $\mathbf{H}_k=\mathbf{U}_k\mathbf{\Lambda}_k\mathbf{V}_k^*$, where $\mathbf{U}_k\in\mathbb{R}^{N_\mathrm{r}\times N_\mathrm{r}}$ and $\mathbf{V}_k\in\mathbb{R}^{N_\mathrm{t}\times N_\mathrm{t}}$ are unitary matrices; $^*$ denotes conjugate transpose; and $\mathbf{\Lambda}_k\in\mathbb{R}^{N_\mathrm{r}\times N_\mathrm{t}}$ is a rectangular diagonal matrix of the ordered singular values, i.e. $\lambda_1\geq \lambda_2\geq\cdots\geq \lambda_{N_\mathrm{t}}>0$ \cite{Tse}. Note that $\mathbf{H}_k=\mathbf{H}_0$ $\forall k$ as discussed so the subscript $k$ can be dropped from the singular values. After \ac{SVD} decoding at the receiver, the $N_\mathrm{t}$-dimensional vector of received symbols on the $k$th subcarrier for $k=0,1,\dots,N_\mathrm{FFT}-1$ becomes:
\begin{equation}\label{eq_3_13}
\tilde{\mathbf{Y}}_k = R_\mathrm{PD}\sqrt{P_\mathrm{elec}}\mathbf{\Lambda}_k\tilde{\mathbf{X}}_k+\tilde{\mathbf{Z}}_k,
\end{equation}
where $\tilde{\mathbf{Z}}_k=\mathbf{U}_k^*\mathbf{Z}_k$. Note that the statistics of the noise vector is preserved under a unitary transformation. Therefore, the $i$th elements of $\tilde{\mathbf{Z}}_k$ and $\mathbf{Z}_k$ have the same variance of $\sigma_i^2$. The received \ac{SNR} per subcarrier for the $i$th link is derived as follows:
\begin{equation}\label{eq_3_15}
\gamma_i = \frac{R_\mathrm{PD}^2\lambda_i^2P_\mathrm{elec}}{\sigma_i^2}.
\end{equation}

\subsection{Aggregate Rate Analysis}\label{sec4_5}
Based on adaptive \ac{QAM} with a given \ac{BER} performance in an \ac{AWGN} channel, a tight upper bound for the number of bits per symbol transmitted by $\text{VCSEL}_i$, for $\mathrm{BER\leq10^{-2}}$ and $0\leq\gamma_i\leq30$~dB, is given by \cite{Goldsmith2}:
\begin{equation}\label{eq_4_2}
	\eta_i = \log_2\left(1+\frac{\gamma_i}{\Gamma}\right),
\end{equation}
where:
\begin{equation}\label{eq_4_4}
	\Gamma = \frac{-\ln\left(5\mathrm{BER}\right)}{1.5},
\end{equation}
represents the \ac{SINR} gap due to the required \ac{BER}. With a symbol rate of $\frac{2B}{N_\mathrm{FFT}}$ for \ac{DCO{-}OFDM}, the achievable rate per subcarrier is $\frac{2B}{N_\mathrm{FFT}}\eta_i$ bit/s. According to $\frac{N_\mathrm{FFT}}{2}-1$ data-carrying subcarriers, the transmission rate for $\text{VCSEL}_i$ is then:
\begin{equation}\label{eq_4_5}
\mathcal{R}_i = \xi B\log_2\left(1+\frac{\gamma_i}{\Gamma}\right),
\end{equation}
where $\xi=\frac{N_\mathrm{FFT}-2}{N_\mathrm{FFT}}$. Hence, the aggregate data rate of the \ac{MIMO}-OFDM system is expressed as:
\begin{equation}\label{eq_4_6}
\mathcal{R} = \sum_{i=1}^{N_\mathrm{t}}\mathcal{R}_i=\xi B\sum_{i=1}^{N_\mathrm{t}}\log_2\left(1+\frac{\gamma_i}{\Gamma}\right).
\end{equation}

\begin{table}[!t]
	\vspace{2pt}
	\centering
	\caption{Simulation Parameters}
	\begin{tabular}{c|l|l}
		\textbf{Parameter} & \textbf{Description}                    & \textbf{Value}  \\ \hline
		$L$                & Link distance                           & $2$ m           \\ 
		$P_\mathrm{t}$     & Transmit power per VCSEL                & $1$ mW          \\ 
		$\lambda$          & Laser wavelength                        & $850$ nm        \\ 
		$w_0$              & Effective waist radius                  & $\geq10$ {\textmu}m \\
		$B$                & System bandwidth           			 & $20$ GHz        \\
		$\mathrm{RIN}$     & Laser noise      						 & $-155$ dB/Hz    \\ 
		$r_\mathrm{PD}$    & PD radius                               & $3$ mm          \\ 
		$A_\mathrm{PD}$    & PD area                                 & $28.3$ mm$^2$   \\
		$R_\mathrm{PD}$    & PD responsivity                         & $0.4$ A/W       \\ 
		$\delta$		   & Inter-element spacing 					 & $6$ mm          \\  
		$R_\mathrm{L}$     & Load resistance                         & $50$ $\Omega$   \\
		$F_\mathrm{n}$     & TIA noise figure			             & $5$ dB          \\
		$\mathrm{BER}$     & Target BER                			     & $10^{-3}$       \\ \hline
	\end{tabular}
	\label{sec5_tbl1}
	\vspace{-20pt}
\end{table}

\section{Numerical Results and Discussions}\label{sec5}
The performance of the \ac{VCSEL}-based \ac{MIMO} \ac{OWC} system is evaluated by using computer simulations and the parameters listed in Table~\ref{sec5_tbl1}, where the \ac{VCSEL} and noise parameters are adopted from \cite{Liu,Szczerba}. Numerical results are presented for the \textit{effective radius} $w_0$ of the beam waist over the range $10~\mu\text{m}\leq w_0\leq100~\mu\text{m}$ with the assumption that there is a convex lens next to each bare \ac{VCSEL} to widen its output beam waist in order to reduce the far field beam divergence. For a link distance of $L=2$~m, the beam spot radius and divergence angle vary from $w(L)=54$~mm and $\theta=1.6^\circ$ to $w(L)=5.4$~mm and $\theta=0.16^\circ$. An optical power per \ac{VCSEL} of $1$~mW is selected on account of eye safety considerations. A laser source is eye-safe if a fraction $\eta$ of the total power of the Gaussian beam entering the eye aperture at the \ac{MHP} for an exposure time of blink reflex is no greater than the \ac{MPE} multiplied by the pupil area $A_\mathrm{pupil}$ \cite{IEC_Std608251}, i.e. $\eta P_\mathrm{t}\leq\mathrm{MPE}\times A_\mathrm{pupil}$. For a given wavelength, the \ac{MPE} value reduces with an increase in the beam waist radius \cite{Henderson}, so the most restrictive case is when $w_0$ is at a maximum. The case of $w_0=100$ {\textmu}m and $\lambda=850$ nm leads to a subtense angle of $\alpha<1.5$ mrad. The subtense angle is measured as the plane angle subtended by the apparent source to the observer's pupil at \ac{MHP}. For $\alpha<1.5$ mrad, the laser source is classified as a point source for which $\eta=1$ and $\mathrm{MPE}=50.8$ W/m$^2$ \cite{Henderson}. For a circular aperture of diameter $7$ mm, $A_\mathrm{pupil}\approx38.5$ mm$^2$, and it follows that $P_\mathrm{t}\leq1.95$ mW for each \ac{VCSEL}. Hence, $P_\mathrm{t}=1$ mW is considered eye-safe.

\subsection{Perfect Alignment}
\subsubsection{Spatial Distribution of SINR}
\begin{figure}[!t]
	\centering
	\subfloat[$w_0=50$ {\textmu}m]
	{\includegraphics[width=0.49\linewidth]{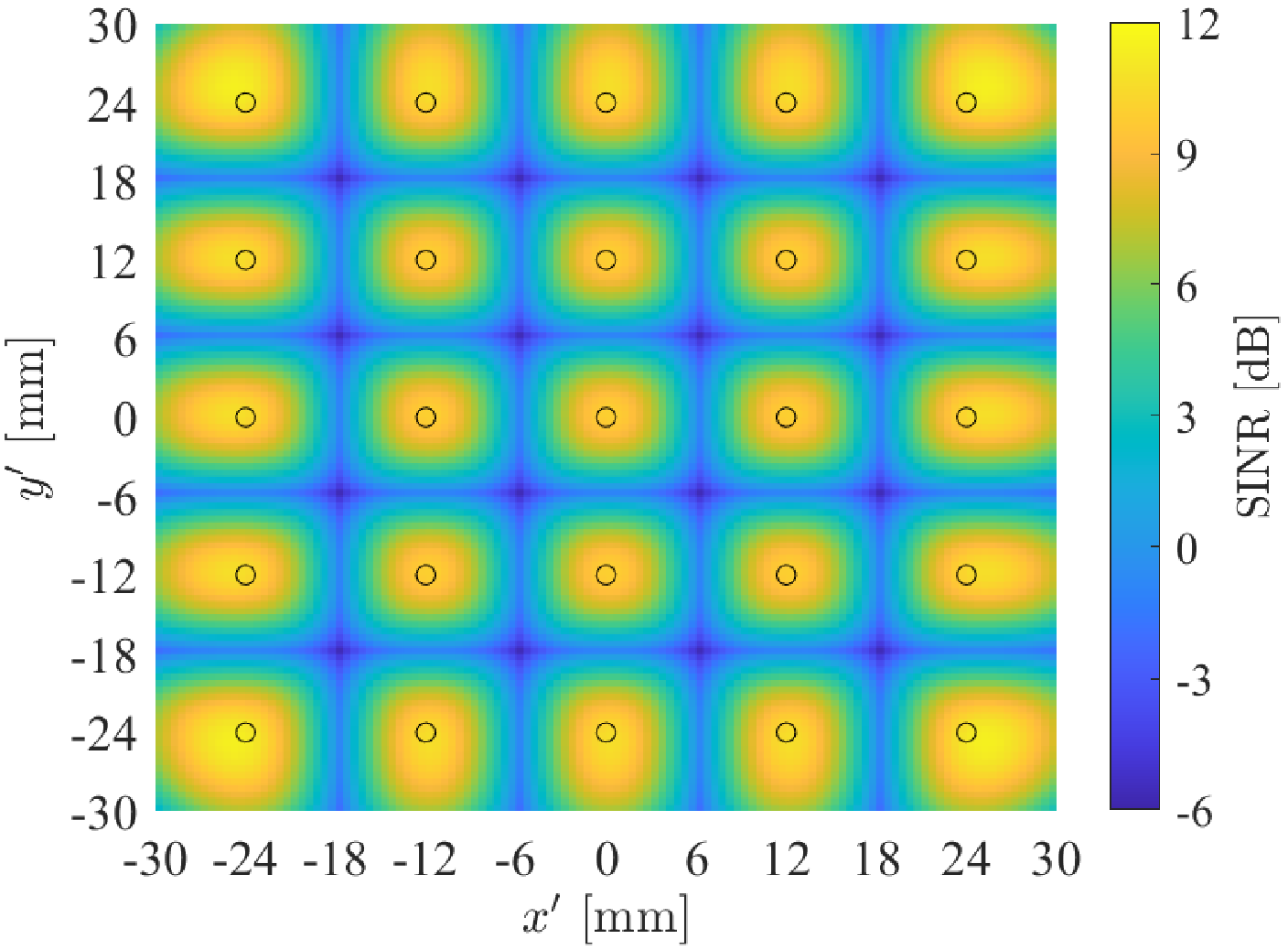}\label{fig_4_2a}}
	\subfloat[$w_0=100$ {\textmu}m]
	{\includegraphics[width=0.49\linewidth]{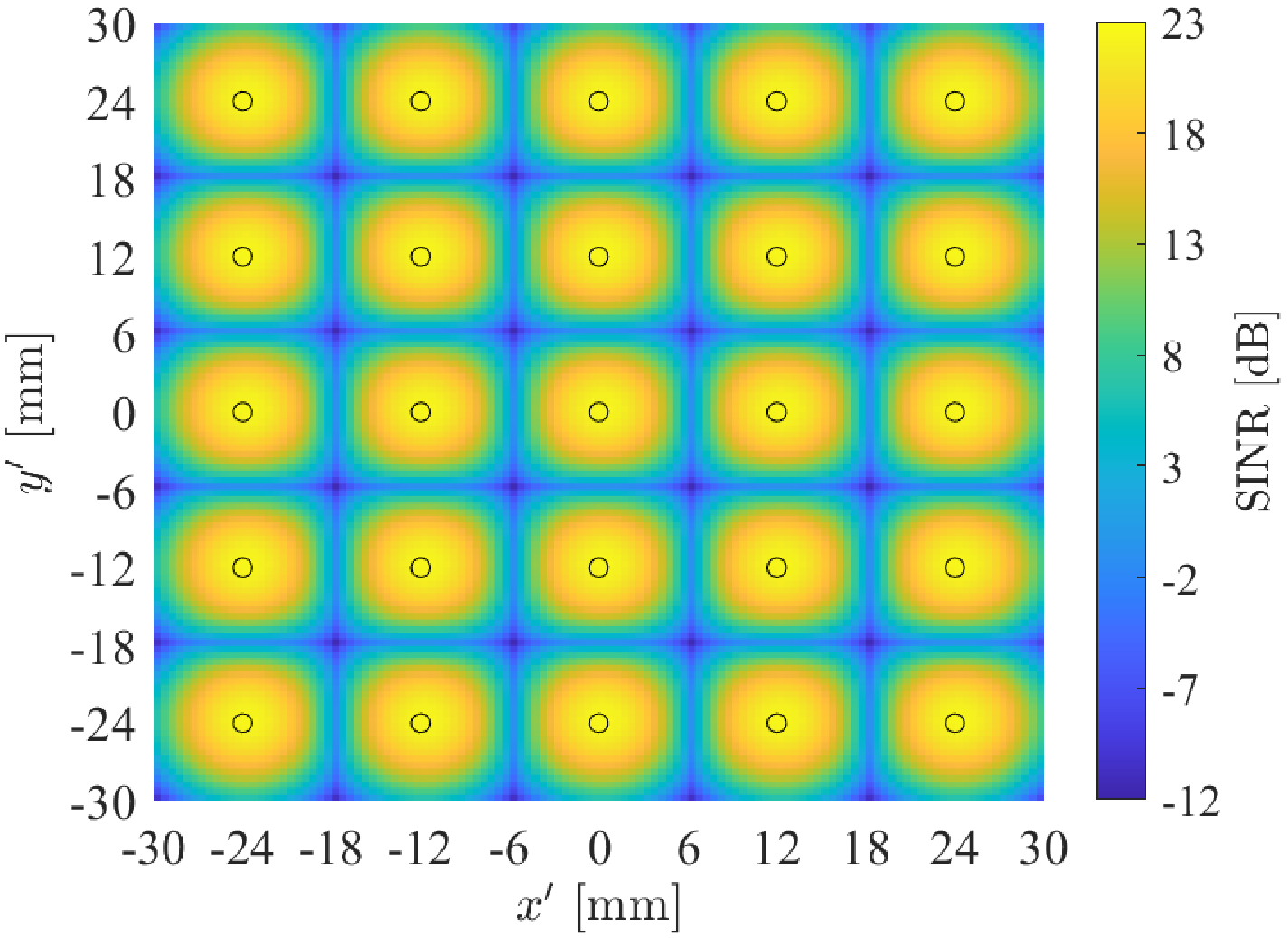}\label{fig_4_2b}}
	\caption{Spatial distribution of SINR on the receiver plane for a $25\times25$ MIMO system.}
	\label{fig_4_2}
	\vspace{-20pt}
\end{figure}

Fig.~\ref{fig_4_2} illustrates the spatial distribution of the received \ac{SINR} on the transverse plane at the receiver for a $5\times5$ \ac{PD} array, for $w_0=50$ {\textmu}m and $w_0=100$ {\textmu}m, representing two cases for the beam spot radius including $w(L)=10.8$ mm and $w(L)=5.4$ mm, respectively. For $w_0=50$ {\textmu}m with a larger beam spot size at the receiver, as shown in Fig.~\subref*{fig_4_2a}, the \ac{SINR} ranges from $-6$ dB to $+12$ dB. A dissimilar distribution of the \ac{SINR} over $25$ different regions of the array is caused by significant overlaps between the neighboring beam spots. It can be observed that in the regions close to the edge of the array, when compared to other regions, there are larger areas over which the highest \ac{SINR} level is experienced, as they are influenced by asymmetric crosstalk effects due to their positions relative to the center of the array. The maximum \ac{MIMO} interference falls on the central region where a small area is covered with the highest \ac{SINR} level. However, on the positions of the \ac{PD} elements as shown by small circles, \ac{SINR} levels are approximately equal, near the maximum value. As shown in Fig.~\subref*{fig_3_2b}, when doubling the effective waist radius to $w_0=100$ {\textmu}m, the value of $w(L)$ is halved and the edge effects are improved, since the beam spot area is quarter of the case where $w_0=50$ {\textmu}m. In this case, the \ac{SINR} is almost evenly distributed on different regions and the \ac{SINR} of $23$ dB is equally achieved at each \ac{PD}.

\subsubsection{Rate vs. Beam Waist}
\begin{figure}[!t]
	\centering
	\includegraphics[width=0.55\linewidth]{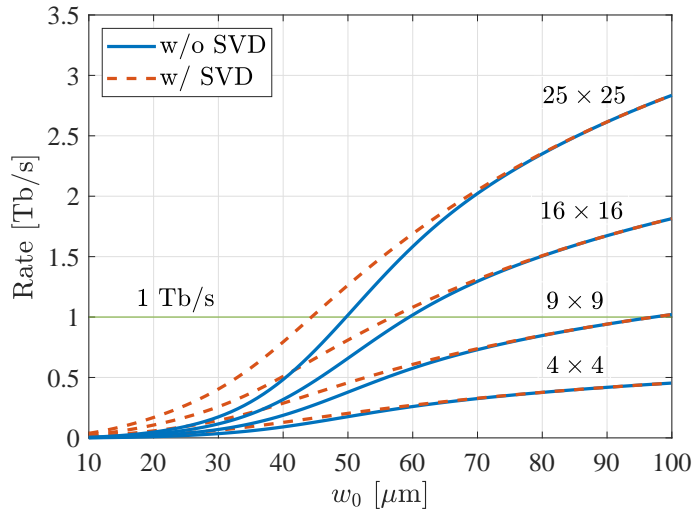}
	\caption{Aggregate data rate of $N_\mathrm{t}\times N_\mathrm{r}$ MIMO OWC system with perfect alignment as a function of the effective waist radius $w_0$ for $N_\mathrm{t}=N_\mathrm{r}=4,9,16,25$ (i.e. $2\times2$, $3\times3$, $4\times4$, $5\times5$ arrays).}
	\label{fig_4_3}
	\vspace{-20pt}
\end{figure}

Fig.~\ref{fig_4_3} demonstrates the aggregate data rate achieved by the proposed \ac{MIMO} system under perfect alignment conditions when $w_0$ is varied from $10$ {\textmu}m to $100$ {\textmu}m. For all \ac{MIMO} realizations under consideration, the aggregate rate monotonically increases for larger values of $w_0$. At the beginning for $w_0=10$ {\textmu}m, the beam spot size at the receiver is very large, i.e. $w(L)=54$ mm. This renders the signal power collected by each \ac{PD} from direct links very low. Besides, there is substantial crosstalk among the incident beams, which severely degrades the performance. The use of \ac{SVD} yields the upper bound performance for the \ac{MIMO} system. When $w_0$ increases, the data rate grows, and so does the gap between the performance of the \ac{MIMO} system without \ac{SVD} and the upper bound. The maximum difference between the performance of the two systems occurs at about $w_0=40$ {\textmu}m. After this point, by increasing $w_0$, the aforementioned gap is rapidly reduced and the data rate for the \ac{MIMO} system without \ac{SVD} asymptotically approaches that with \ac{SVD}. The right tail of the curves in Fig.~\ref{fig_4_3} indicates the noise-limited region for $w_0\geq80$ {\textmu}m, whereas $w_0<80$ {\textmu}m represents the crosstalk-limited region. Also, $4\times4$, $9\times9$, $16\times16$ and $25\times25$ systems, respectively, attain $0.454$ Tb/s, $1.021$ Tb/s, $1.815$ Tb/s and $2.835$ Tb/s, for $w_0=100$ {\textmu}m. In order to achieve a target data rate of $1$ Tb/s, $9\times9$, $16\times16$ and $25\times25$ systems, respectively, require the beam waist radii of $98$ {\textmu}m, $60$ {\textmu}m and $50$ {\textmu}m. This target is not achievable by a $4\times4$ system for $w_0\leq100$ {\textmu}m.

\begin{figure}[!t]
	\centering
	\subfloat[\centering Horizontal displacement:\newline $x_\mathrm{DE}=r_\mathrm{DE}$ and $\phi_\mathrm{a}=\psi_\mathrm{a}=0$]
	{\hspace{-0.6cm}\includegraphics[width=0.37\linewidth]{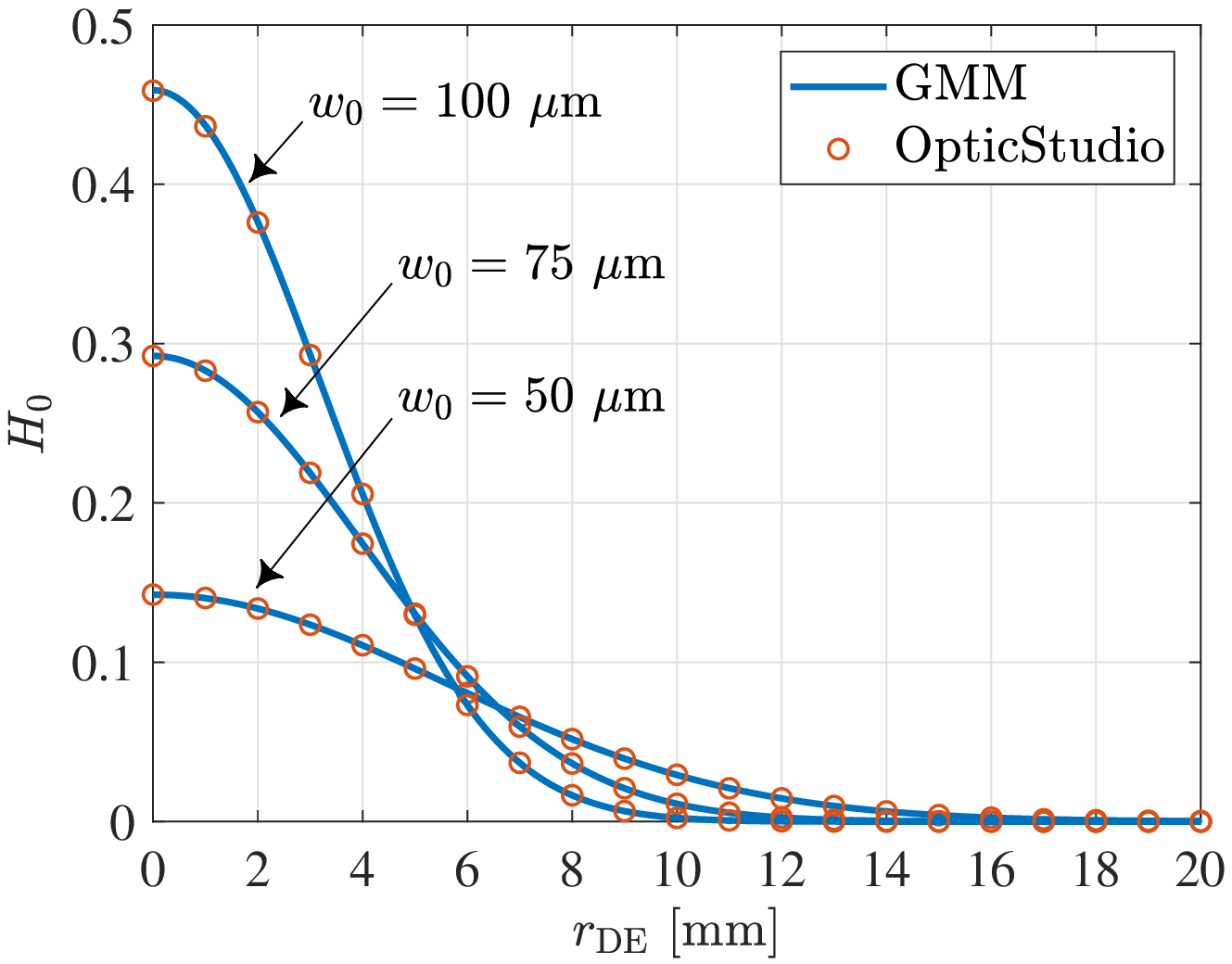}\label{fig_4_4a}\hspace{-0.45cm}}
	\subfloat[\centering Transmitter azimuth angle error:\newline $x_\mathrm{DE}=0$ and $\psi_\mathrm{a}=0$]
	{\includegraphics[width=0.37\linewidth]{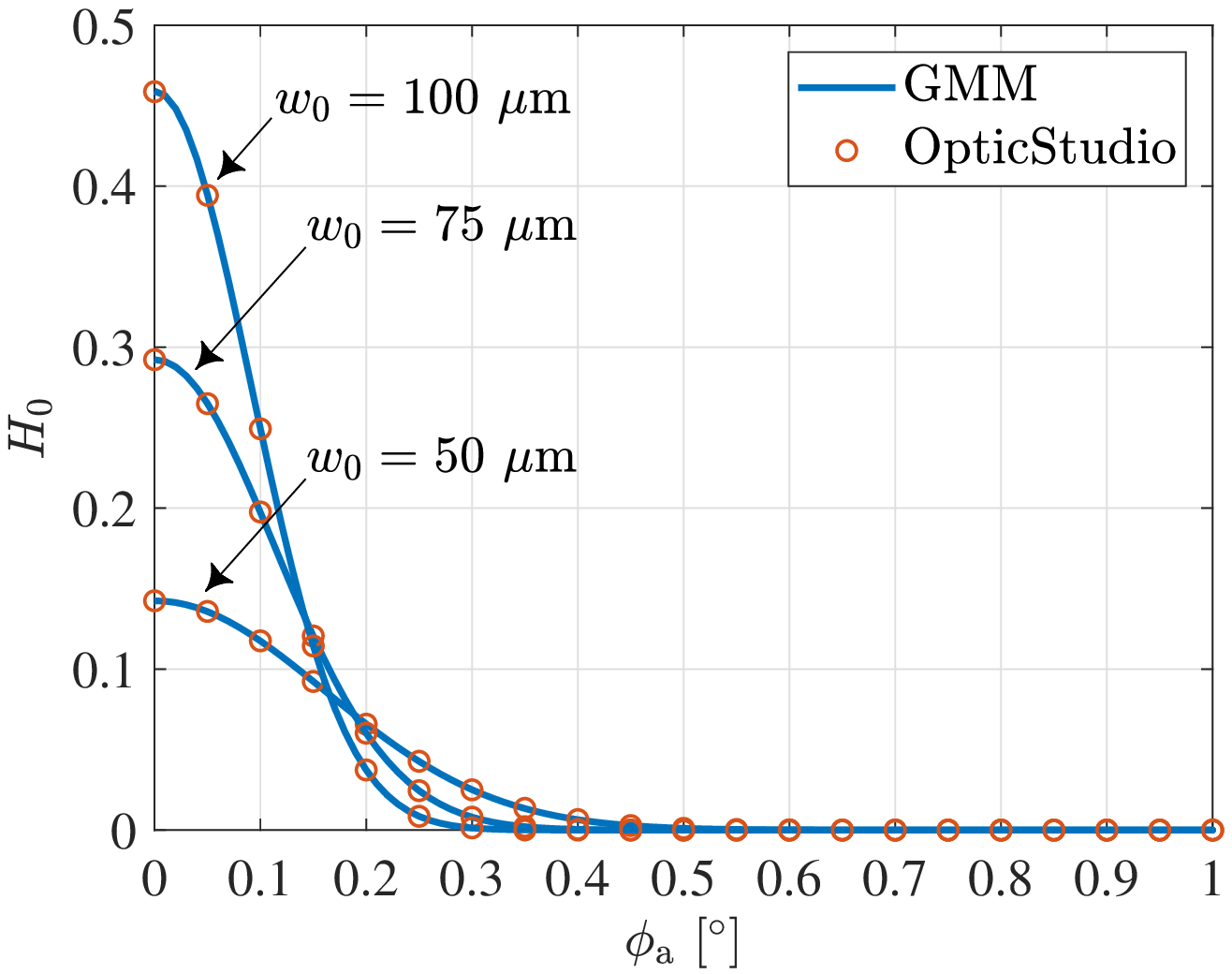}\label{fig_4_4b}\hspace{-0.45cm}}
	\subfloat[\centering Receiver azimuth angle error:\newline $x_\mathrm{DE}=0$ and $\phi_\mathrm{a}=0$]
	{\includegraphics[width=0.37\linewidth]{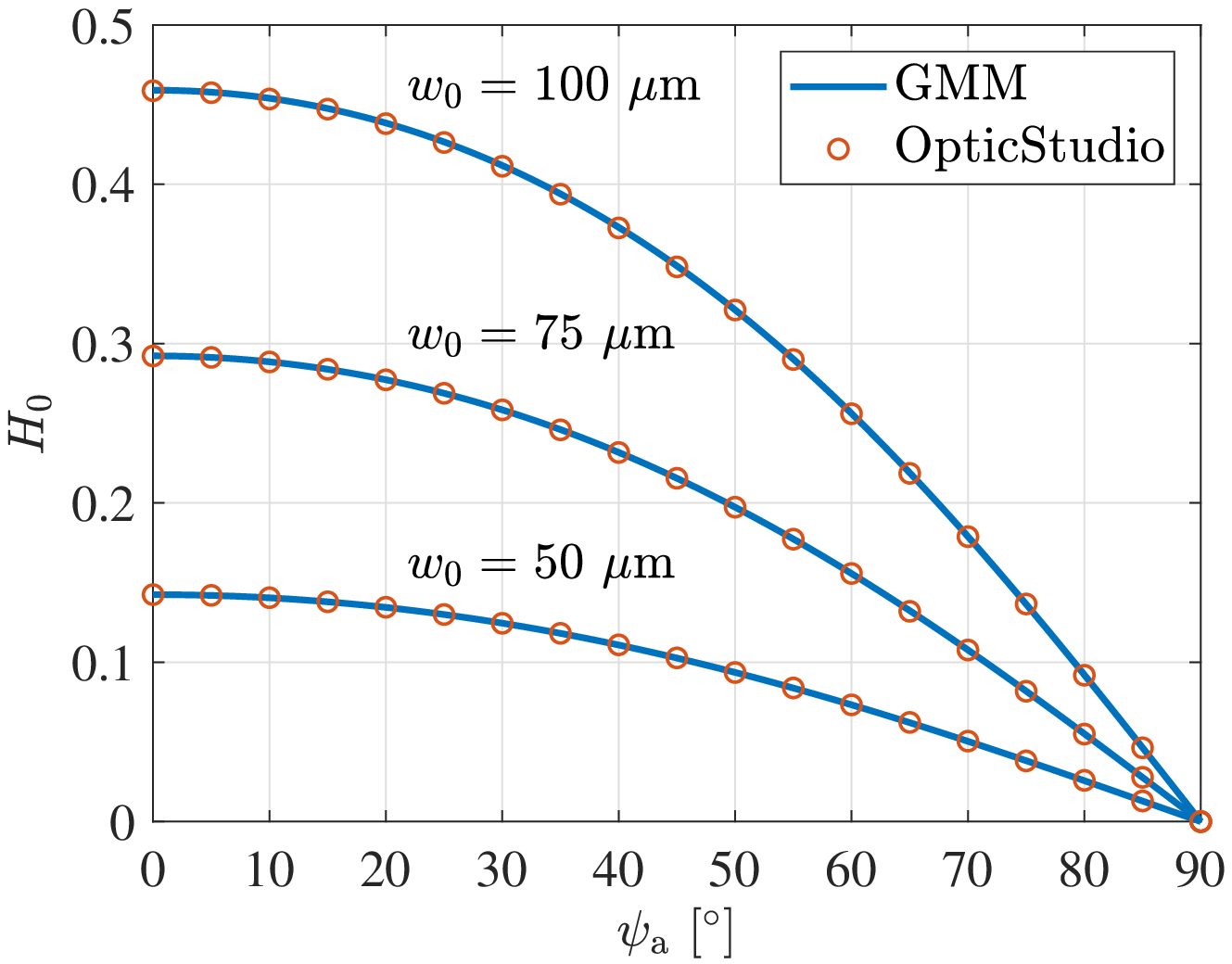}\label{fig_4_4c}\hspace{-0.6cm}}\\
	\subfloat[\centering Horizontal displacement:\newline $x_\mathrm{DE}=-r_\mathrm{DE}$, $\phi_\mathrm{a}=0.1^\circ$ and $\psi_\mathrm{a}=10^\circ$]
	{\hspace{-0.6cm}\includegraphics[width=0.37\linewidth]{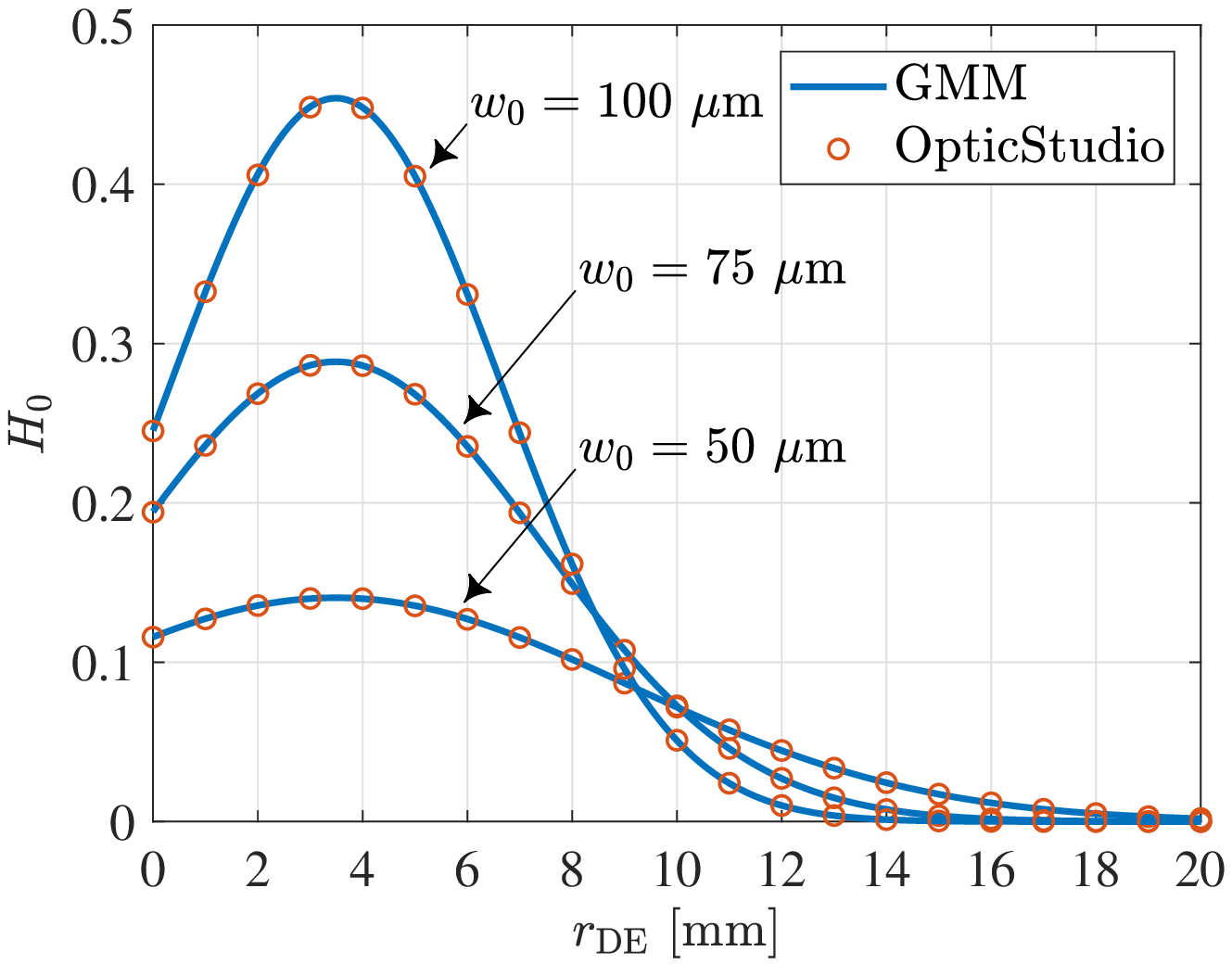}\label{fig_4_4d}\hspace{-0.45cm}}
	\subfloat[\centering Transmitter azimuth angle error:\newline $x_\mathrm{DE}=-2$ mm and $\psi_\mathrm{a}=10^\circ$]
	{\includegraphics[width=0.37\linewidth]{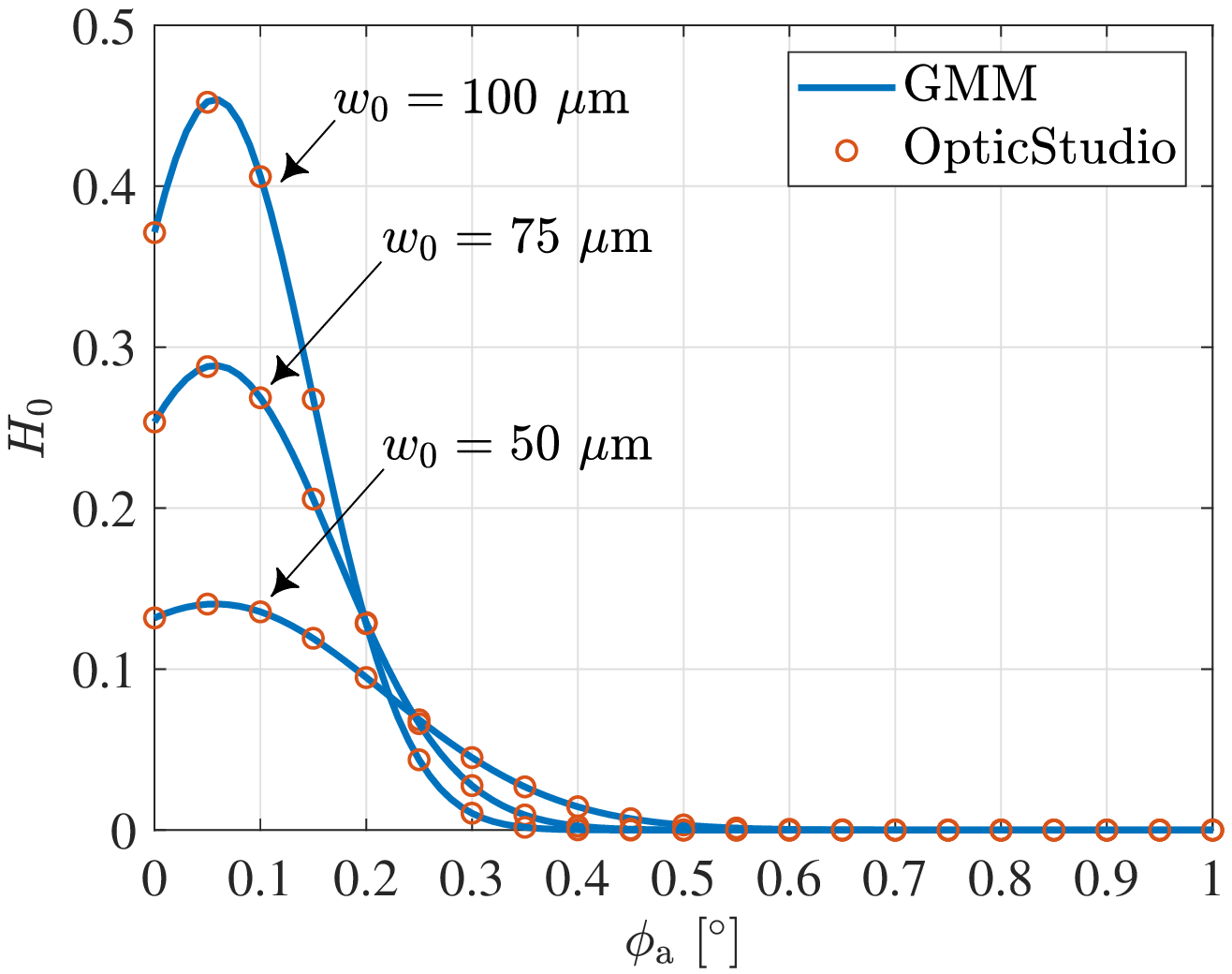}\label{fig_4_4e}\hspace{-0.45cm}}
	\subfloat[\centering Receiver azimuth angle error:\newline $x_\mathrm{DE}=-2$ mm and $\phi_\mathrm{a}=0.1^\circ$]
	{\includegraphics[width=0.37\linewidth]{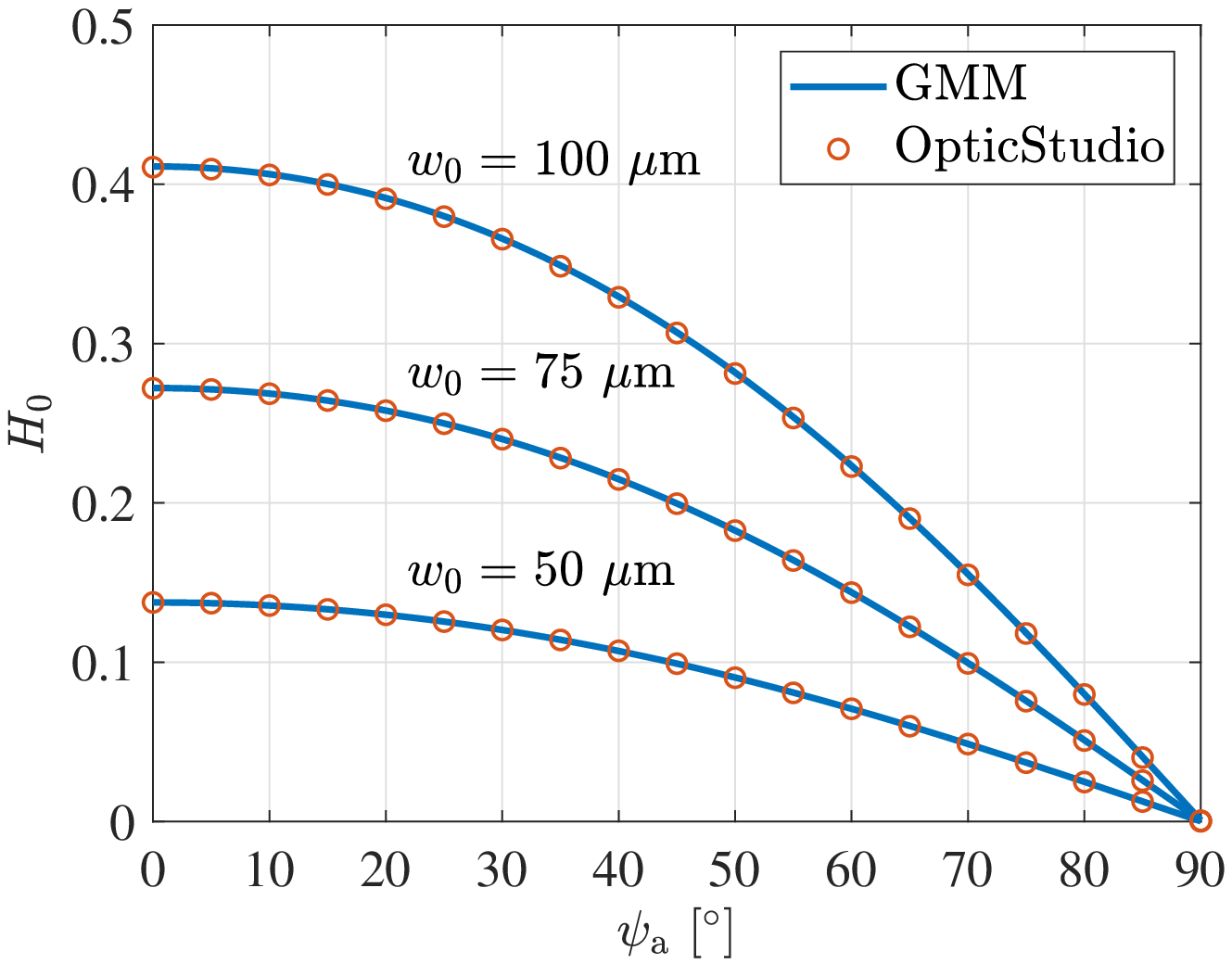}\label{fig_4_4f}\hspace{-0.6cm}}
	\caption{Comparison of the analytical results based on the GMM and the simulation results from OpticStudio for the SISO channel gain. For all cases, $y_\mathrm{DE}=0$ and $\phi_\mathrm{e}=\psi_\mathrm{e}=0$.}
	\label{fig_4_4}
	\vspace{-20pt}
\end{figure}

\subsection{GMM Verification}
The \ac{GMM} of the \ac{SISO} channel developed in Section~\ref{sec3} is the underlying foundation for the \ac{MIMO} misalignment modeling presented in Section~\ref{sec4}. Therefore, its accuracy needs to be verified with a dependable benchmark. A powerful commercial optical design software by Zemax, known as OpticStudio \cite{OpticStudio}, is used for this purpose. Empirical data is collected by running extensive simulations based on non-sequential ray tracing in OpticStudio. Fig.~\ref{fig_4_4} presents a comparison between the results of the \ac{GMM} and those computed by using OpticStudio for different values of the beam waist radius. Without loss of generality, results are presented by assuming that the radial displacement error of the receiver relative to the transmitter is solely comprised of its horizontal component and that the orientation angle error at both sides of the link only involves the azimuth component, i.e. for $y_\mathrm{DE}=0$ and $\phi_\mathrm{e}=\psi_\mathrm{e}=0$. Six example scenarios are inspected. The first set of results include (a) horizontal displacement for $x_\mathrm{DE}=r_\mathrm{DE}$ and $\phi_\mathrm{a}=\psi_\mathrm{a}=0$, (b) azimuth angle error at the transmitter for $x_\mathrm{DE}=0$ and $\psi_\mathrm{a}=0$, and (c) azimuth angle error at the receiver for $x_\mathrm{DE}=0$ and $\phi_\mathrm{a}=0$. The aim of these scenarios is to confirm the \ac{GMM} accuracy when the link undergoes one type of misalignment at a time. The second set of results represents (d) horizontal displacement for $x_\mathrm{DE}=-r_\mathrm{DE}$ and $\phi_\mathrm{a}=0.1^\circ$ and $\psi_\mathrm{a}=10^\circ$, (e) transmitter azimuth angle error for $x_\mathrm{DE}=-2$ mm and $\psi_\mathrm{a}=10^\circ$, and (f) receiver azimuth angle error for $x_\mathrm{DE}=-2$ mm and $\phi_\mathrm{a}=0.1^\circ$. These scenarios are intended to showcase the \ac{GMM} accuracy when a combination of radial displacement and orientation angle errors comes about simultaneously. Note that in scenarios (d), (e) and (f), horizontal displacement increases in the negative direction of the $x'$ axis on the receiver plane. As observed from Figs.~\subref*{fig_4_4d} and \subref*{fig_4_4e}, the channel gain exhibits a peak, which corresponds to the moment that the beam spot center falls onto the \ac{PD}. As shown in Fig.~\ref{fig_4_4}, the \ac{GMM} accurately predicts the channel gain results obtained by OpticStudio simulations and there is an excellent match between them in all cases. This corroborates the high reliability of the proposed analytical modeling framework.

\subsection{Impact of Misalignment}
\begin{figure}[!t]
	\centering
	\subfloat[Config. I ($N_\mathrm{r}=25$)]
	{\hspace{-0.6cm}\includegraphics[width=0.41\linewidth]{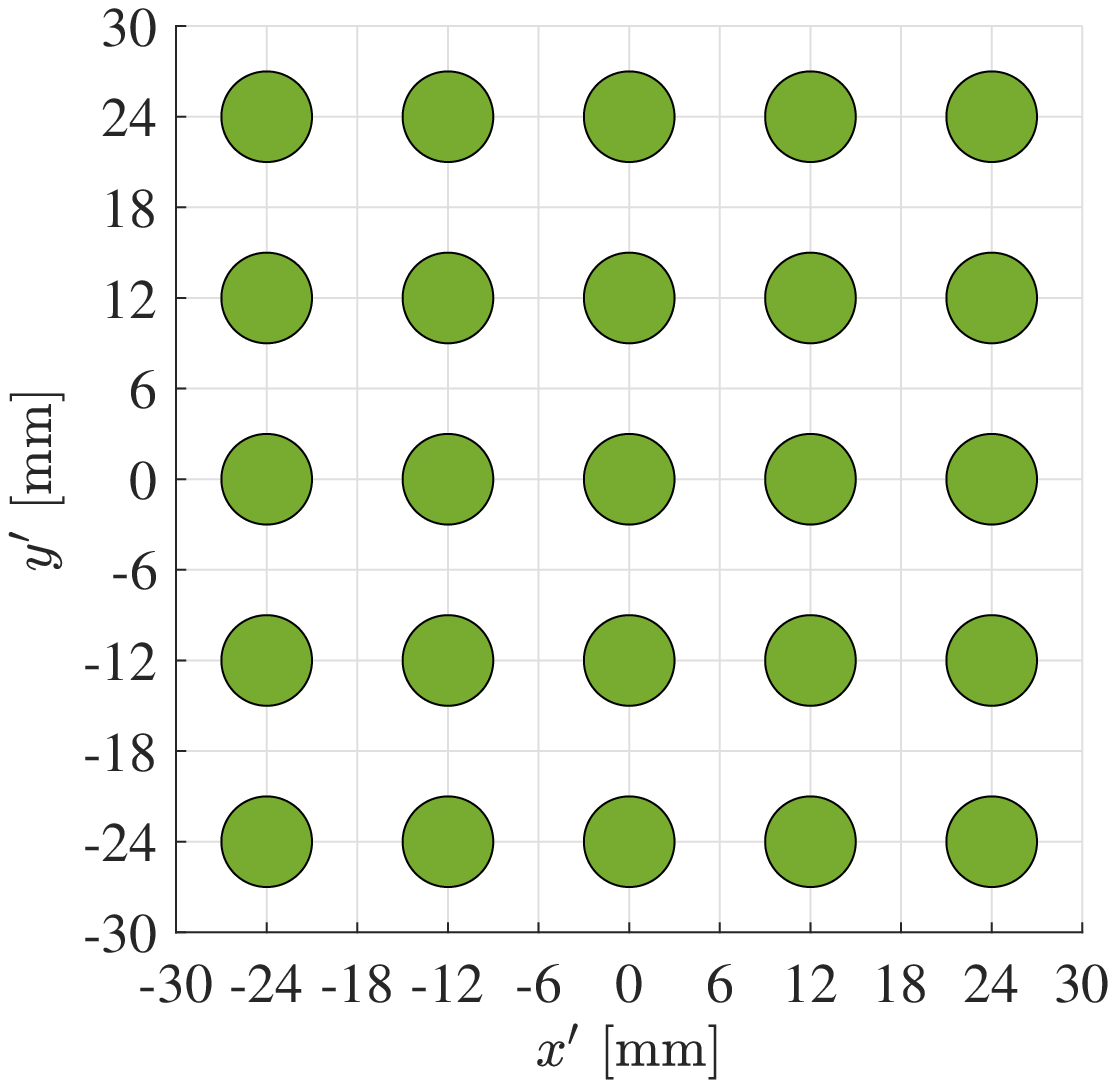}\label{fig_4_6a}\hspace{-1cm}}
	\subfloat[Config. II ($N_\mathrm{r}=41$)]
	{\includegraphics[width=0.41\linewidth]{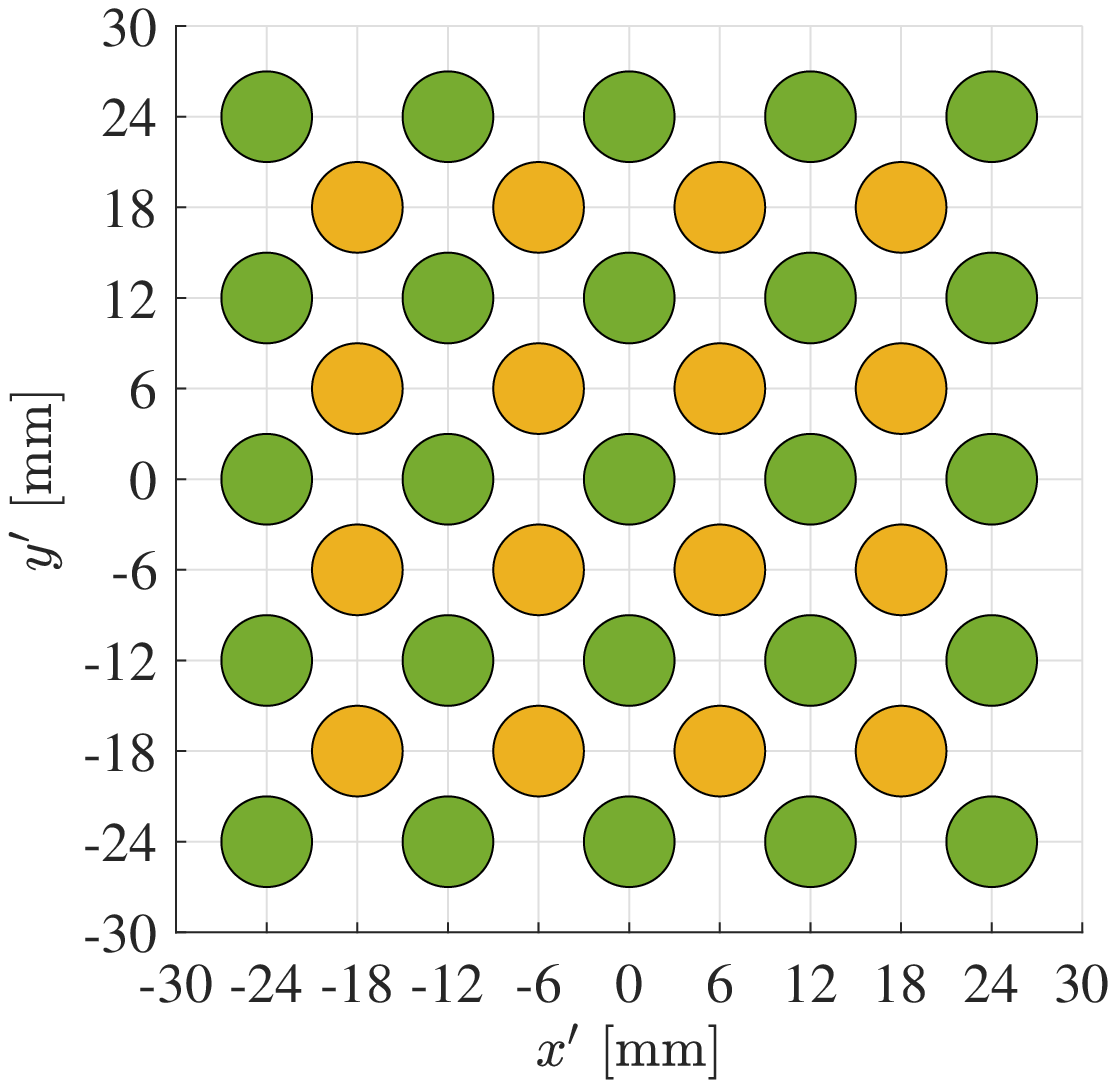}\label{fig_4_6b}\hspace{-1cm}}
	\subfloat[Config. III ($N_\mathrm{r}=81$)]
	{\includegraphics[width=0.41\linewidth]{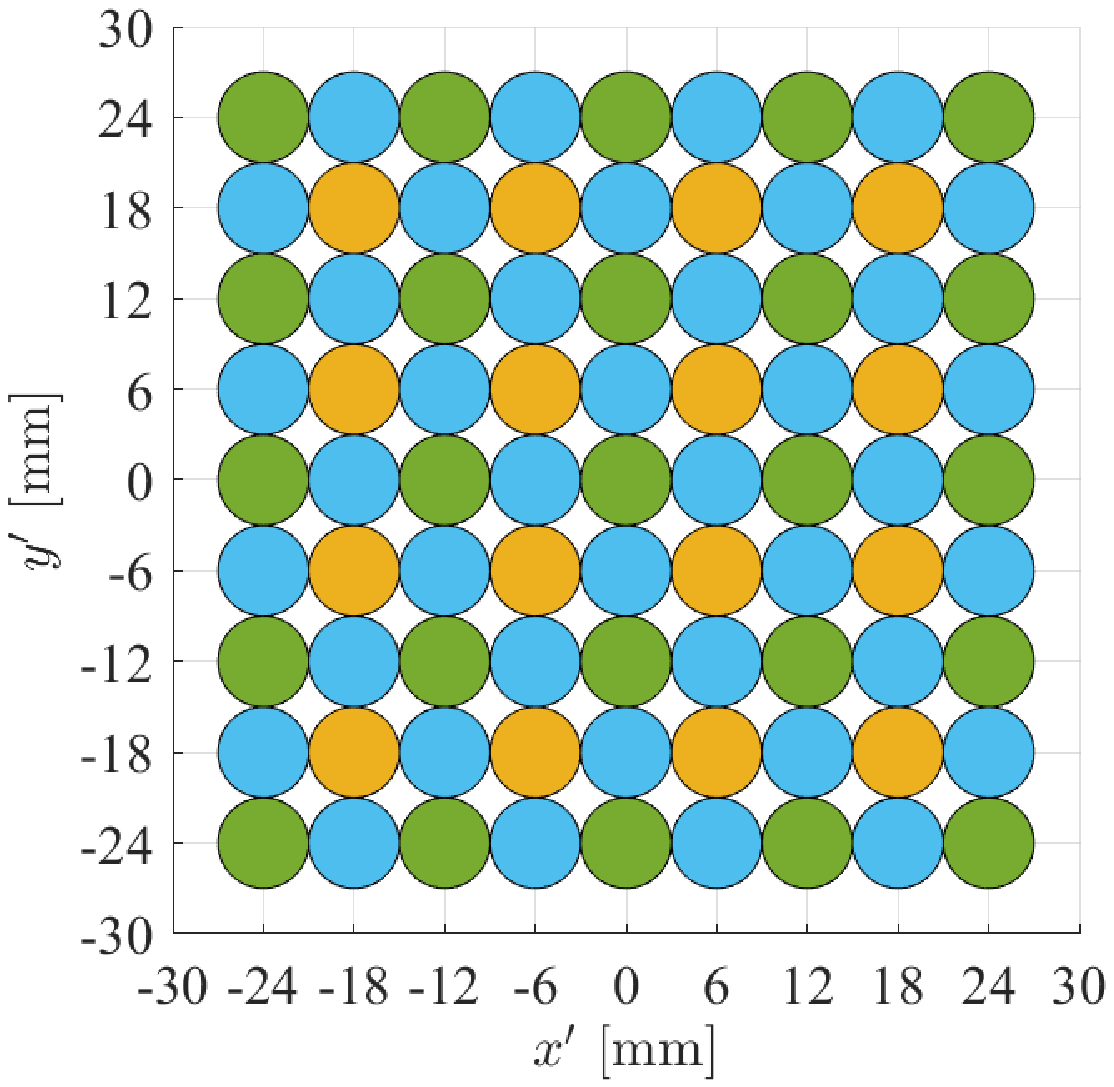}\label{fig_4_6c}\hspace{-0.6cm}}
	\caption{Configurations of the PD array in $N_\mathrm{t}\times N_\mathrm{r}$ MIMO OWC system with $N_\mathrm{t}=25$.}
	\label{fig_4_6}
	\vspace{-20pt}
\end{figure}

The impact of misalignment is studied for an $N_\mathrm{t}\times N_\mathrm{r}$ \ac{MIMO} system with $N_\mathrm{t}=25$ (i.e. $5\times5$ \ac{VCSEL} array) and $w_0=100$ {\textmu}m, as this is the case where the maximum data rate can be achieved, though such very low divergence beams are prone to misalignment errors. To make use of the \ac{SVD} architecture, three configurations of the \ac{PD} array with different number of elements as shown in Fig.~\ref{fig_4_6} are examined, where $N_\mathrm{r}=25,41,81$ in Configs. I, II and III, respectively. For the \ac{PD} array structure described in Section~\ref{sec4_1}, the \ac{FF} can be calculated as $\mathrm{FF}=\frac{N_\mathrm{r}A_\mathrm{PD}}{W^2}$. The corresponding \ac{FF} value for Configs. I, II and III is $\mathrm{FF}=20\%,32\%,64\%$.

\subsubsection{Radial Displacement Error}
\begin{figure}[!t]
	\centering
	\subfloat[Horizontal displacement ($x_\mathrm{DE}=r_\mathrm{DE}$ and $y_\mathrm{DE}=0$)]
	{\includegraphics[width=0.49\linewidth]{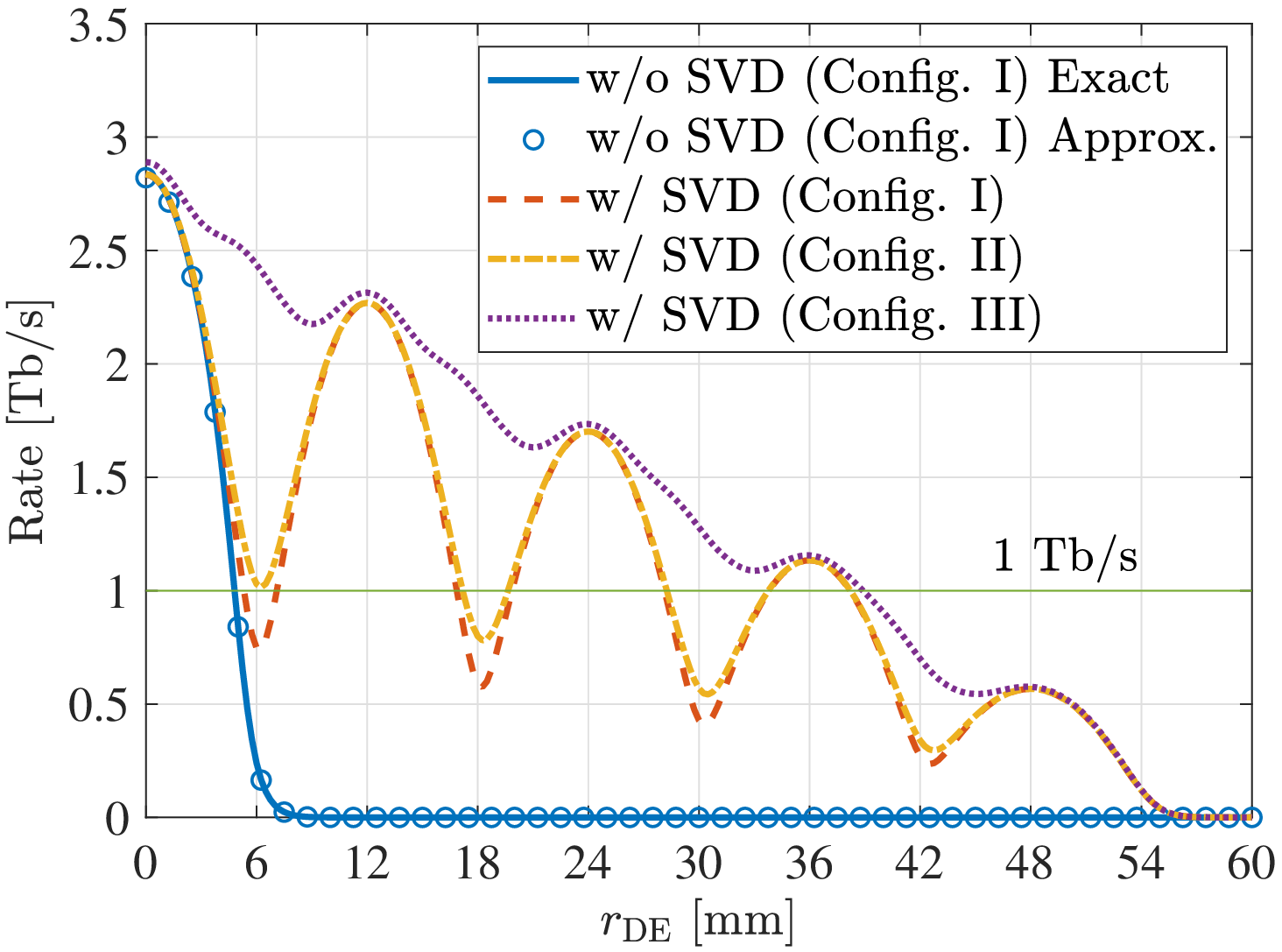}\label{fig_4_5a}}
	\subfloat[Diagonal displacement ($x_\mathrm{DE}=y_\mathrm{DE}=\frac{r_\mathrm{DE}}{\sqrt{2}}$)]
	{\includegraphics[width=0.49\linewidth]{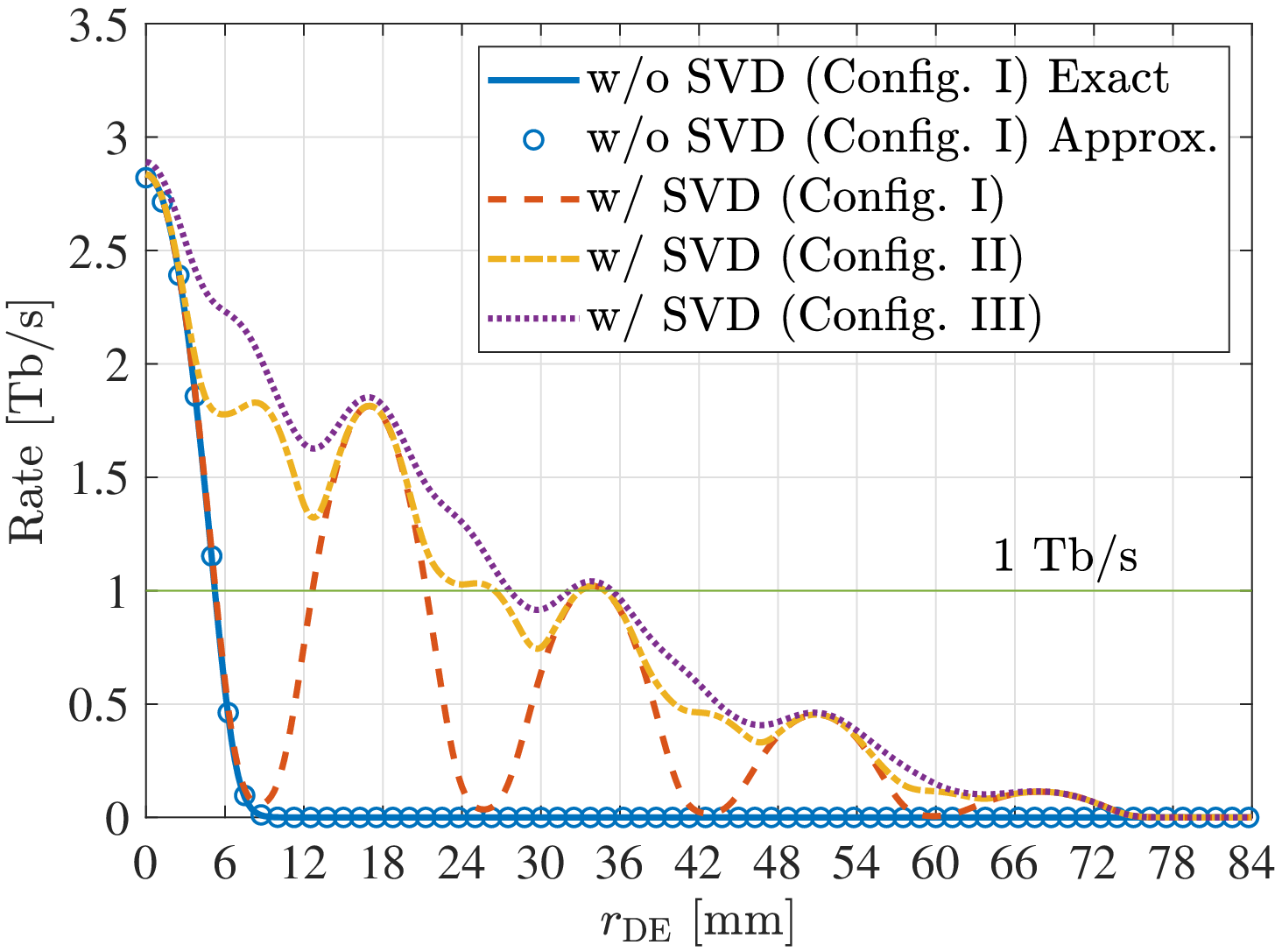}\label{fig_4_5b}}
	\caption{Aggregate data rate as a function of the radial displacement error.}
	\label{fig_4_5}
	\vspace{-20pt}
\end{figure}

Fig.~\ref{fig_4_5} plots the aggregate data rate of the system with respect to the radial displacement error. `Exact' and `Approx.' labels refer to the exact \ac{MIMO} \ac{GMM} procedure from Section~\ref{sec4_2} and the approximate formula in \eqref{eq_A_2} used for computing the \ac{MIMO} channel gains for the case without \ac{SVD}. The results of the two approaches perfectly match with each other for both scenarios shown in Figs.~\ref{fig_4_5a} and \ref{fig_4_5b}. In Fig.~\subref*{fig_4_5a}, the radial displacement takes place along one of the two lateral dimensions of the \ac{PD} array relative to the \ac{VCSEL} array, i.e. horizontally. Note that vertical displacement is similar to this case due to the symmetry in the square lattice of the arrays. It is observed from Fig.~\subref*{fig_4_5a} that without \ac{SVD}, the data rate rapidly drops to zero after $r_\mathrm{DE}=6$ mm. When using \ac{SVD} for Config. I, the rate performance exhibits a decaying oscillation with exactly five peaks, each of which corresponds to an event that a given number of columns of the \ac{VCSEL} array are aligned to those of the \ac{PD} array. The use of Config. II slightly improves the valleys because of the second lattice of \acp{PD} placed on the vertices of the first lattice. This keeps the aggregate data rate above $1$ Tb/s for $r_\mathrm{DE}\leq17$ mm. By comparison, significant performance improvements are attained with \ac{SVD} and Config. III in which $81$ \acp{PD} are densely packed onto the receiver array, thus heightening the chance to acquire strong eigenmodes from the \ac{MIMO} channel as $r_\mathrm{DE}$ increases. This comes at the expense of higher hardware complexity for the \ac{CSI} estimation and the \ac{SVD} computation for a $25\times81$ \ac{MIMO} setup. When the radial displacement occurs diagonally as shown in Fig.~\subref*{fig_4_5b}, the amplitude of oscillations under Config. I becomes larger and the extent of improvements by Config. II is more pronounced. Also, Config. III consistently retains the highest performance.

\subsubsection{Transmitter Orientation Error}
\begin{figure}[!t]
	\centering
	\subfloat[$\phi_\mathrm{e}=0$]
	{\includegraphics[width=0.49\linewidth]{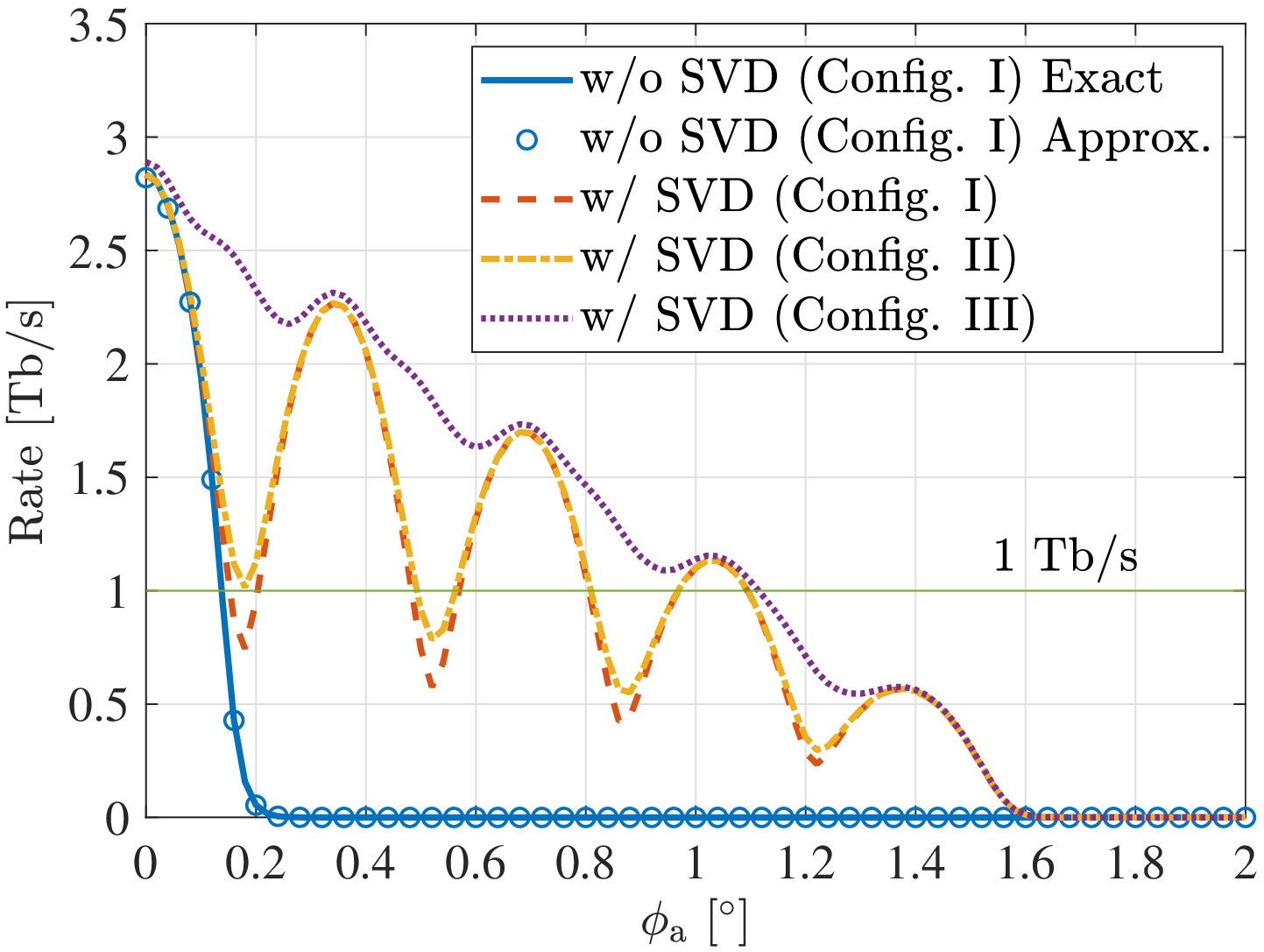}\label{fig_4_7a}}
	\subfloat[$\phi_\mathrm{e}=\phi_\mathrm{a}$]
	{\includegraphics[width=0.49\linewidth]{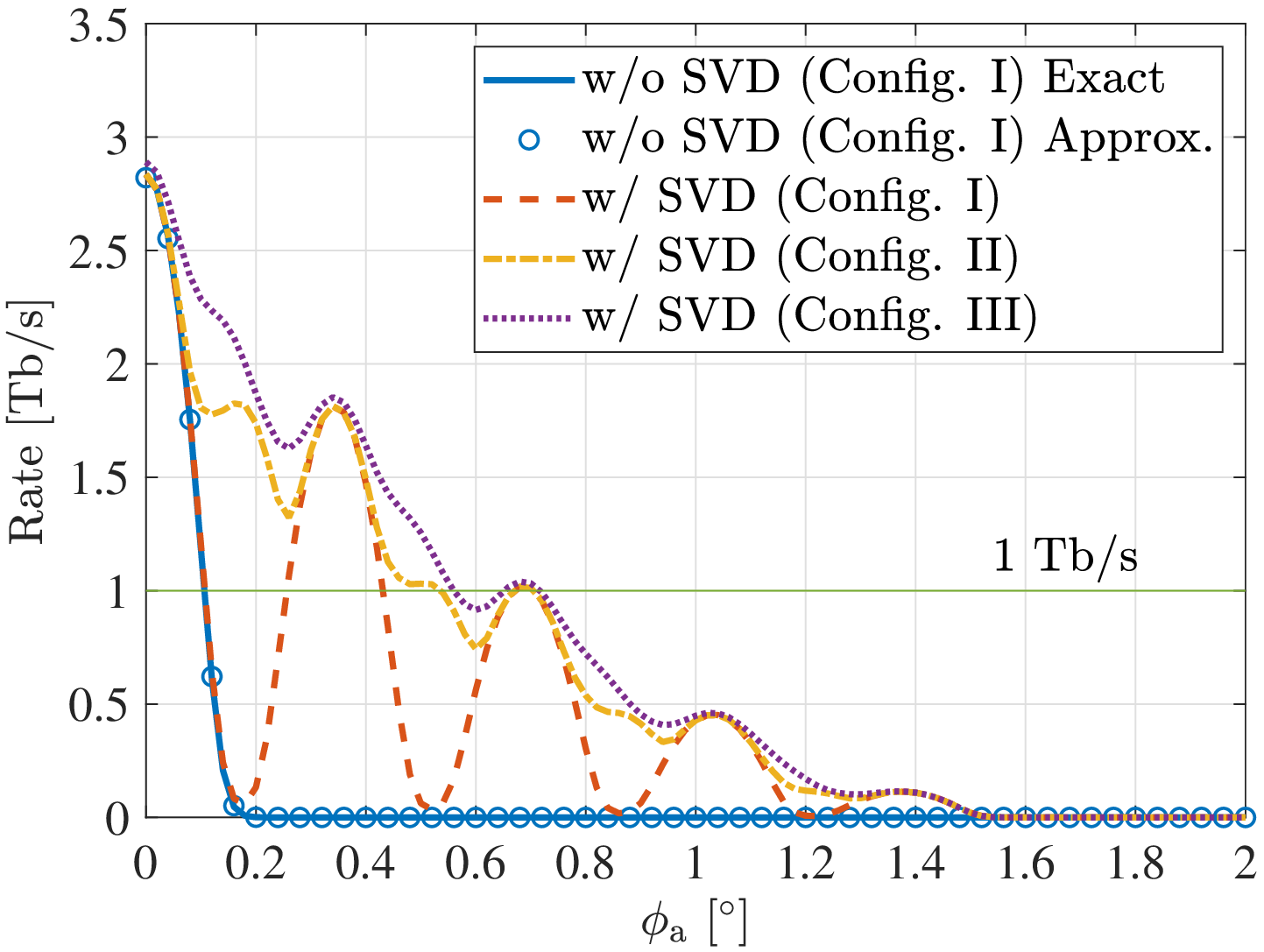}\label{fig_4_7b}}
	\caption{Aggregate data rate as a function of the orientation angle error at the transmitter.}
	\label{fig_4_7}
	\vspace{-20pt}
\end{figure}

Fig.~\ref{fig_4_7} presents the aggregate data rate of the system as a function of the orientation angle error at the transmitter. For the system without \ac{SVD}, the results evaluated by using the approximate expression in \eqref{eq_A_7} perfectly match with those obtained based on the \ac{MIMO} \ac{GMM} from Section~\ref{sec4_2} for both cases shown in Figs.~\ref{fig_4_5a} and \ref{fig_4_5b}. It is evident how sensitive the system performance is with respect to the transmitter orientation error such that a small error of about $1.7^\circ$ is enough to make the data rate zero. This is because the transmitter is $2$ m away from the receiver, and hence small deviations in its orientation angle are translated into large displacements of the beam spots on the other end. In Fig.~\subref*{fig_4_7a}, the orientation error happens only in the azimuth angle $\phi_\mathrm{a}$ by assuming $\phi_\mathrm{e}=0$. The results in Fig.~\subref*{fig_4_7a} have similar trends as those in Fig.~\subref*{fig_4_5a}, except for their different scales in the horizontal axis. In fact, an azimuth angle error of $1.7^\circ$ is equivalent to a horizontal displacement error of $60$ mm. Consequently, $\phi_\mathrm{a}=1.7^\circ$ causes the beam spots of the \acp{VCSEL} to completely miss the receiver array. Likewise, Fig.~\subref*{fig_4_7b} can be paired with Fig.~\subref*{fig_4_5b}. Fig.~\subref*{fig_4_7b} represents the case where the orientation error comes about in both azimuth and elevation angles equally, which produces the same effect as the diagonal displacement error as shown in Fig.~\subref*{fig_4_5b}. Therefore, the transmitter orientation error can be viewed as an equivalent radial displacement error, if the beam spot size at the receiver array is sufficiently small, as formally established in Section~\ref{sec4_3}.

\begin{figure}[!t]
	\centering
	\subfloat[$\psi_\mathrm{e}=0$]
	{\includegraphics[width=0.49\linewidth]{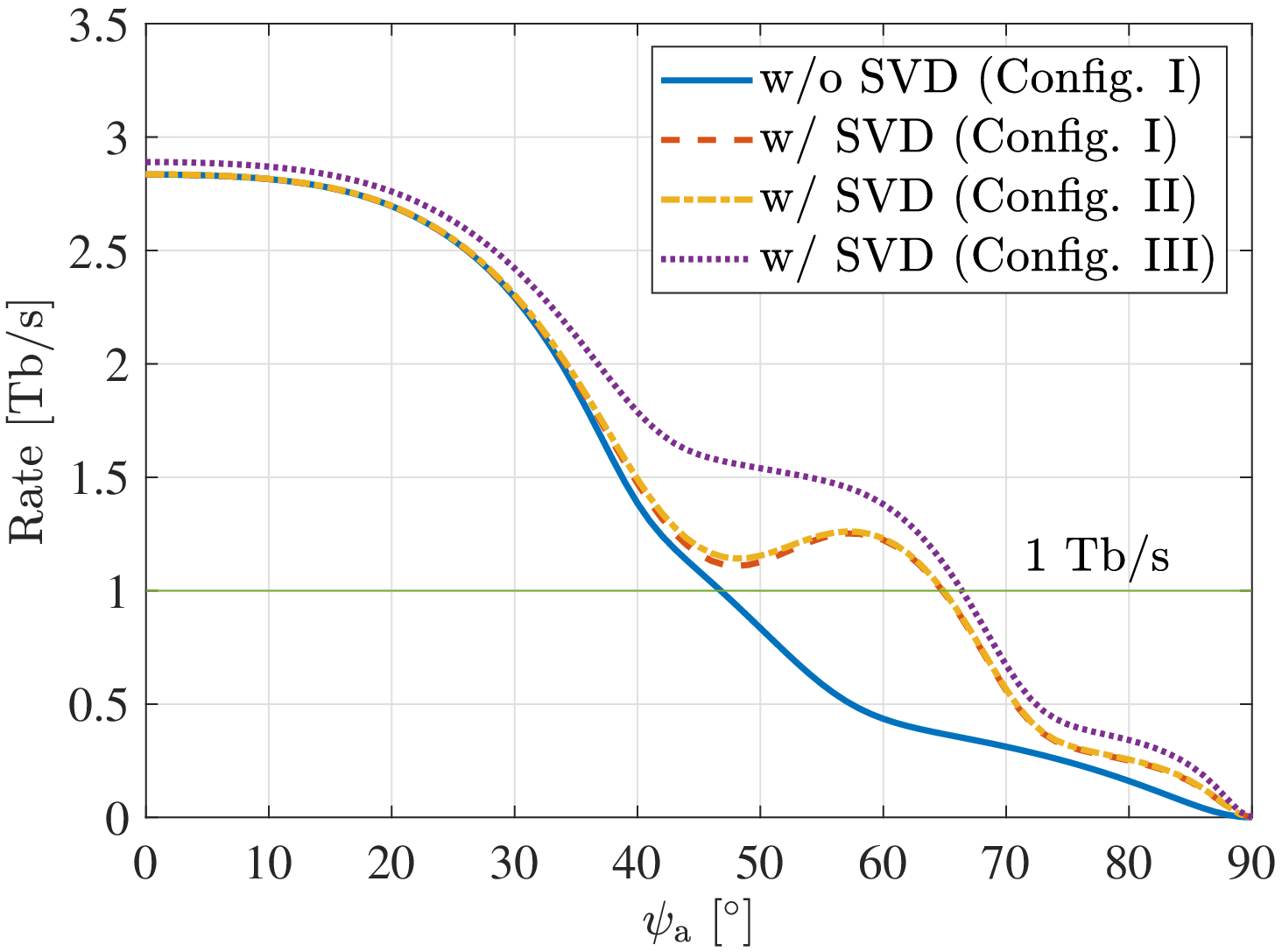}\label{fig_4_8a}}
	\subfloat[$\psi_\mathrm{e}=\psi_\mathrm{a}$]
	{\includegraphics[width=0.49\linewidth]{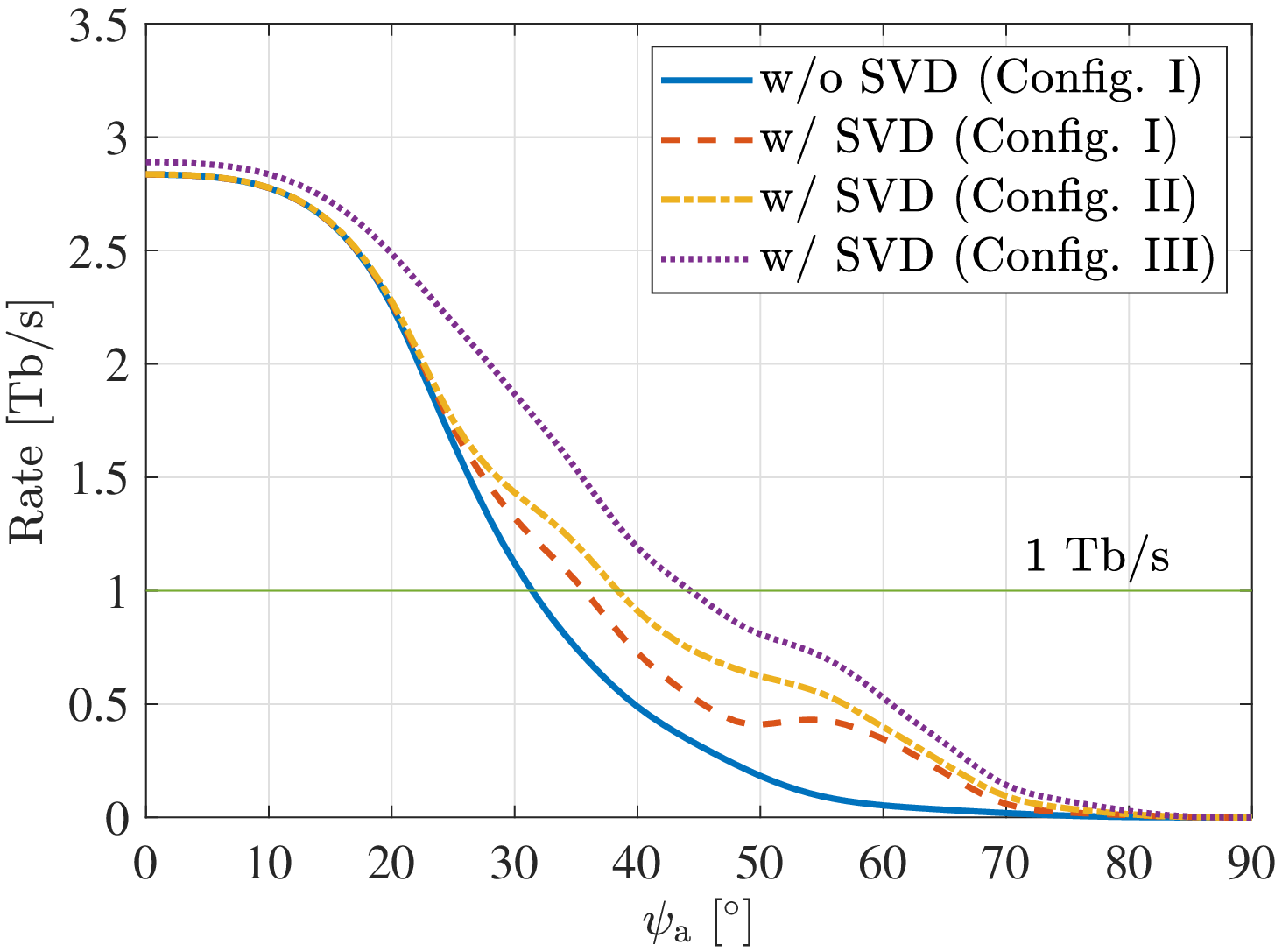}\label{fig_4_8b}}
	\caption{Aggregate data rate as a function of the orientation angle error at the receiver.}
	\label{fig_4_8}
	\vspace{-20pt}
\end{figure}

\subsubsection{Receiver Orientation Error}
Fig.~\ref{fig_4_8} shows the aggregate data rate when the orientation angle error at the receiver is variable. It can be clearly seen that the \ac{MIMO} system is significantly more tolerant against the receiver misalignment as compared to the transmitter misalignment in terms of the orientation angle error. In Fig.~\subref*{fig_4_8a}, the azimuth angle $\psi_\mathrm{a}$ is varied between $0$ and $90^\circ$ while the elevation angle is fixed at $\psi_\mathrm{e}=0$. It is observed that even without \ac{SVD}, the rate is above $1$ Tb/s over a wide range of $\psi_\mathrm{a}$, i.e. for $0\leq\psi_\mathrm{a}\leq46^\circ$. The use of \ac{SVD} gives an almost equal performance for Configs. I and II, providing a noticeable improvement with respect to the case without \ac{SVD} by maintaining the rate above $1$ Tb/s for $0\leq\psi_\mathrm{a}\leq65^\circ$. Since the size of the \acp{PD} is smaller than the size of the beam spots, small rotations of the \ac{PD} array about its axes have a marginal effect on the system performance, unless they are sufficiently large to alter the distribution of the received optical power on the \ac{PD} array. Also, the performance of Config. III is slightly better than Configs. I and II. In Fig.~\subref*{fig_4_8b}, where $\psi_\mathrm{e}=\psi_\mathrm{a}$, the performance without \ac{SVD} crosses $1$ Tb/s at $\psi_\mathrm{a}=31^\circ$. This occurs at $\psi_\mathrm{a}=36^\circ,39^\circ,44^\circ$ for Configs. I, II and III.

\subsubsection{Performance-Complexity Tradeoff}
The use of Config. III leads to a significantly smoother curve of the aggregate data rate as shown in Figs.~\ref{fig_4_5} and \ref{fig_4_7}. However, since the hardware complexity is mainly due to the \ac{SVD} computation, increasing the number of elements of the receiver array heightens the hardware complexity too. The time complexity associated with computing the full \ac{SVD} for the $N_\mathrm{r}\times N_\mathrm{t}$ matrix of the \ac{MIMO} channel based on the \textit{R}-\ac{SVD} algorithm is $\mathcal{O}(4N_\mathrm{r}^2N_\mathrm{t}+22N_\mathrm{t}^3)$ \cite{Golub}. Thus, the computational complexity for Configs. I, II and III, respectively, is $\mathcal{O}(4\times10^5)$, $\mathcal{O}(5\times10^5)$ and $\mathcal{O}(10^6)$. Therefore, Config. III requires $2.5$ times higher computational power than Config. I. To elaborate on the performance gain, in the case of radial displacement as shown in Fig.~\ref{fig_4_5}, the first crossover of the aggregate rate from the $1$ Tb/s threshold takes place at horizontal displacements $r_\mathrm{DE}=5.3,17.2,38.8$ mm for Configs.~I, II and III, respectively. Compared with Config. I, Config. III provides more than $7$ times wider room for misalignment tolerance while incurring $2.5$ times higher complexity. This underlines the point that it is worthwhile to use Config. III for the system implementation.

\section{Conclusions}\label{sec6}
A \ac{VCSEL}-based \ac{MIMO} \ac{OWC} system using \ac{DCO{-}OFDM} and spatial multiplexing techniques was designed and elaborated. The fundamental problem of the link misalignment was supported by extensive analytical modeling and the generalized model of misalignment errors was derived for both \ac{SISO} and \ac{MIMO} link configurations. Under perfect alignment conditions, data rates of $\geq1.016$ Tb/s are achievable with a $25\times25$ \ac{MIMO} system without \ac{SVD} over a link distance of $2$ m for a beam waist radius of $w_0\geq50$ {\textmu}m, equivalent to a beam divergence angle of $\theta\leq0.3^\circ$, while fulfilling the eye safety constraint. The same system setup attains $\geq1.264$ Tb/s data rates when \ac{SVD} is applied. The use of $w_0=100$ {\textmu}m (i.e. $\theta=0.16^\circ$) renders beam spots on the receiver array almost nonoverlapping and the aforementioned $25\times25$ system delivers $2.835$ Tb/s with or without \ac{SVD}. In fact, when the link is perfectly aligned, the use of \ac{SVD} is essentially effective in the crosstalk-limited regime. The derived \ac{GMM} was used to study the effect of different misalignment errors on the system performance. Under radial displacement error or orientation angle error at the transmitter, the performance of \ac{MIMO} systems with \ac{SVD} shows a declining oscillation behavior with increase in the error value. For a $25\times25$ system using $w_0=100$ {\textmu}m, the aggregate rate stays above the $1$ Tb/s level for horizontal displacements of up to $r_\mathrm{DE}=17$ mm ($0.28$ relative to the array side length). The performance remains over $1$ Tb/s for an orientation error of $\phi\leq0.8^\circ$ in the azimuth angle of the transmitter. In the presence of the receiver orientation error, the aggregate rate is maintained above $1$ Tb/s over a wide range of $\psi\leq65^\circ$ for the azimuth angle of the receiver. The results indicate that the orientation angle error at the transmitter is the most impactful type of misalignment. They also confirm that the impact of misalignment is alleviated by using a receiver array with densely packed \ac{PD} elements, improving the system tolerance against misalignment errors. This is especially pronounced for the radial displacement and orientation angle error at the transmitter. Future research involves extended system modeling and performance evaluation under practical design limitations including multi-mode output profile of \acp{VCSEL}, frequency-selective modulation response of \acp{VCSEL} and \acp{PD}, and receiver optics. An interesting application of the proposed Tb/s \ac{OWC} backhaul system is wireless inter-rack communications in high-speed data center networks, which brings an avenue for future research.

\vspace{-20pt}
\appendix
\section{Verifying the Approximations in \eqref{eq_A_2} and \eqref{eq_A_7}}\label{sec_A}
In this appendix, the aim is to verify the accuracy of the approximations applied to the \ac{MIMO} \ac{GMM} in Section~\ref{sec4_3} to derive \eqref{eq_A_2} and \eqref{eq_A_7}. For clarity, the \ac{SISO} channel gain between $\text{VCSEL}_j$ and $\text{PD}_i$ is evaluated for $x_i=y_i=0$. Due to the circular symmetry of the beam, it is assumed that the radial displacement is located along the $x'$ axis on the receiver plane such that $x_\mathrm{DE}=r_\mathrm{DE}$ and $y_\mathrm{DE}=0$, and the corresponding channel gain is denoted by $H_0(r_\mathrm{DE})$. Similarly, it can be assumed that only the azimuth component $\phi_\mathrm{a}$ of the transmitter orientation angle is non-zero so that $\phi_\mathrm{e}=0$, and the resulting channel gain is represented by $H_0(\phi_\mathrm{a})$. The exact values are numerically computed based on \eqref{eq_2_6}, by using \eqref{eq_A_8}, \eqref{eq_A_3} and \eqref{eq_A_4} in Section~\ref{sec4_3}.

\begin{table}[!t]
	\centering
	\caption{NMSE between the exact and approximate values for $H_0(r_\mathrm{DE})$ and $H_0(\phi_\mathrm{a})$}
	\begin{tabular}{c|c|c|c|c|c}
		$\frac{w(L)}{r_\mathrm{PD}}$ & $1$     & $2$     & $3$ & $4$ & $5$ \\ \hline
		$\text{NMSE}(H_0(r_\mathrm{DE}))$ & $6.3534\times10^{-4}$ & $5.7037\times10^{-5}$ & $1.4768\times10^{-5}$ & $5.1984\times10^{-6}$ & $2.2401\times10^{-6}$ \\
		$\text{NMSE}(H_0(\phi_\mathrm{a}))$ & $6.1092\times10^{-4}$ & $5.7647\times10^{-5}$ & $1.4911\times10^{-5}$ & $5.2373\times10^{-6}$ & $2.2532\times10^{-6}$ \\ \hline
	\end{tabular}
	\label{secA_tbl1}
	\vspace{-20pt}
\end{table}

\begin{figure}[!t]
	\centering
	\subfloat[Radial displacement]
	{\includegraphics[width=0.49\linewidth]{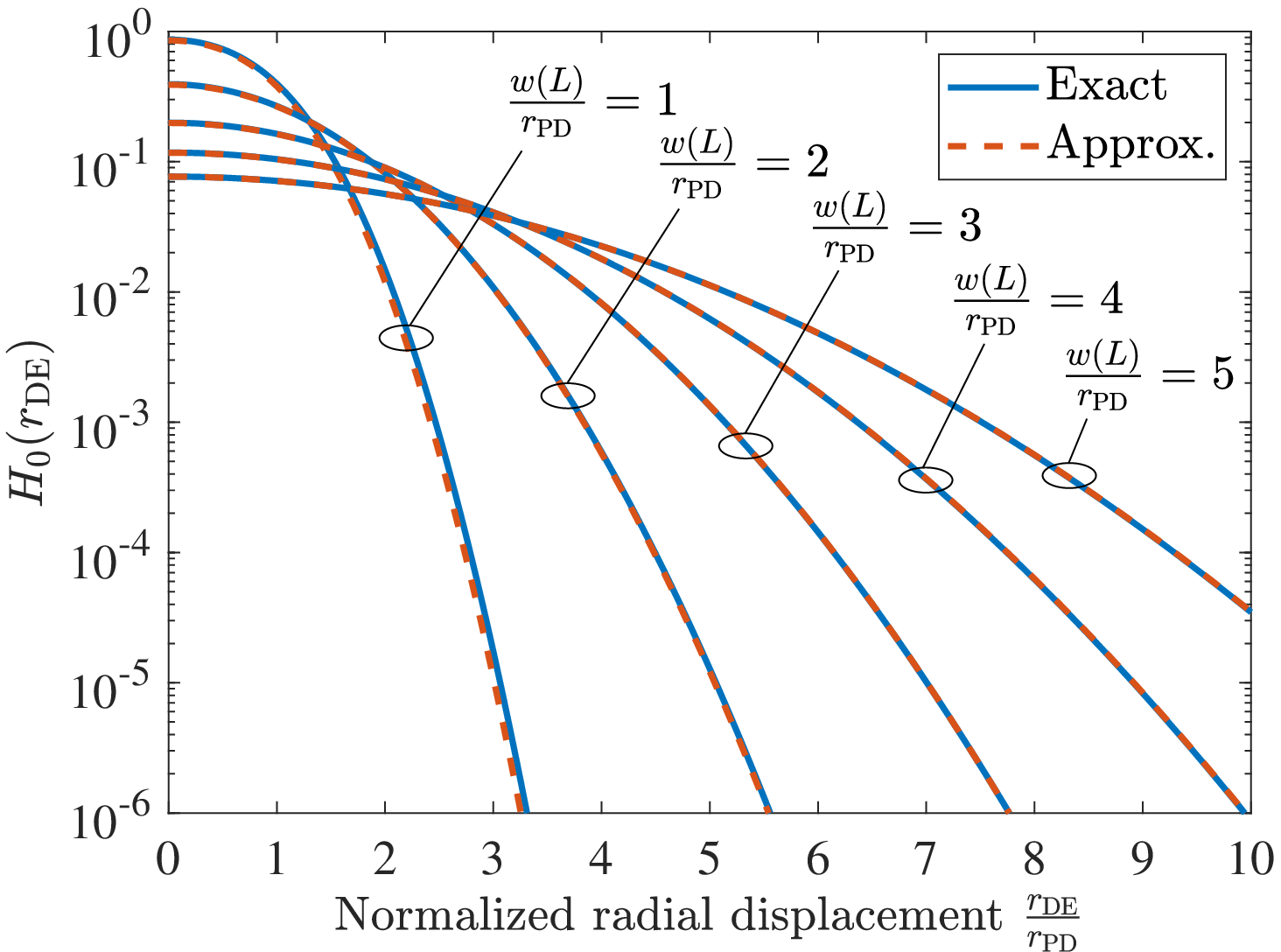}\label{fig_A_1a}}
	\subfloat[Transmitter orientation angle]
	{\includegraphics[width=0.49\linewidth]{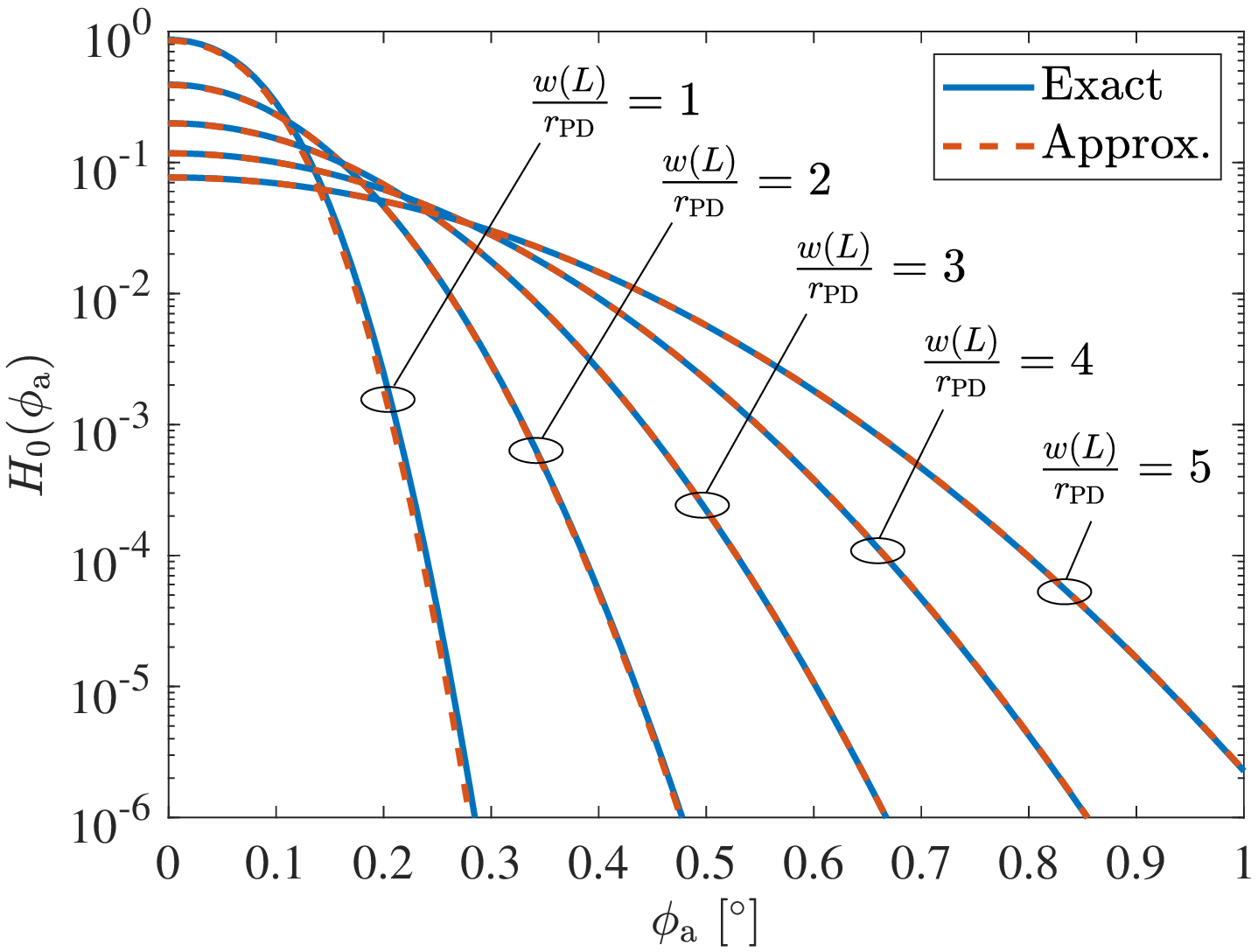}\label{fig_A_1b}}
	\caption{Comparison between the exact and approximate expressions of $H_0(r_\mathrm{DE})$ and $H_0(\phi_\mathrm{a})$.}
	\label{fig_A_1}
	\vspace{-20pt}
\end{figure}

Fig.~\ref{fig_A_1} demonstrates the comparison between the exact and approximate expressions for both radial displacement and orientation angle error at the transmitter. The horizontal axis represents the normalized radial displacement $\frac{r_\mathrm{DE}}{r_\mathrm{PD}}$ and the results are shown for different values of the ratio $\frac{w(L)}{r_\mathrm{PD}}$. It can be seen that the proposed approximations excellently conform with the exact values especially when $\frac{w(L)}{r_\mathrm{PD}}>1$. In the case where the beam spot size is equal to the \ac{PD} size, the approximate expressions slightly underestimate the exact values in a logarithmic scale. As a measure of accuracy, \ac{NMSE} between the exact and approximate expressions are calculated in Table~\ref{secA_tbl1}. By increasing $\frac{w(L)}{r_\mathrm{PD}}$, \ac{NMSE} rapidly decreases for both cases and $\text{NMSE}<10^{-3}$ always holds. Hence, the underlying approximations are highly accurate for $\frac{w(L)}{r_\mathrm{PD}}\geq1$. Note that this conclusion is independent of the link distance as the results are presented in terms of the beam spot size at the receiver. The smallest beam spot size considered in Section~\ref{sec5} is $w(L)=5.4$ mm for $w_0=100$ {\textmu}m, in which case $\frac{w(L)}{r_\mathrm{PD}}=1.8$ and $\text{NMSE}=7.9055\times10^{-5}$.

\newpage

\bibliographystyle{IEEEtran}
\bibliography{IEEEabrv,List_of_References}

\newpage

\begin{IEEEbiography}
	[{\includegraphics[width=1in,height=1.25in,clip,keepaspectratio]{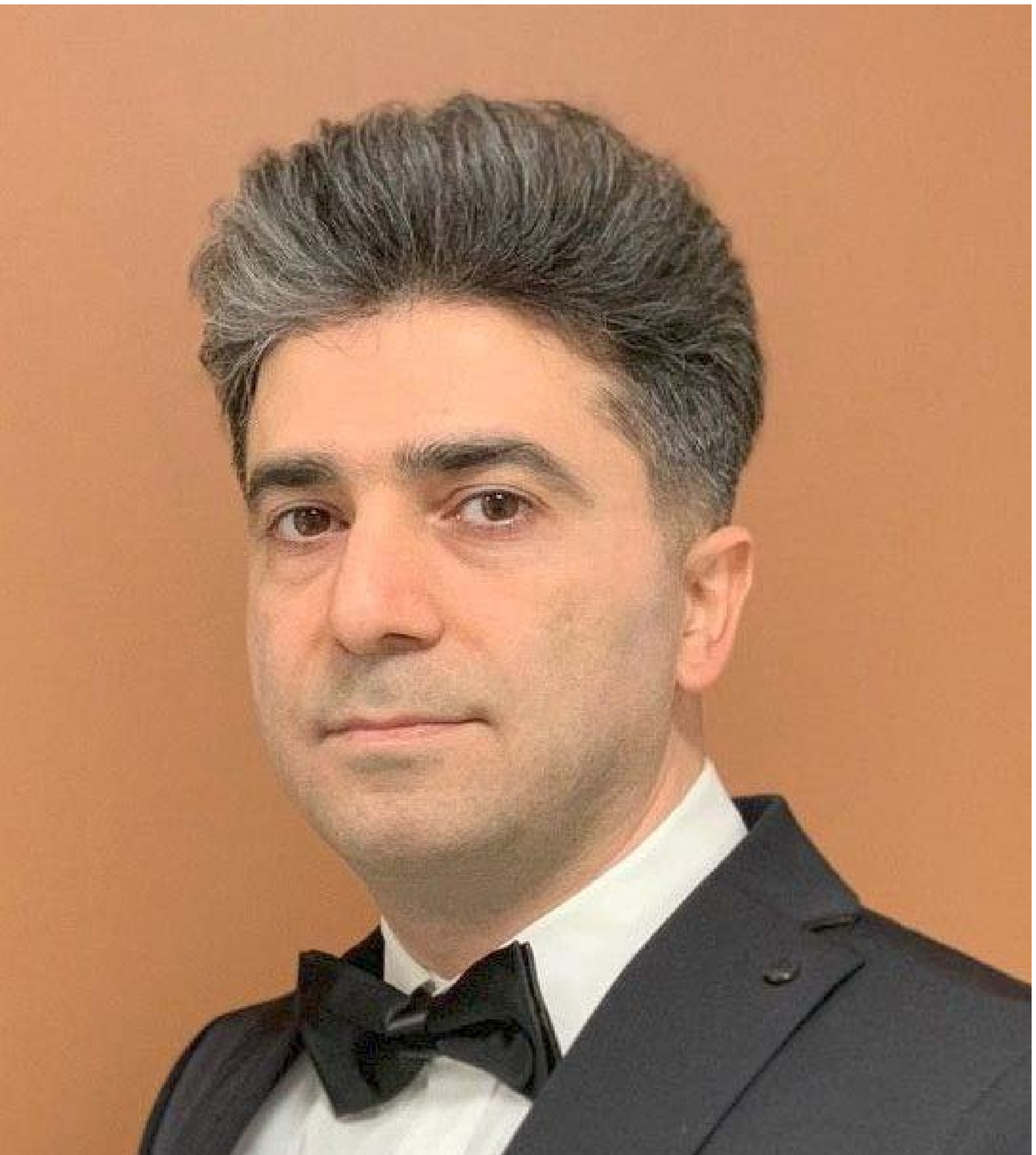}}]{Hossein Kazemi} received his Ph.D. degree in Electrical Engineering from the University of Edinburgh, UK in 2019. He also received an M.Sc. degree in Electrical Engineering (Microelectronic Circuits) from Sharif University of Technology, Tehran, Iran in 2011, and an M.Sc. degree (Hons) in Electrical Engineering (Wireless Communications) from Ozyegin University, Istanbul, Turkey in 2014. He is currently a Research Associate at the LiFi Research and Development Centre, University of Strathclyde, Glasgow, UK. His main research interests include wireless communications and optical wireless communications.
\end{IEEEbiography}

\vspace{-2cm}

\begin{IEEEbiography}
	[{\includegraphics[width=1in,height=1.25in,clip,keepaspectratio]{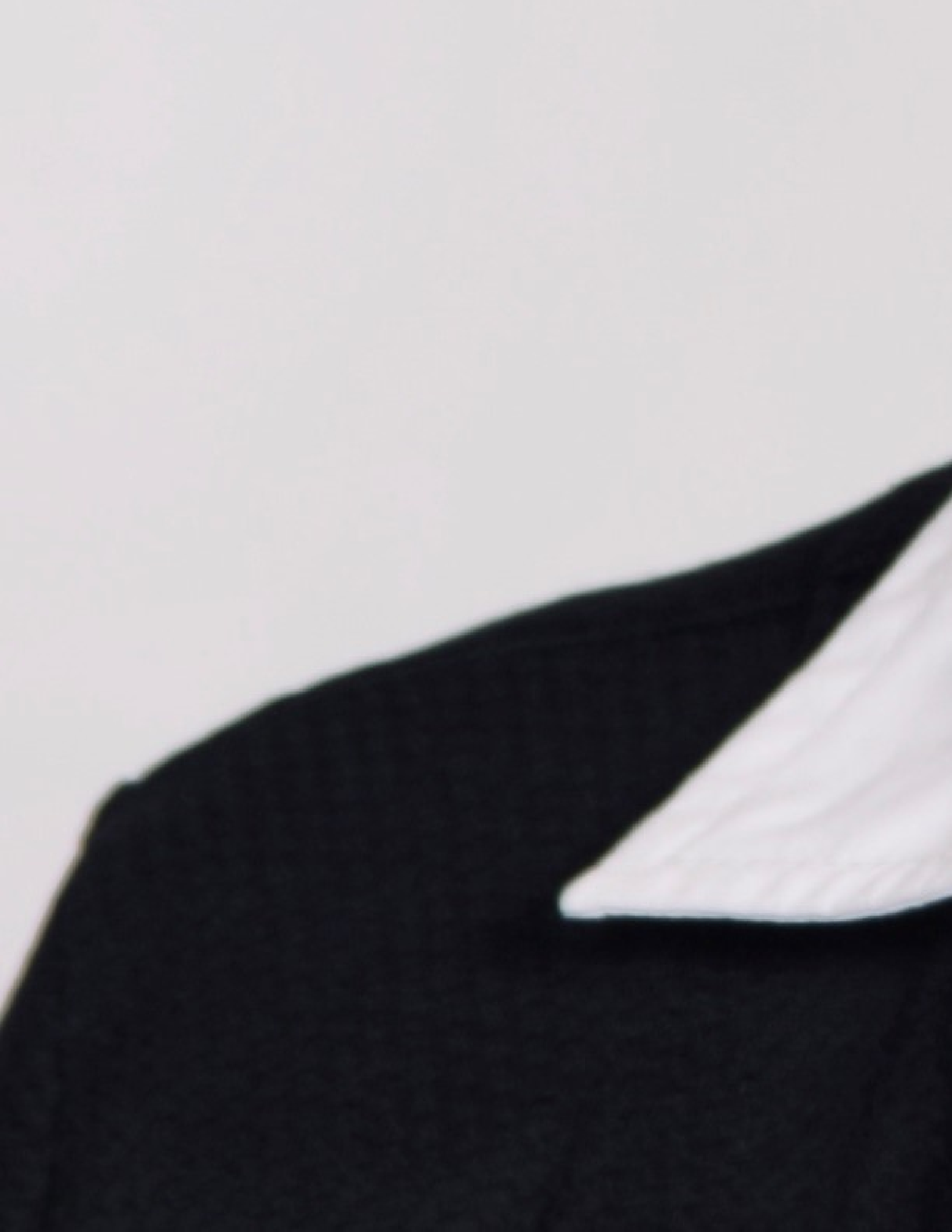}}]{Elham Sarbazi} received her Ph.D. degree in Electrical Engineering from the University of Edinburgh, UK, in 2019. She also received her B.Sc. degree in Electrical and Computer Engineering from the University of Tehran, Iran, in 2011, and her M.Sc. degree in Electrical Engineering from Ozyegin University, Istanbul, Turkey, in 2014. She is currently a Research Associate at the LiFi Research and Development Centre, University of Strathclyde, Glasgow, UK. Her main research interests include optical wireless communications, visible light communications and photon-counting optical receivers.
\end{IEEEbiography}

\vspace{-2cm}

\begin{IEEEbiography}
	[{\includegraphics[width=1in,height=1.25in,clip,keepaspectratio]{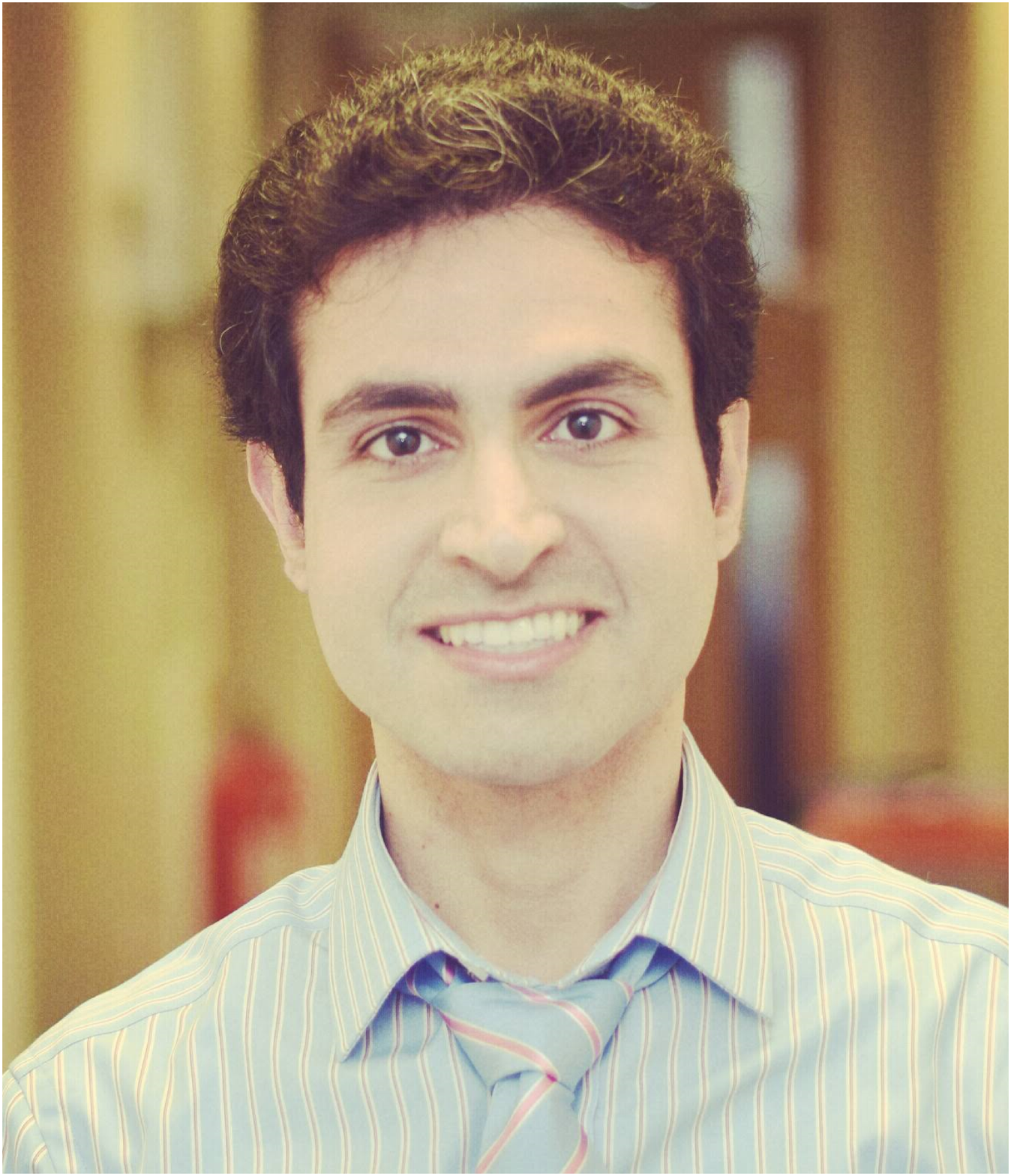}}]{Mohammad Dehghani Soltani} received the M.Sc. degree from the Department of Electrical Engineering, Amirkabir University of Technology, Tehran, Iran, in 2012, and the Ph.D. degree in Electrical Engineering from the University of Edinburgh, UK, in 2019. During his M.Sc., he was studying wireless communications, MIMO coding and low complexity design of MIMO-OFDM systems. He worked for two years in the telecommunication industry in  Iran. He is currently a Research Associate with the Institute for Digital Communications (IDCOM) at the University of Edinburgh. His current research interests include mobility and handover management in wireless cellular networks, optical wireless communications, visible light communications, and LiFi.
\end{IEEEbiography}

\vspace{-2cm}

\begin{IEEEbiography}
	[{\includegraphics[width=1in,height=1.25in,clip,keepaspectratio]{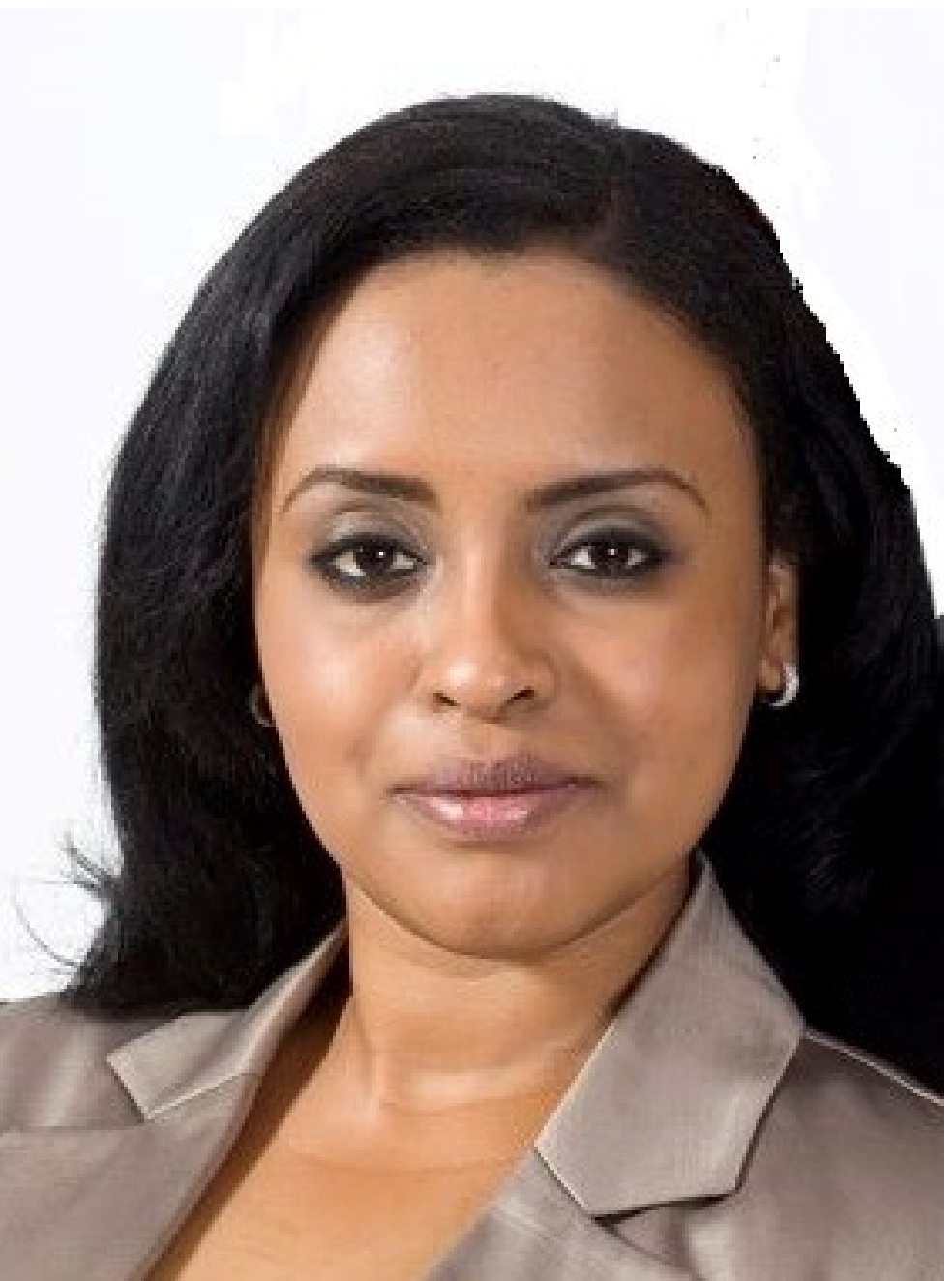}}]{Taisir E. H. El-Ghorashi} received the B.S. degree (first-class Hons.) in Electrical and Electronic Engineering from the University of Khartoum, Khartoum, Sudan, in 2004, the M.Sc. degree (with distinction) in Photonic and Communication Systems from the University of Wales, Swansea, UK, in 2005, and the PhD degree in Optical Networking from the University of Leeds, Leeds, UK, in 2010. She is currently a Lecturer in optical networks in the School of Electronic and Electrical Engineering, University of Leeds. Previously, she held a Postdoctoral Research post at the University of Leeds (2010– 2014), where she focused on the energy efficiency of optical networks investigating the use of renewable energy in core networks, green IP over WDM networks with datacenters, energy efficient physical topology design, energy efficiency of content distribution networks, distributed cloud computing, network virtualization and big data. In 2012, she was a BT Research Fellow, where she developed energy efficient hybrid wireless-optical broadband access networks and explored the dynamics of TV viewing behavior and program popularity. The energy efficiency techniques developed during her postdoctoral research contributed 3 out of the 8 carefully chosen core network energy efficiency improvement measures recommended by the GreenTouch consortium for every operator network worldwide. Her work led to several invited talks at GreenTouch, Bell Labs, Optical Network Design and Modelling conference, Optical Fiber Communications conference, International Conference on Computer Communications, EU Future Internet Assembly, IEEE Sustainable ICT Summit and IEEE 5G World Forum and collaboration with Nokia and Huawei.
\end{IEEEbiography}

\begin{IEEEbiography}
	[{\includegraphics[width=1in,height=1.25in,clip,keepaspectratio]{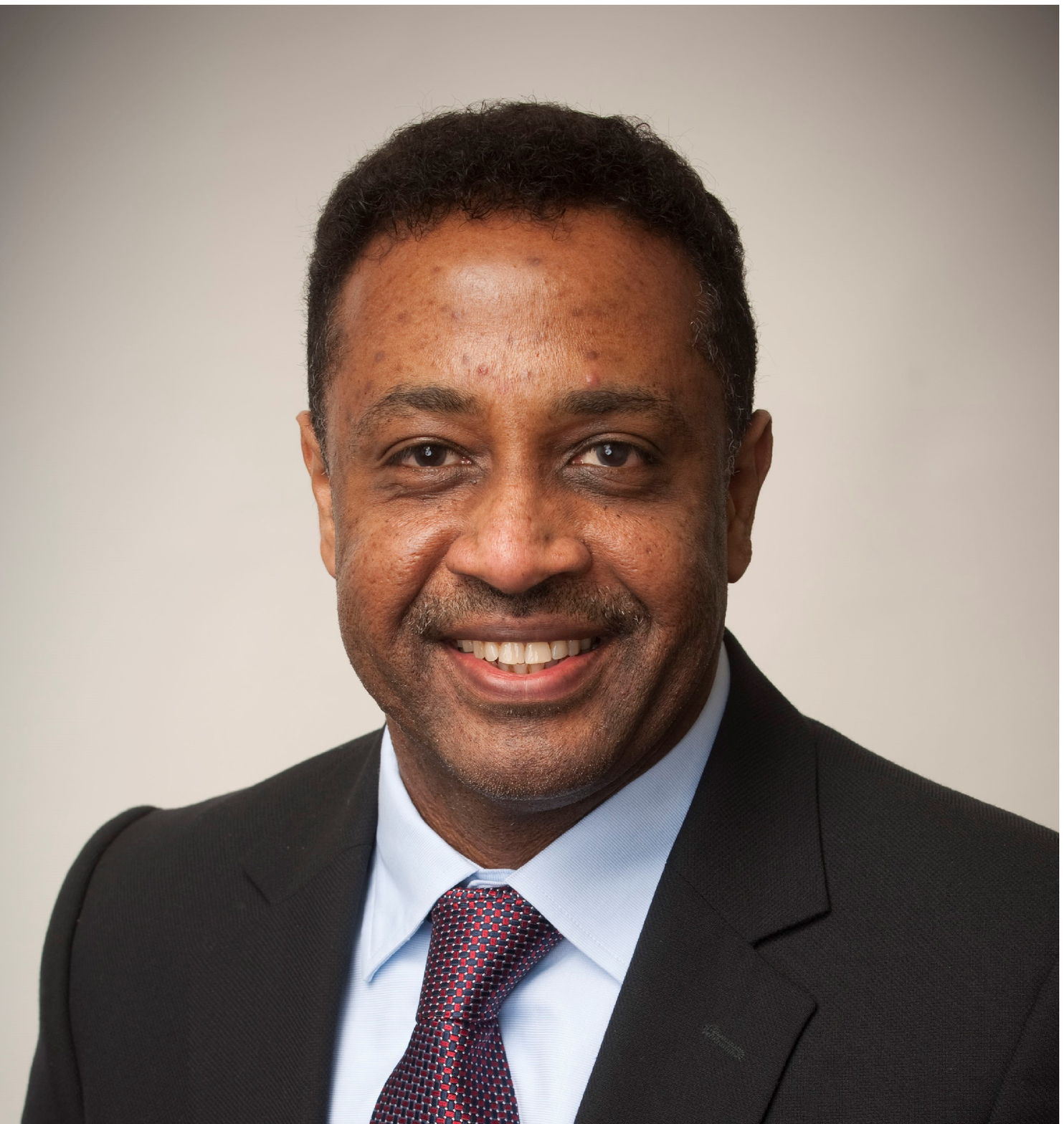}}]{Professor Jaafar Elmirghani} is Fellow of IEEE, Fellow of the IET, Fellow of the Institute of Physics and is the Director of the Institute of Communication and Power Networks and Professor of Communication Networks and Systems within the School of Electronic and Electrical Engineering, University of Leeds, UK. He joined Leeds in 2007 having been full professor and chair in Optical Communications at the University of Wales Swansea 2000-2007. 
	He received the BSc in Electrical Engineering, First Class Honours from the University of Khartoum in 1989 and was awarded all 4 prizes in the department for academic distinction. He received the PhD in the synchronization of optical systems and optical receiver design from the University of Huddersfield UK in 1994 and the DSc in Communication Systems and Networks from University of Leeds, UK, in 2012. He co-authored Photonic Switching Technology: Systems and Networks, (Wiley) and has published over 550 papers.
	He was Chairman of the IEEE UK and RI Communications Chapter and was Chairman of IEEE Comsoc Transmission Access and Optical Systems Committee and Chairman of IEEE Comsoc Signal Processing and Communication Electronics (SPCE) Committee. He was a member of IEEE ComSoc Technical Activities Council’ (TAC), is an editor of IEEE Communications Magazine and is and has been on the technical program committee of 42 IEEE ICC/GLOBECOM conferences between 1995 and 2022 including 21 times as Symposium Chair. He was founding Chair of the Advanced Signal Processing for Communication Symposium which started at IEEE GLOBECOM’99 and has continued since at every ICC and GLOBECOM. Prof. Elmirghani was also founding Chair of the first IEEE ICC/GLOBECOM optical symposium at GLOBECOM’00, the Future Photonic Network Technologies, Architectures and Protocols Symposium. He chaired this Symposium, which continues to date. He was the founding chair of the first Green Track at ICC/GLOBECOM at GLOBECOM 2011, and is Chair of the IEEE Sustainable ICT Initiative, a pan IEEE Societies Initiative responsible for Green ICT activities across IEEE, 2012-present. He has given over 90 invited and keynote talks over the past 15 years.
	He received the IEEE Communications Society 2005 Hal Sobol award for exemplary service to meetings and conferences, the IEEE Communications Society 2005 Chapter Achievement award, the University of Wales Swansea inaugural ‘Outstanding Research Achievement Award’, 2006, the IEEE Communications Society Signal Processing and Communication Electronics outstanding service award, 2009, a best paper award at IEEE ICC’2013, the IEEE Comsoc Transmission Access and Optical Systems outstanding Service award 2015 in recognition of “Leadership and Contributions to the Area of Green Communications”, the GreenTouch 1000x award in 2015 for “pioneering research contributions to the field of energy efficiency in telecommunications", the IET 2016 Premium Award for best paper in IET Optoelectronics, shared the 2016 Edison Award in the collective disruption category with a team of 6 from GreenTouch for their joint work on the GreenMeter, the IEEE Communications Society Transmission, Access and Optical Systems technical committee 2020 Outstanding Technical Achievement Award for outstanding contributions to the “energy efficiency of optical communication systems and networks”. He was named among the top 2\% of scientists in the world by citations in 2019 and 2020 in Elsevier Scopus, Stanford University database which includes the top 2\% of scientists in 22 scientific disciplines and 176 sub-domains. He was elected Fellow of IEEE for “Contributions to Energy-Efficient Communications,” 2021.
	He is currently an Area Editor of IEEE Journal on Selected Areas in Communications series on Machine Learning for Communications, an editor of IEEE Journal of Lightwave Technology, IET Optoelectronics and Journal of Optical Communications, and was editor of IEEE Communications Surveys and Tutorials and IEEE Journal on Selected Areas in Communications series on Green Communications and Networking. He was Co-Chair of the GreenTouch Wired, Core and Access Networks Working Group, an adviser to the Commonwealth Scholarship Commission, member of the Royal Society International Joint Projects Panel and member of the Engineering and Physical Sciences Research Council (EPSRC) College. 
	He has been awarded in excess of £30 million in grants to date from EPSRC, the EU and industry and has held prestigious fellowships funded by the Royal Society and by BT. He was an IEEE Comsoc Distinguished Lecturer 2013-2016. He was PI of the £6m EPSRC Intelligent Energy Aware Networks (INTERNET) Programme Grant, 2010-2016 and is currently PI of the EPSRC £6.6m Terabit Bidirectional Multi-user Optical Wireless System (TOWS) for 6G LiFi, 2019-2024. He leads a number of research projects and has research interests in communication networks, wireless and optical communication systems.
\end{IEEEbiography}

\vspace{-2.21cm}

\begin{IEEEbiography}
	[{\includegraphics[width=1in,height=1.25in,clip,keepaspectratio]{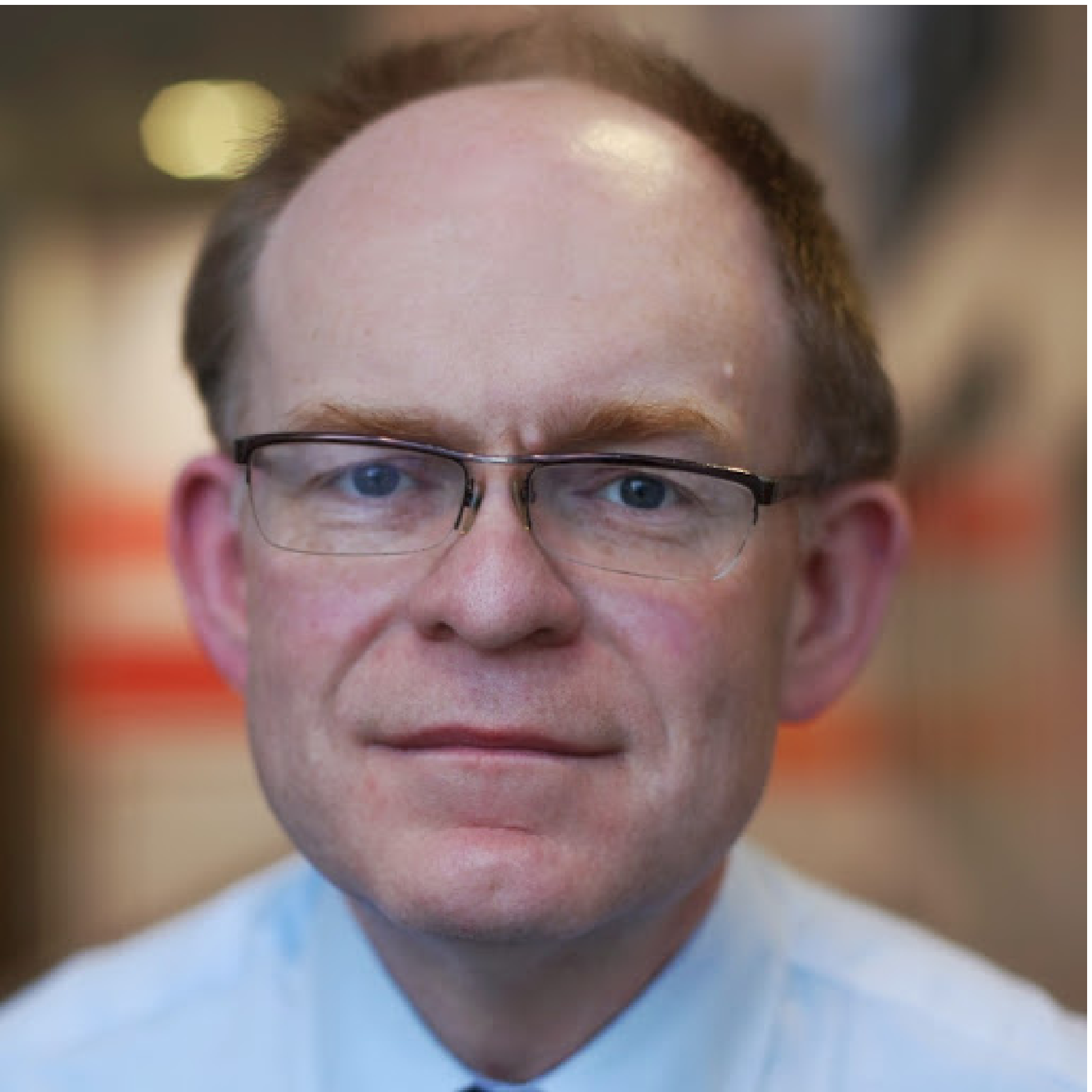}}]{Richard V. Penty} received the Ph.D. degree in engineering from the University of Cambridge, Cambridge, U.K., in 1989, for his research on optical fiber devices for signal processing applications, where he was a Science and Engineering Research Council Information Technology Fellow researching on all optical nonlinearities in waveguide devices. He is currently Professor of Photonics and a Deputy Vice Chancellor at the University of Cambridge, having previously held academic posts with the University of Bath and the University of Bristol. He has authored over 900 refereed journal and conference papers. His research interests include high speed optical communications systems, photonic integration, optical switching, and quantum communication systems. He is a fellow of the Royal Academy of Engineering and the IET.
\end{IEEEbiography}

\vspace{-2.21cm}

\begin{IEEEbiography}
	[{\includegraphics[width=1in,height=1.25in,clip,keepaspectratio]{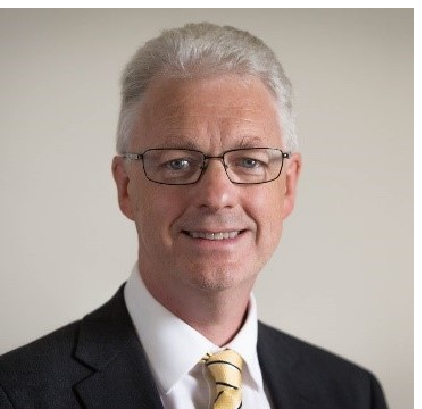}}]{Professor Ian White} joined the University of Bath from the University of Cambridge where he had been Master of Jesus College, van Eck Professor of Engineering, Deputy Vice-Chancellor and Head of Photonics Research in the Electrical Division of the Department of Engineering. 
	Ian gained his BA and PhD degrees from the University of Cambridge (in 1980 and 1984) and was then appointed a Research Fellow and Assistant Lecturer at the University of Cambridge before becoming Professor of Physics at the University of Bath in 1991. 
	In 1996 he moved to the University of Bristol as Professor of Optical Communications and became Head of the Department of Electrical and Electronic Engineering in 1998.  He returned to the University of Cambridge in October 2001 and, in 2005, became Head of the School of Technology and subsequently Chair.  He left the School of Technology to take up the position of Pro-Vice-Chancellor for Institutional Affairs in 2010.   
	Professor White’s research interests are in photonics, including optical data communications and laser diode-based devices. He is a Fellow of the Institute of Electrical and Electronic Engineers (IEEE), the UK Royal Academy of Engineering, and the Institution of Engineering and Technology. He was a Member of the Board of Governors of the IEEE Photonics Society (2008-2012). He is an Editor-in-Chief of Electronics Letters and Nature Microsystems and Nanoengineering.
\end{IEEEbiography}

\vspace{-2.21cm}

\begin{IEEEbiography}
	[{\includegraphics[width=1in,height=1.25in,clip,keepaspectratio]{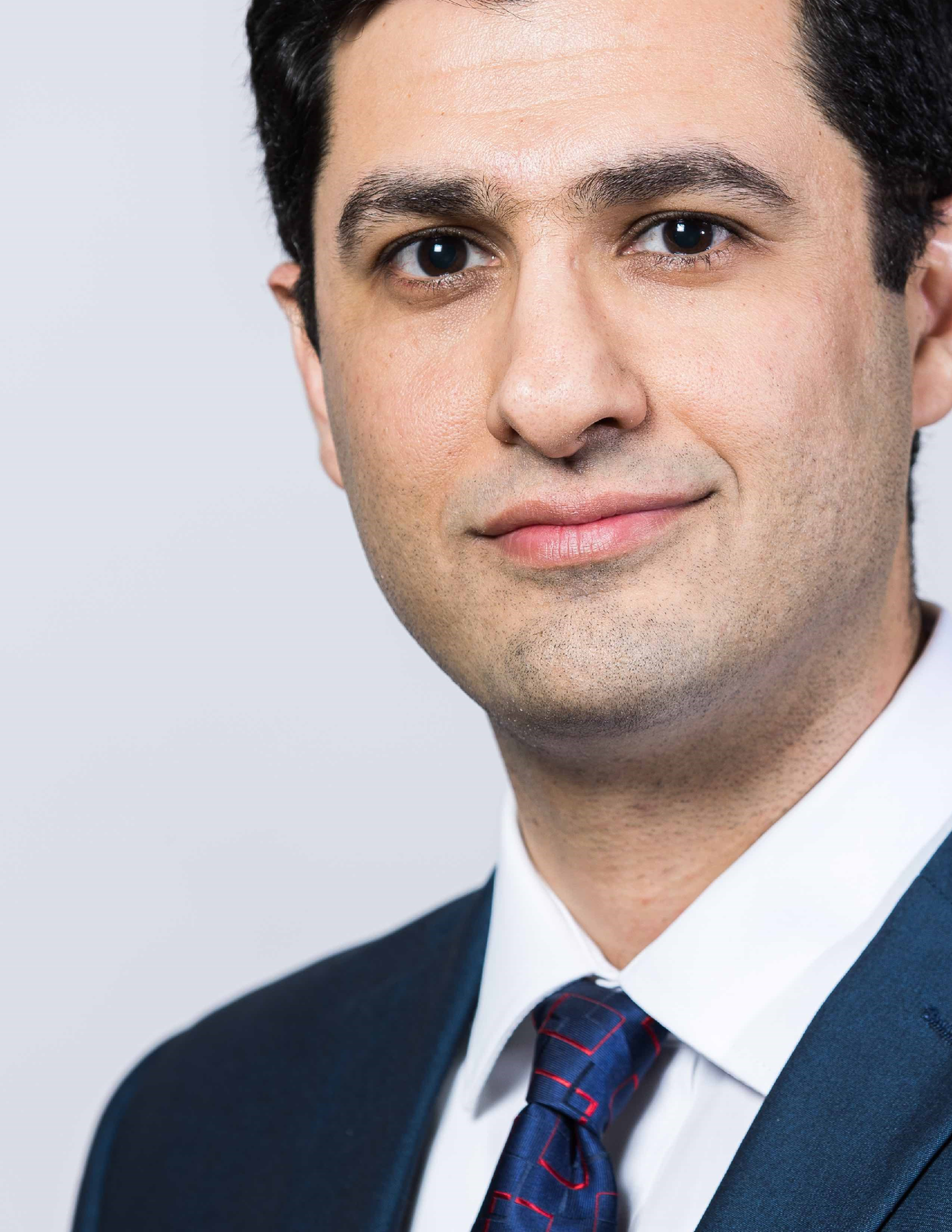}}]{Majid Safari} received his Ph.D. degree in Electrical and Computer Engineering from the University of Waterloo, Canada in 2011. He also received his B.Sc. degree in Electrical and Computer Engineering from the University of Tehran, Iran, in 2003, M.Sc. degree in Electrical Engineering from Sharif University of Technology, Iran, in 2005. He is currently a Reader in the Institute for Digital Communications at the University of Edinburgh. Before joining Edinburgh in 2013, He held postdoctoral fellowship at McMaster University, Canada. Dr. Safari is currently an associate editor of IEEE Transactions on Communications and was the TPC co-chair of the 4th International Workshop on Optical Wireless Communication in 2015. His main research interest is the application of information theory and signal processing in optical communications including fiber-optic communication, free-space optical communication, visible light communication, and quantum communication.
\end{IEEEbiography}

\vspace{-2.21cm}

\begin{IEEEbiography}
	[{\includegraphics[width=1in,height=1.25in,clip,keepaspectratio]{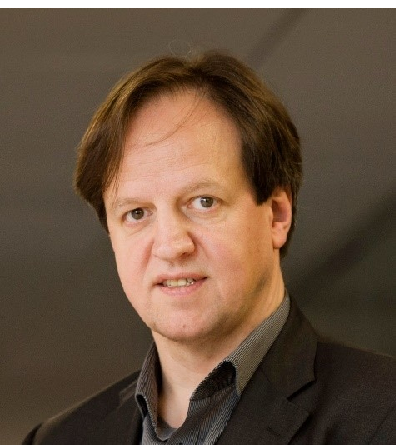}}]{Professor Harald Haas} received the Ph.D. degree from The University of Edinburgh in 2001. He is a Distinguished Professor of Mobile Communications at The University of Strathclyde/Glasgow, Visiting Professor at the University of Edinburgh and the Director of the LiFi Research and Development Centre. Prof Haas set up and co-founded pureLiFi. He currently is the Chief Scientific Officer. He has co-authored more than 600 conference and journal papers. He has been among the Clarivate/Web of Science highly cited researchers between 2017-2021. Haas’ main research interests are in optical wireless communications and spatial modulation which he first introduced in 2006. In 2016, he received the Outstanding Achievement Award from the International Solid State Lighting Alliance. He was the recipient of IEEE Vehicular Society James Evans Avant Garde Award in 2019. In 2017 he received a Royal Society Wolfson Research Merit Award. He was the recipient of the Enginuity The Connect Places Innovation Award in 2021. He is a Fellow of the IEEE, the Royal Academy of Engineering (RAEng), the Royal Society of Edinburgh (RSE) as well as the Institution of Engineering and Technology (IET).
\end{IEEEbiography}

\end{document}